\documentclass[preprint]{aastex}
\usepackage{emulateapj5,times,mathptm}

\begin{document}

\title{Ten Million Degree Gas in M~17 and the Rosette Nebula:
X-ray Flows in Galactic H{\sc II} Regions}

\author{Leisa K. Townsley\altaffilmark{1},  
Eric D. Feigelson\altaffilmark{1}\altaffilmark{2}, Thierry Montmerle\altaffilmark{2}, 
Patrick S. Broos\altaffilmark{1}, You-Hua Chu\altaffilmark{3}, and 
Gordon P. Garmire\altaffilmark{1}}

\altaffiltext{1}{Department of Astronomy \& Astrophysics, 525 Davey
Laboratory, Pennsylvania State University, University Park, PA 16802}
\altaffiltext{2}{Service d'Astrophysique, Centre d'\'Etudes de Saclay,
91191 Gif-sur-Yvette, France}
\altaffiltext{3}{Astronomy Department, University of Illinois at
Urbana-Champaign, 1002 West Green Street, Urbana, IL 61801}

\begin{abstract}
 
We present the first high-spatial-resolution X-ray images of two
high-mass star forming regions, the Omega Nebula (M~17) and the Rosette
Nebula (NGC~2237--2246), obtained with the {\it Chandra X-ray
Observatory} Advanced CCD Imaging Spectrometer (ACIS) instrument.  The
massive clusters powering these H{\sc II} regions are resolved at the
arcsecond level into $>$900 (M~17) and $>$300 (Rosette) stellar sources
similar to those seen in closer young stellar clusters.  However, we
also detect soft diffuse X-ray emission on parsec scales that is
spatially and spectrally distinct from the point source population.
The diffuse emission has luminosity $L_x \simeq 3.4 \times
10^{33}$~ergs~s$^{-1}$ in M~17 with plasma energy components at $kT
\simeq 0.13$ and $\simeq 0.6$~keV (1.5 and 7~MK), while in Rosette it
has $L_x \simeq 6 \times 10^{32}$~ergs~s$^{-1}$ with plasma energy
components at $kT \simeq 0.06$ and $\simeq 0.8$~keV (0.7 and 9~MK).
This extended emission most likely arises from the fast O-star winds
thermalized either by wind-wind collisions or by a termination shock
against the surrounding media.  We establish that only a small portion
of the wind energy and mass appears in the observed diffuse X-ray
plasma; in these blister H{\sc II} regions, we suspect that most
of it flows without cooling into the low-density interstellar
medium.  These data provide compelling observational evidence that
strong wind shocks are present in H{\sc II} regions.

\end{abstract}

\keywords{HII regions $-$ open clusters and associations: individual
(NGC~2244, NGC~6618) $-$ stars: activity $-$ stars: early-type $-$
stars: high mass $-$ stars:  pre$-$main-sequence $-$ X-rays: stars}

\section{Introduction \label{sec:intro}}

One of the great successes in astrophysics during the mid-20th century
was the development of a detailed explanation for H{\sc II} regions and
other bright emission-line nebulae \citep{Menzel37, Stromgren39,
Aller56}.  A hot star with strong ultraviolet radiation produces an
expanding ionization front which spreads into the surrounding medium.
Early studies concentrated on the ionization front, its associated
shock, and various other considerations such as magnetic fields,
turbulence, dynamical instabilities, and radiation pressure on dust
particles \citep{Mathews69}.

These efforts omitted the potential effects of the powerful winds of OB
stars, which were not known until the discovery of ultraviolet P Cyngi
profiles by \citet{Morton67}.  Winds make a relatively small
contribution to the overall energy budget of H{\sc II} regions:  the
wind mechanical luminosity ranges from 0.3\% (B0 V) to 2\% (O4 V) of
the ionizing continuum luminosity \citep{Panagia73, Howarth89}.
However, they can easily dominate the dynamics of the nebula by
evacuating a hot cavity with characteristic temperatures of $T \sim
10^7-10^8$ K, and producing two-fluid shocks which lie within the
ionization shocks in the cooler $T \sim 10^4$ K H{\sc II} region
\citep[e.g.][]{Pikelner68, Dyson72, Weaver77}.  See \citet{Lamers99}
for details of these models and a historical review of stellar wind studies.

In the interstellar bubble model developed by \citet{Weaver77}, a
single $10^6$ yr old O7 star with a terminal wind velocity $v_w =
2000$~km~s$^{-1}$ and mass loss rate $\dot{M} = 1 \times
10^{-6}$~M$_\odot$~yr$^{-1}$, surrounded by a uniform interstellar
medium of density $n_o = 1$~cm$^{-3}$, will have an innermost region
where the wind freely expands and cools, followed by a hot $T \sim
10^6$~K region starting at the wind termination shock 6~pc from the
star and extending to an outer shock at 27~pc.  This X-ray emitting
region occupies most of the volume of the bubble and consists mostly of
evaporated interstellar material rather than wind material.  Outside
this hot region lies a thin shell of swept-up interstellar matter at $T
\simeq 10^4$~K, beyond which lies the much larger standard photoionized
H{\sc II} region.

The fate of OB winds in H{\sc II} regions may not be as simple as
described in the Weaver et al.\ interstellar bubble model.  The
wind-swept bubble can be larger or smaller than the ionized H{\sc II}
region, and nonadiabatic dissipation due to conduction, mixing,
turbulence and magnetic fields at the shock fronts may affect the
properties of the coronal region \citep{Capriotti01}.  Another
possibility is that the winds are slowed by the entrainment of dense
clumps of interstellar matter before the termination shock.  In such
mass-loaded stellar winds, extensively studied by Dyson and his
collaborators \citep[and references therein]{Pittard01}, the flow is
heated to X-ray temperatures closer to the star and the outer shock is
weaker.  Finally, it is possible that no X-ray emission will emerge at
all if the surrounding medium is too fragmented \citep{McKee84} or wind
energy is slowly dissipated in a thick turbulent mixing layer
\citep{Kahn90}. 

Despite three decades of theoretical discussions of hot cavities
produced by OB stellar winds, only a few detailed predictions of
observable X-ray properties of the shocked wind have been made.
\citet{Dorland87} investigated the dissipation of a $v_w =
1000$~km~s$^{-1}$ O star wind, including conduction processes,
predicting an X-ray to wind kinetic luminosity ratio $L_x/L_w \sim 10^{-5}$
and X-ray temperatures around 2~keV (23~MK) with a nonthermal hard-energy
tail.  \citet{Comeron97} presents a two-dimensional hydrodynamical
model of an O star residing near the edge of an interstellar cloud,
showing the evolution of X-ray morphology as the wind cavity produces a
blow-out into the diffuse interstellar medium.  \citet{Strickland99}
give a hydrodynamical calculation for a very young ($t \sim 10^4$~yr) O
star completely embedded in a dense cloud and find that, while most of
the emission from the wind termination shock is in the EUV band, the
wind-swept interior emits X-rays up to several keV in temperature.
\citet{Canto00} present a 3-dimensional hydrodynamical calculation of a
rich ensemble of O stars allowing for mass loading, wind collisions, and
shocks, but without an encounter with a surrounding cloud.  This
results in an inhomogeneous X-ray structure with $L_x \simeq 6 \times
10^{35}$~ergs~s$^{-1}$ and $T \simeq 2 \times 10^7$~K concentrated in
the inner $r \simeq 0.3$~pc of the stellar cluster.  

On the observational side, until very recently there had been no
unambiguous report of diffuse X-ray emission from H{\sc II} regions
although it has been detected around some isolated Wolf-Rayet stars,
within planetary nebulae, and in starburst superbubbles \citep[see
reviews by][]{MacLow00, Chu02a, Chu02b, Chu03}.  Using the {\it
Einstein Observatory}, \citet{Seward82} found extended soft X-rays over
tens of parsecs in the Carina Nebula with $L_x \simeq 2 \times
10^{35}$~ergs~s$^{-1}$, in addition to a score of individual OB/W-R
stellar sources.  However, this is a large and old star formation
region with many supergiants, so it is possible that this extended
emission is dominated by past supernova remnants.  \citet{Leahy85}
suggested a diffuse origin for emission within the Rosette Nebula seen
with {\it Einstein}, further discussed by \citet{Dorland87}, but a
later higher-resolution {\it ROSAT} study concluded that this
unresolved component was due to the integrated emission from many lower
mass pre-main sequence stars \citep{Berghofer02}.  In an {\it Einstein}
survey of LMC H{\sc II} regions, \citet{Chu90} found several with X-ray
emission at levels around $10^{35}-10^{36}$~ergs~s$^{-1}$, an order of
magnitude higher than expected from the wind shocked bubble, which they
attribute to off-center supernova remnants hitting the outer shells. 

The difficulties encountered in past efforts to detect the $T \sim
10^6-10^7$ K gas in H{\sc II} region wind-blown bubbles arise from the
complex environment in the X-ray band.  A high-mass star forming region
$1-10$~pc in extent will have thousands of lower-mass pre-main sequence
stars each emitting $10^{28}-10^{31}$~ergs~s$^{-1}$, individual OB and
Wolf-Rayet stars each emitting $10^{30}-10^{34}$~ergs~s$^{-1}$, possibly
supernova remnants and superbubbles emitting
$10^{34}-10^{37}$~ergs~s$^{-1}$, and possibly X-ray binary systems from
past episodes of star formation each emitting
$10^{33}-10^{40}$~ergs~s$^{-1}$ \citep{Feigelson01}.  These
observational difficulties are largely overcome with the {\it Chandra
X-ray Observatory}, where the mirrors give unprecedented
subarcsecond spatial resolution and the ACIS instrument has the very
low background necessary to study faint diffuse emission.  {\it
Chandra} thus provides the best opportunity to date to resolve the
extended, low surface brightness wind-blown bubbles from the plethora
of compact stellar X-ray sources expected in high-mass star forming
regions.  Several reports of wind-generated diffuse X-ray emission from
H{\sc II} regions are thus emerging from recent {\it Chandra} observations.

We report here on {\it Chandra} studies of two bright, nearby ($d \sim
1.5$~kpc), and well-studied H{\sc II} regions: M~17 and the Rosette
Nebula.  In M~17, we show that the {\it Chandra} image reveals strong
diffuse X-rays both within and expanding asymmetrically from the
stellar cluster, in addition to several hundred young stars.
Preliminary results for M~17 were presented in \citet{Townsley03}.  In
the Rosette, we see relatively faint X-ray emission suffusing the
hollow core of the H{\sc II} gas, together with a rich stellar
population\footnote{See the press release at 
\url{http://chandra.harvard.edu/press/01\_releases/press\_090601wind.html}.}.
These observations, and the astrophysical analysis derived from them,
represent the strongest observational evidence to date for X-ray
emission from the wind-blown cavities of H{\sc II} regions; the
morphological and spectral properties of the hot plasma should
constrain some of the theoretical uncertainties concerning the fate of
OB winds and the resulting evolution of H{\sc II} regions.

The paper starts with an overview of each H{\sc II} region
(\S\ref{review.sec}) and then presents our observations and results
from data analysis (\S\ref{observations.sec}). The complex morphologies
of the X-ray images are reviewed in \S\ref{morph.sec} and the physical
nature of the emitting gas is derived in \S\ref{diffuse.sec}.  Several
possible origins of the diffuse emission are considered in
\S\ref{origins.sec} and the wind shock model is discussed in more
detail in \S\ref{winds_elab.sec}.  The two regions studied here are
placed into a wider context in \S\ref{hmsfr.sec}. The paper ends with
general comments about the fate of O star winds in H{\sc II} regions
(\S\ref{conclusions.sec}).

\section{Observational Background on the H{\sc II} Regions \label{review.sec}}

\subsection{M~17 \label{review_M17.sec}}

The H{\sc II} region M~17 (= Omega Nebula = W~38, at
$(l,b)=(15.1,-0.7)$) is visually very bright and has been extensively
studied for two centuries.  It is illuminated by the massive stellar
cluster NGC~6618, the center of which consists of a ring of seven O stars
$\sim 1\arcmin$ in diameter located behind substantial obscuring
material, with $A_V \sim 8$~mag \citep{Hanson97}.  The cluster core
radius is $\sim 5.5\arcmin$ (2.6~pc) and its outer radius is $\sim
12.5\arcmin$ (5.8~pc) \citep{Lynga87}.  The earliest O stars are an
O4--O4 binary called Kleinmann's Anonymous Star \citep{Kleinmann73}.
The nebula is the second brightest thermal radio source in the sky
(after Orion A) and is situated at the edge of a massive and dense
molecular cloud; its appearance varies significantly as a function of
wavelength because of spatially complex extinction.  

\begin{figure}
\centering
\vspace*{-0.3in}
\includegraphics[width=0.35\textwidth]{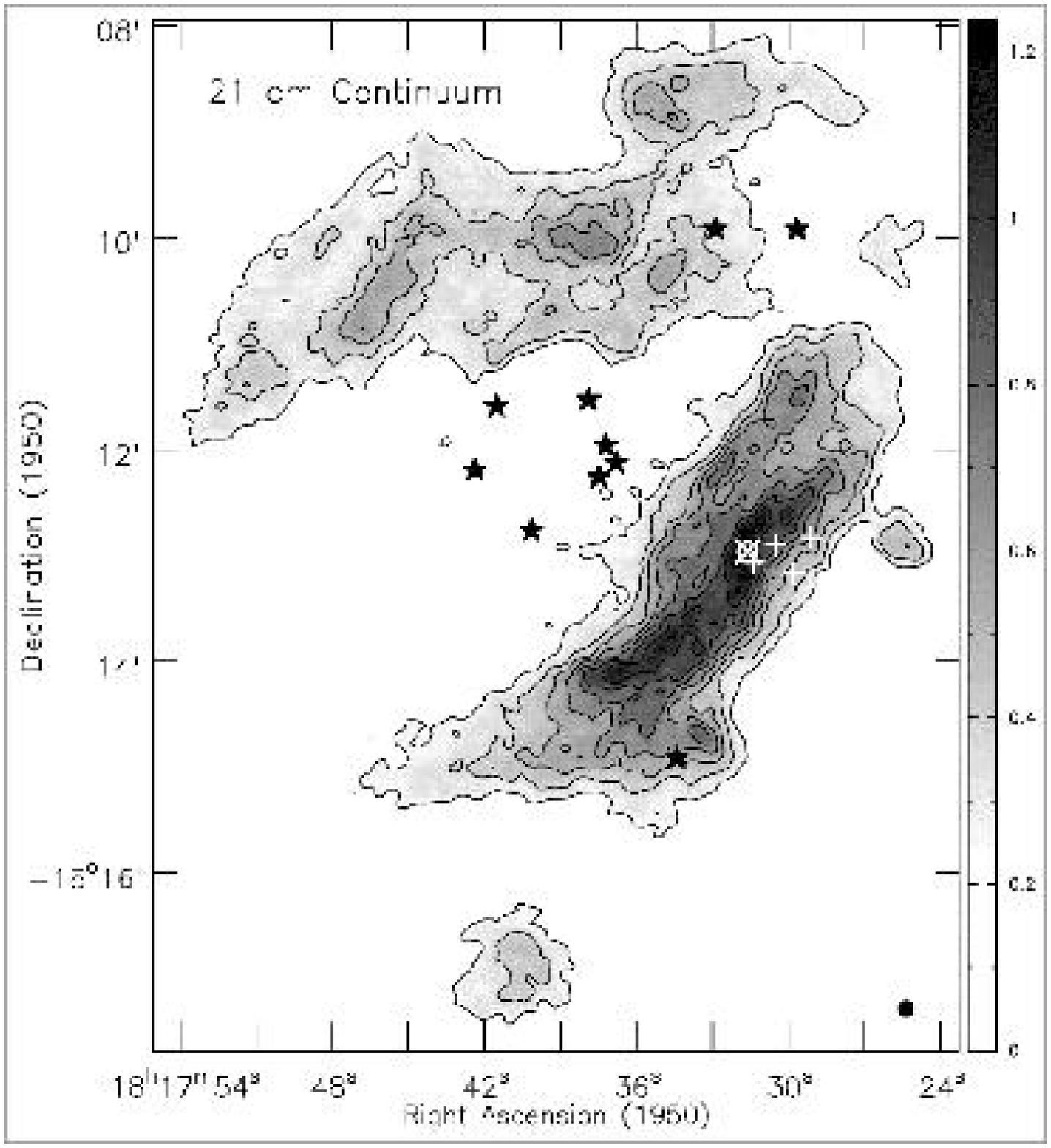}
\hspace{0.6in}
\includegraphics[width=0.45\textwidth]{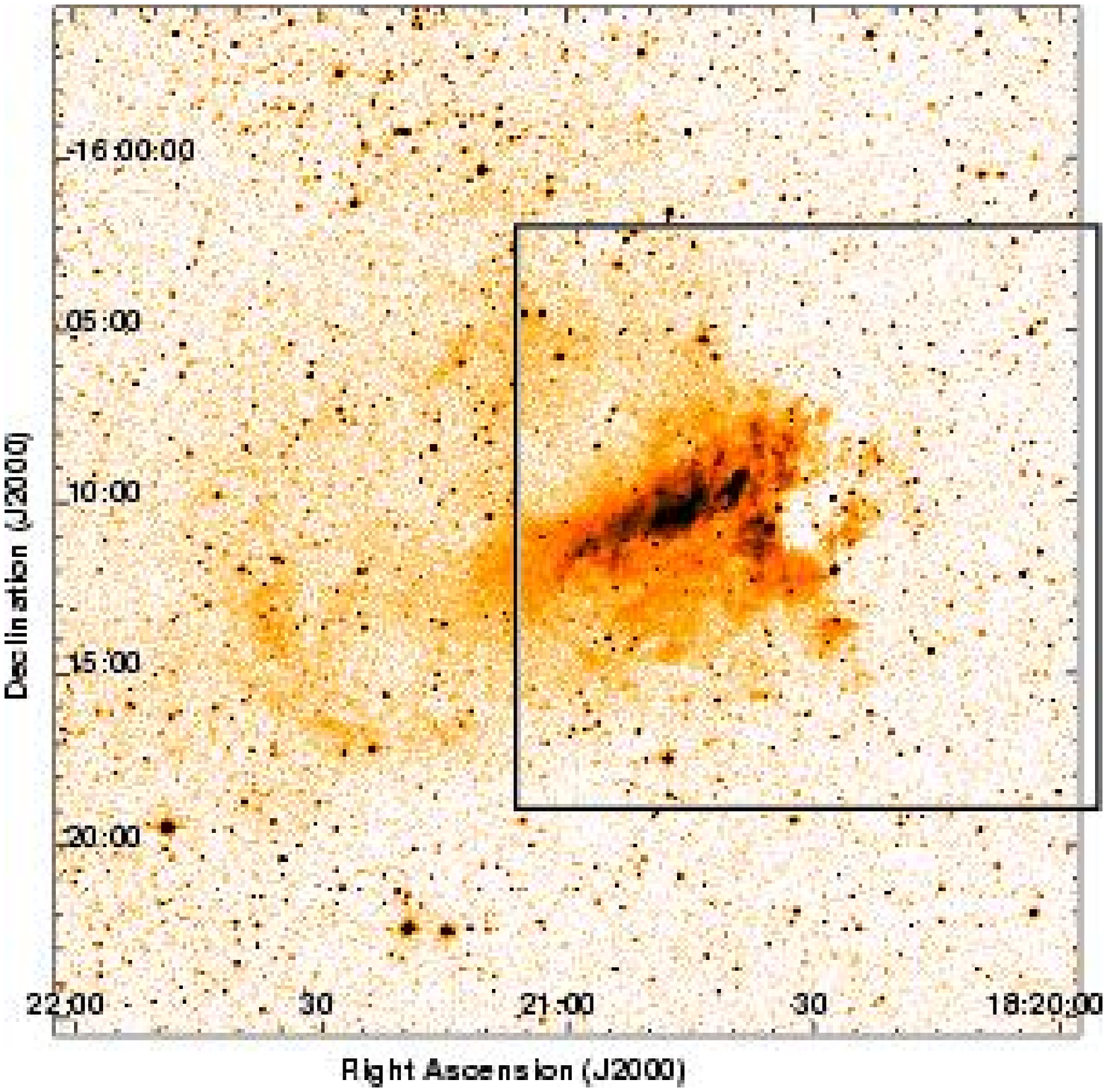}
\hspace*{-0.3in}
\includegraphics[width=0.40\textwidth]{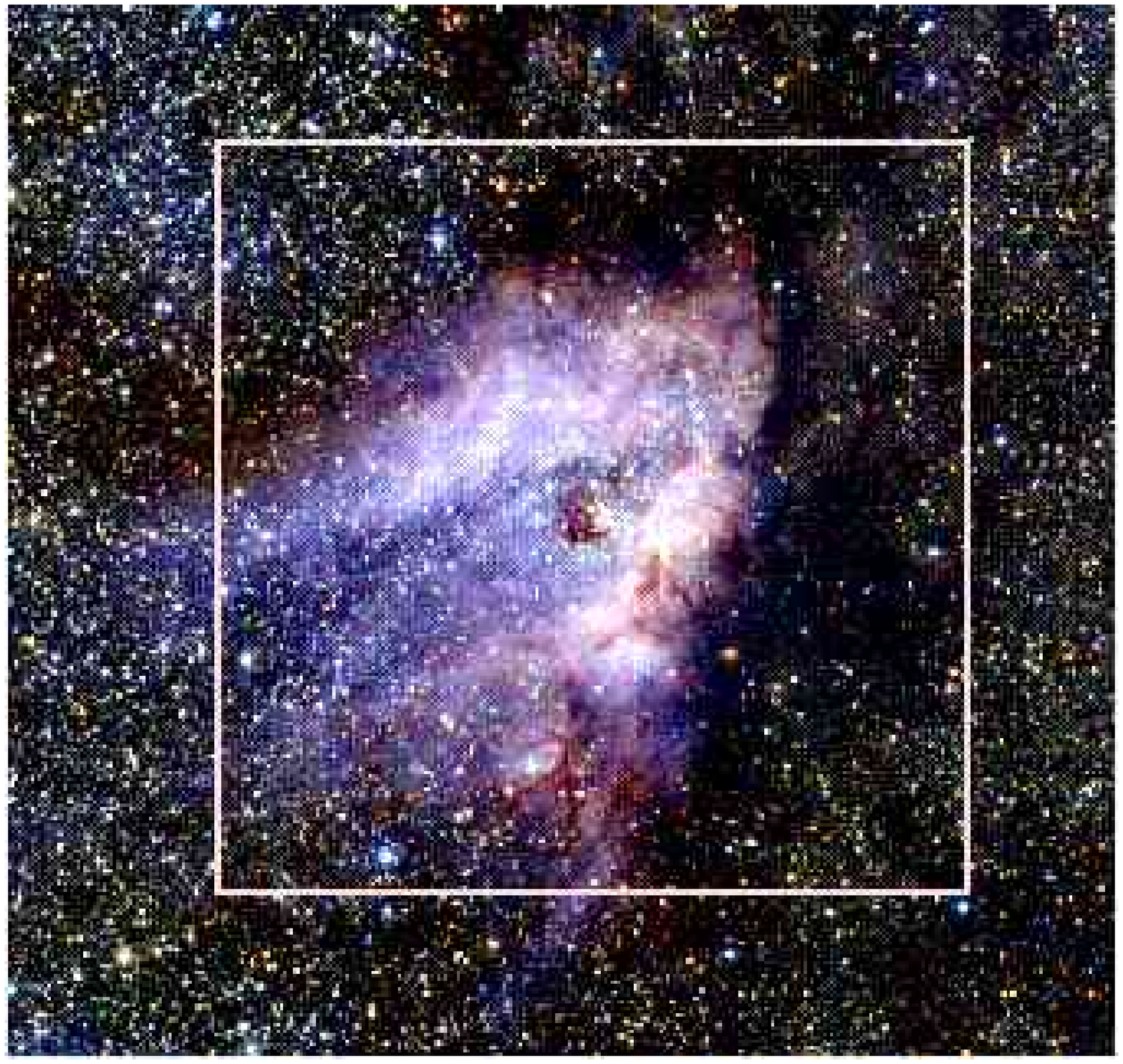}
\hspace{0.6in}
\includegraphics[width=0.45\textwidth]{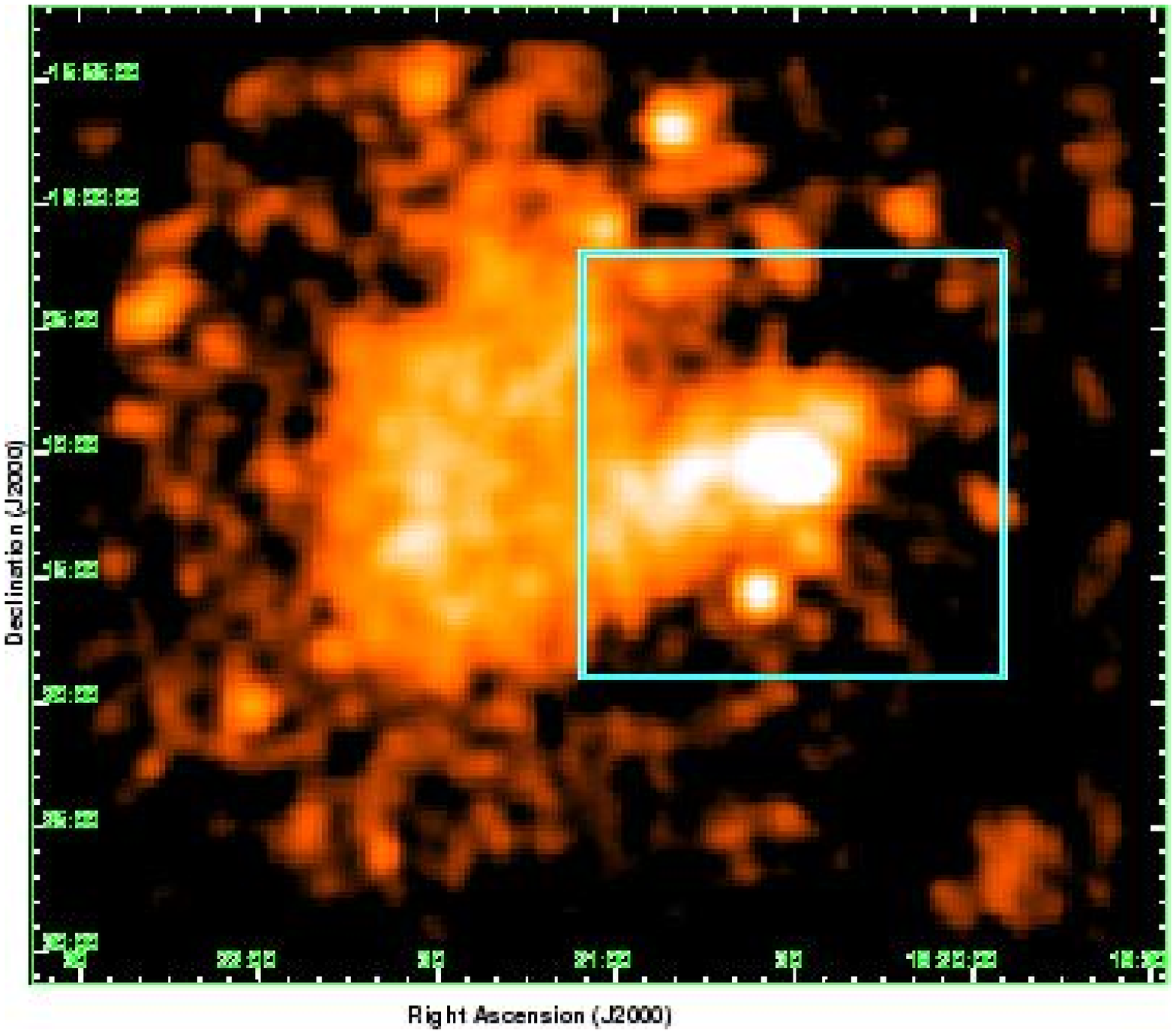}
\caption{Multiwavelength views of M~17: 
(a) Radio continuum map showing H{\sc II} gas with respect to the
brightest O stars (filled stars), the embedded ultracompact H{\sc II}
region (circle) and nearby molecular masers (plusses and cross), from
\citet{Brogan01}. 
(b) Digitized Sky Survey image showing the stellar field and H$\alpha$
emission from the H{\sc II} region, with the approximate location of
the $\sim 17\arcmin \times 17\arcmin$ ACIS-I array in our observation
outlined in black.
(c) $JHK$ image of the central region (roughly $23\arcmin \times
25\arcmin$) from the 2MASS survey, showing the NGC~6618 stellar cluster
and ionized gas surrounded by the dark molecular cloud (Atlas Image
courtesy of 2MASS/UMass/IPAC-Caltech/NASA/NSF); again the approximate
location of the ACIS-I array is overlaid.
(d) {\it ROSAT} PSPC X-ray image of M~17 and vicinity, with the
approximate location of the ACIS-I array outlined in blue. 
In these and
all following images, North is up and East to the left unless otherwise
noted.
\label{fig:m17_comp}}
\end{figure}

A global geometric model of the M~17 complex is presented by
\citet[][their Figure 4]{Tsivilev99}; they note its similarity to
the pressure-driven bubble model of \citet{Clayton85}, who in turn note
that the highly asymmetric bubble they see agrees with the models of
\citet{Dyson77} and \citet{Rozyczka85} for the formation of a
wind-blown bubble in the density gradient present at the edge of a
molecular cloud.  \citet{Tsivilev99} describe a bowl-shaped region
$\sim 2$~pc across and expanding westward into the molecular cloud,
with a much larger, unobscured optical H{\sc II} region spreading into
the low-density medium at the eastern edge of the cloud.  The complex
resembles the Orion Nebula/KL region seen nearly edge-on
\citep{Meixner92}: the bowl-shaped ionization blister is eroding the
edge of the clumpy molecular cloud and triggering massive star
formation, as evidenced by an ultra-compact H{\sc II} region and luminous
protostars.  Velocity studies show an ionized shell in the middle of
the nebula with a diameter of 8~pc ($\sim 17\arcmin$)
\citep{Clayton85}.  The ionization front encounters the molecular cloud
at two photodissociation regions, known as the northern and southern
bars, about $20\arcmin$ in length and at an apparent angle of $\sim
45^\circ$ to each other.  These dominate the radio continuum and
emission line images of the region (Figure~\ref{fig:m17_comp}a).
The bowl is oriented somewhat past edge-on to our line-of-sight, with
the blow-out moving away from us.

In the radio band unaffected by extinction, low resolution and low
frequency studies with the Very Large Array show an optically thick,
bright region associated with bright H$\alpha$ emission truncated
abruptly to the west.  This is embedded within an optically thin region
extending over $30\arcmin \times 40\arcmin$ ($\sim 14 \times 19$~pc)
which represents the full extent of ionized material
\citep{Subrahmanyan96}.  The electron temperature is $T_e \sim
8000$~K.  At higher resolution and frequencies, where the H{\sc II} gas
is optically thin, the emission is dominated by the bars
\citep[Figure~\ref{fig:m17_comp}a;][]{Felli84, Brogan01}.  The gas in
the bars has electron density $n_e \sim 10^3$~cm$^{-3}$ but is not
uniform, with clumps of higher density.  The Lyman continuum flux
necessary to excite the bars is consistent with the ultraviolet
emission from the O4--B0 stellar cluster, confirming a simple
photoionization origin for the radio-emitting gas.

In the visual band, the {\it Digital Sky Survey} (DSS) red image is
dominated by H$\alpha$ nebulosity from the northern bar, but emission
from the southern bar and ionizing cluster is largely absorbed
(Figure~\ref{fig:m17_comp}b).  At low surface brightness, the DSS image
shows H$\alpha$ extending over $\sim 20\arcmin \times 20\arcmin$ with
features closely correlated with the low frequency radio structures
where extinction is low; particularly noticeable are a few large
arc-shaped features.  Optical spectroscopy has shown high-excitation
optical emission lines rarely seen in other H{\sc II} regions
\citep{Glushkov98}.  Near-infrared images suffer less obscuration and
clearly display line emission from the bars seen in the radio, as well
as the exciting stellar cluster (Figure~\ref{fig:m17_comp}c).  Recent
deep near-infrared observations reveal a circular cavity in the gas
around the exciting cluster as well as many cluster members and
embedded young stars in the surrounding dark cloud \citep{Ando02,
Jiang02}.

In the X-ray band, M~17 was observed with {\it ROSAT} for 6.7~ks; the
findings have just recently been published \citep[][hereafter
Dunne03]{Dunne03}.  We show the {\it ROSAT} proportional counter (PSPC)
`full band' (0.1--2.4~keV) image, slightly smoothed with a Gaussian
kernel, in Figure~\ref{fig:m17_comp}d; contours from this exposure are
superposed on a new H$\alpha$ image in Dunne03 (their Figure 1).  The
X-ray image shows strong emission around the core of the ionizing
cluster and an elongated structure extending $20\arcmin$ to the
east-southeast, parallel to the northern bar and filling the optically
thick part of the DSS image.  Extended, fainter emission matching the
diffuse H$\alpha$ emission is visible towards the east and north; no
limb-brightening is seen.  Dunne03 found $kT = 0.66$--0.78~keV ($T =
7.7$--9.1~MK) and $N_H = 1$--$5 \times 10^{20}$~cm$^{-2}$ for this
emission, with electron densities $n_e \sim 0.06$--$0.09$~cm$^{-3}$ and
a total luminosity (in the 0.1--2.4~keV band) of $L_x \sim 2 \times
10^{33}$~ergs~s$^{-1}$.  For the area of bright X-ray emission close to
the ionizing cluster, the {\it ROSAT} spectrum was not well-fit, but
Dunne03 estimate $L_x \sim 1 \times 10^{33}$~ergs~s$^{-1}$ for this
region.  

Dunne03 note that the region is too young to have produced a supernova
remnant and interpret the X-ray emission as hot gas filling a
superbubble blown by the OB star winds, but note that the {\it ROSAT}
X-ray luminosity for this region is three orders of magnitude lower
than the predictions of standard wind-blown bubble models.  They
conclude that the discrepancy may be due in part to the inclusion of
heat conduction across the interface between the hot X-ray emitting
region and the cool swept-up shell in those models, whereas the strong
magnetic field in M~17 \citep{Brogan01} may suppress this heat
conduction.  They thus consider a bubble without heat conduction, but
can only reproduce the observations by assuming substantial mixing of
cold nebular gas with the shocked stellar winds; they further require
the stellar winds to be clumpy \citep{Moffat94} and have reduced mass
loss rates in order to get reasonable agreement between the models and
their observations.

An observation with the {\it ASCA} satellite, which has a broad point
spread function, shows emission coincident with the cluster core with
0.5--10~keV luminosity $L_x \sim 10^{33-34}$~ergs~s$^{-1}$, with high
plasma energy ($kT > 3$~keV, or $T > 35$~MK) and absorption ($N_H =
1$--$5 \times 10^{22}$~cm$^{-2}$) \citep{Matsuzaki99}.  These authors
interpreted the emission as a point source associated with the single
O4--O4 binary system, but we establish here that their spectrum is
comprised of emission from many hundreds of cluster members.

On the western side, all tracers of warm and hot gas are truncated by a
wall of dense, cold molecular material which includes the dense cores
known as M17 Southwest and M17 North; these exhibit many other tracers
of current massive star formation.  These include: the Kleinmann-Wright
object, an embedded $L_{bol} \simeq 2600$~L$_\odot$ Herbig B2e star;
M17-UC1, an extremely small (diameter $\sim 1000$~AU) and dense
ultra-compact region powered by a very young massive binary system with
a luminosity of several thousand L$_\odot$; the even more luminous, but
radio-faint, protostar IRS 5; and a deeply embedded millimeter-infrared
source \citep{Johnson98, Henning98, Chini00, Nielbock01, Kassis02}.
There is a strong suspicion that shocks from the blister H{\sc II}
region have compressed the cloud and triggered this star formation
activity.  The exact structural relationships between the cold and warm
gaseous components are not certain;  for example, neutral-hydrogen
clumps may be mixed with H{\sc II} gas on the edge of the molecular
cloud \citep{Clayton85}.

Only the most massive members of the young NGC~6618 stellar cluster
exciting the nebula have been characterized, due to the comparatively
high extinction.  Near-infrared imagery and spectroscopy reveal an
embedded cluster of about 100 stars earlier than B9 \citep{Lada91,
Hanson97}; see Table~\ref{tab:m17_OB} in Appendix A for a list of O--B3
stars.  These studies did not cover the entire cluster, so even more
early stars may be present.  This is substantially richer than the
Orion Nebula Cluster which has only 8 stars between O6 and B9
\citep{Hillenbrand97}.  \citet{Hanson97} report the age of the cluster,
obtained from isochronal fitting, to be $\sim 1$~Myr or younger.  Many
of the stars in M~17 show infrared excesses attributed to disks, and a
considerable fraction are close binaries \citep{Garcia01}.  The
extinction is patchy, ranging from $A_V = 3$ to 15 with an average of 8
magnitudes.  From main sequence fitting of the OB stars, the distance
to M~17 has been measured to be $1.3^{+0.4}_{-0.2}$~kpc
\citep{Hanson97} and $1.6 \pm 0.3$~kpc \citep{Nielbock01},
significantly closer that the traditional value of $\simeq 2.2$~kpc
based on the Galactic rotation curve.  We adopt here a distance of
1.6~kpc.

In summary, prior to the present study, M~17 was known to have four
major morphological components:  (1) An inner, bright photoionized
nebula with a hollow, conical shape, with gas concentrated in the
northern and southern bars.  The gas in the bars, photoexcited by the
early OB stars, has $T_e \sim 8000$~K and $n_e \sim 10^3$~cm$^{-3}$.
(2) An outer, fainter, optically thin ionized region, also with
extended soft X-ray emission seen by {\it ROSAT}, with $T_x \sim 8$~MK
and $n_e \sim 0.1$~cm$^{-3}$.  This outer region lies largely outside
the field of view of our {\it Chandra} observation.  (3) Dense, cold
molecular gas surrounding the two bars with active star formation,
including new rich stellar clusters.  (4) The open cluster NGC~6618
which excites the nebula.

\subsection{Rosette \label{review_Rosette.sec}}

The Rosette Nebula (= NGC~2237--2246, at $(l,b)=(206.4,-2.0)$) is
powered by the open cluster NGC~2244, the youngest cluster within the
larger Mon~OB2 association.  It is a blister H{\sc II} region lying on
the edge of the giant Rosette Molecular Cloud (RMC) which in turn lies
within a larger H~I cloud (see \citet{Perez91} for a review of the
large-scale star formation environment).  Due to the orientation with
respect to the line of sight, the extinction to the stars and nebula is
lower than for M~17, generally only $A_V \sim 1$--2~mag \citep{Celnik86}.

\begin{figure}
\includegraphics[width=0.45\textwidth]{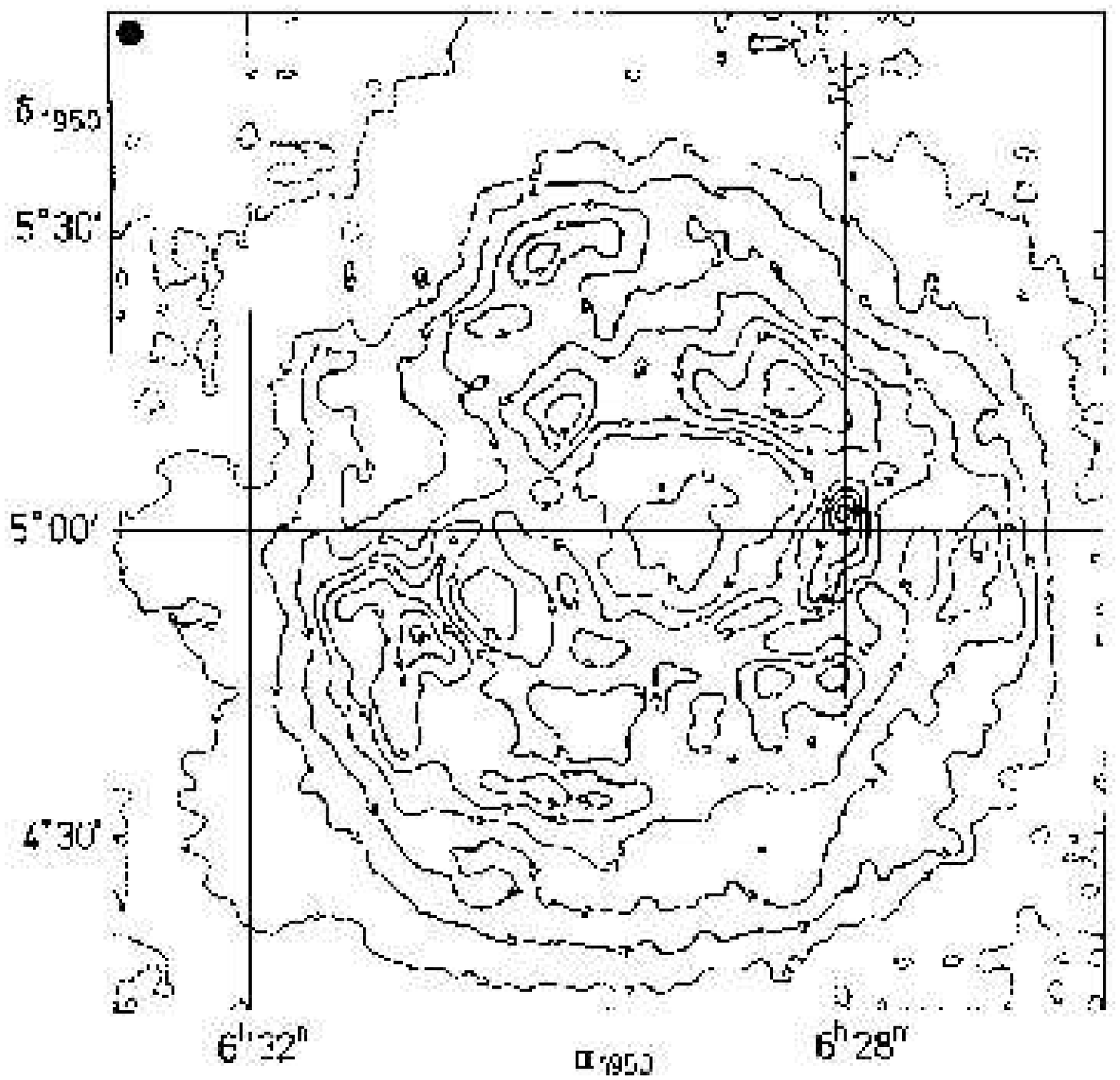}
\hspace{0.25in}
\includegraphics[width=0.45\textwidth]{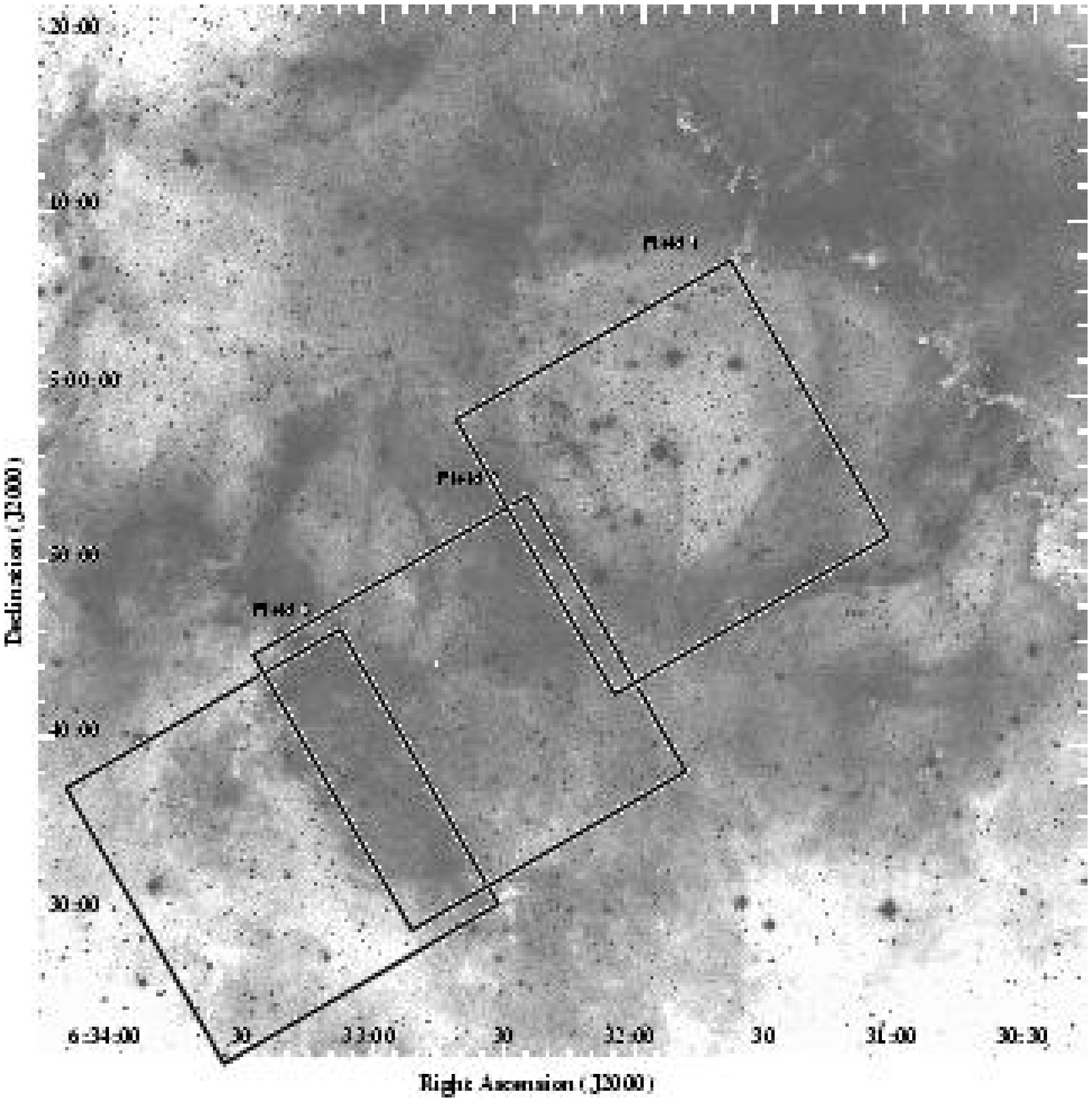}
\includegraphics[width=0.45\textwidth]{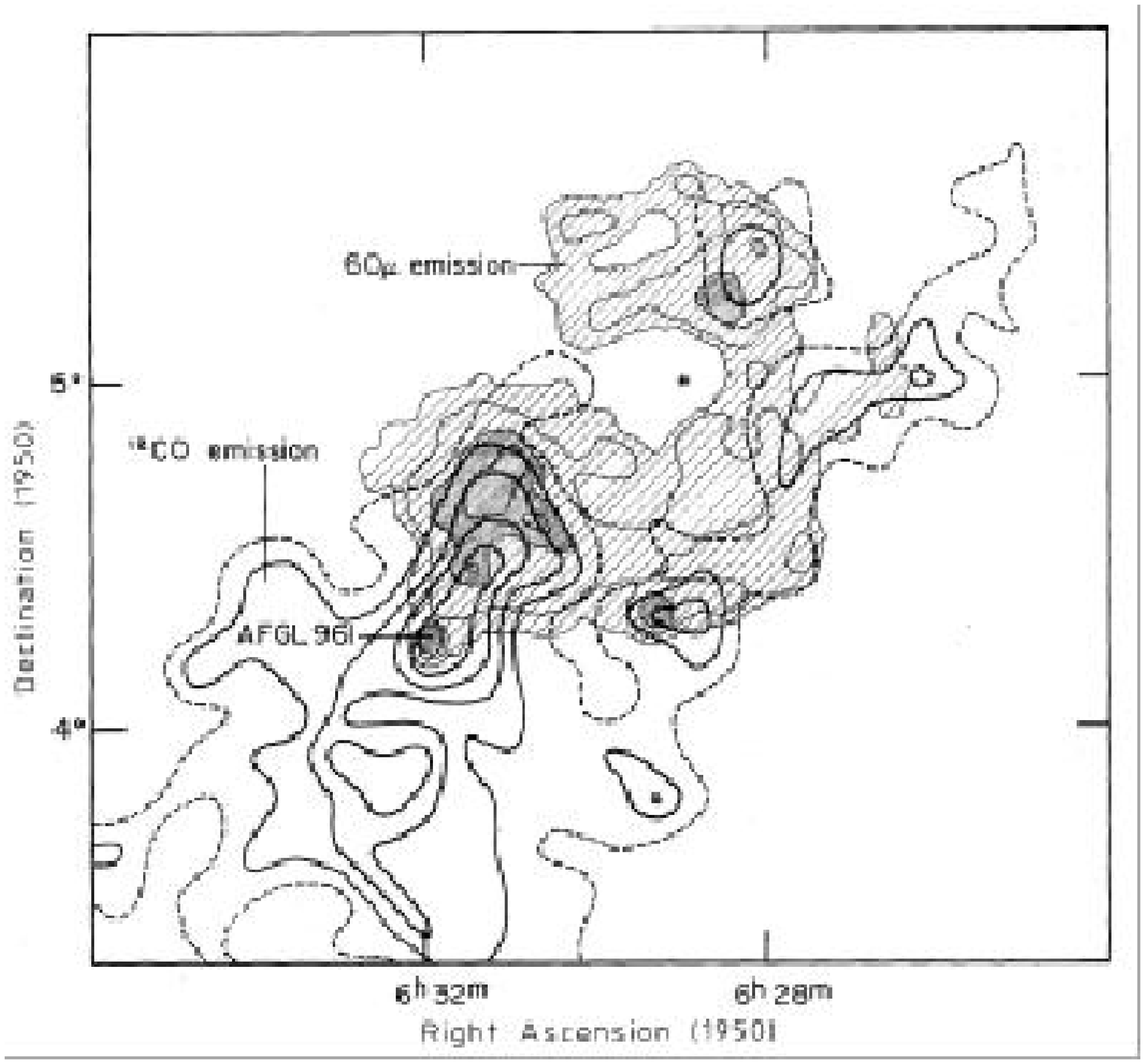}
\hspace{0.5in}
\includegraphics[width=0.45\textwidth]{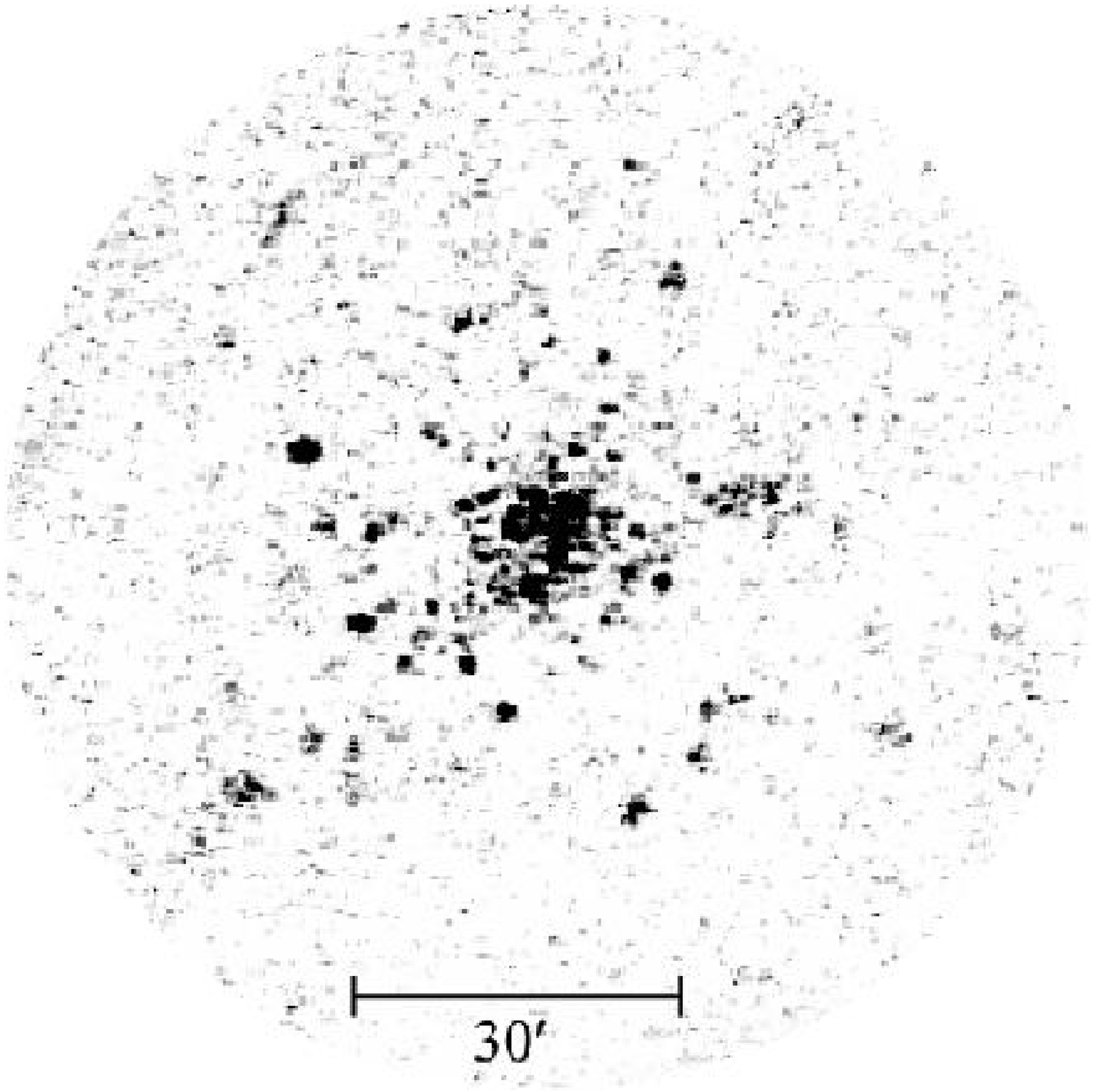}
\caption{\footnotesize Multiwavelength views of the Rosette Nebula and Rosette
Molecular Cloud (RMC):  
(a) Radio continuum map showing H{\sc II} gas, from \citet{Celnik85}; 
(b) Digitized Sky Survey image of the Rosette Nebula with overlays
showing the approximate locations of the first three ACIS-I fields; 
(c) contour map of molecular CO in the RMC with 60$\mu$m {\it IRAS} emission
superposed and the central cluster marked by an asterisk, from \citet{Cox90}; 
(d) {\it ROSAT} PSPC X-ray image of the open cluster NGC~2244, from \citet{Berghofer02}.
\label{fig:rosette_comp}}
\end{figure}

The Rosette is morphologically similar to M~17 in that the nebular
emission is edge-brightened, though in the case of Rosette it resembles
a torus (in projection) rather than two bars.  At both radio and visual
wavelengths (Figures~\ref{fig:rosette_comp}a and b), the Rosette H{\sc
II} region appears as a thick circular ring of ionized gas with an
outer diameter $\sim 80\arcmin$ ($\sim 33$~pc) and an inner hole $\sim
30\arcmin$ ($\sim 12$~pc) in extent \citep{Celnik85}.  The {\it IRAS} data
show a similar annular morphology extending beyond the H{\sc II} region
into the photodissociation region and surrounding molecular cloud 
(see Figure~\ref{fig:rosette_comp}c, from \citet{Cox90}).  Researchers have
variously suggested that this annular or `hollow' H{\sc II} morphology
arises from evacuation by the OB stellar winds or by the depletion of
gas into newly-formed stars \citep{Mathews66, Fountain79, Leahy85,
Dorland86, McCullough00}. The shape of the nebula is generally
considered to be spherical, but it may equally well be cylindrical.

The H{\sc II} region was modeled by \citet{Celnik85} to explain the
radial distribution of brightness temperature derived from radio
continuum observations.  In these data, $T_e = 5800$~K and showed no
radial gradient.  The favored model consists of a central spherical
cavity surrounded by nested spherical shells of varying density
(maximum $n_e \sim 12$~cm$^{-3}$), with a conical hole penetrating the
shells and oriented approximately along the line of sight.  This model
is further developed by \citet{Celnik86} into a three-dimensional model
of the entire RMC and Rosette Nebula complex.  In this model, the H{\sc
II} region is located on the near side of the molecular cloud complex;
it and the molecular clouds are rotating about the center of the
complex.  The conical hole penetrating the shells of the H{\sc II}
region is oriented roughly $30^{\circ}$ west of the line of sight,
toward the nearest edge of the molecular cloud.

In a recent paper, \citet{Tsivilev02} describe radio recombination line
observations of Rosette, giving $T_e = 7980 \pm 580$~K.  They build a
model of the region, using a constant, average $T_e = 6400$~K and
assuming that the region consists of concentric shells, similar to the
\citet{Celnik85} model.  The main new feature of the \citet{Tsivilev02}
model is the inclusion of dense clumps of material in the shells.  To
match the data, their model required two shells:  an outer one with
uniform density $n_e = 15$~cm$^{-3}$, similar to the \citet{Celnik85}
model, and an inner one with high-density clumps ($n_e \sim
10^3$--$10^4$~cm$^{-3}$).

The RMC, extending a degree to the east (see
Figure~\ref{fig:rosette_comp}c), is a clumpy, elongated giant molecular
cloud with no evidence for cloud compression or star formation close to
the H{\sc II} region \citep{Williams95}.  The densest parts of the
cloud are populated with embedded near-infrared star clusters without
massive ($L_{IR} \sim 10^3$ L$_\odot$) members \citep{Phelps97,
Schneider98}, although massive young B stars such as AFGL~961 are also
seen in the {\it IRAS} data \citep{Cox90}.

The NGC~2244 cluster exciting the nebula has a population intermediate
between that of M~17 and the Orion Nebula Cluster, with 31 stars
between O4 V and B3 V (Table~\ref{tab:rosette_OB} in Appendix A).  It
has a core radius of $\sim 2$~pc ($\sim 4.9\arcmin$), an outer radius
of $\sim 8$~pc ($\sim 19.5\arcmin$), and low stellar density, thus it
shows substantial contamination from field stars \citep{Nilakshi02}.
Curiously, the earliest exciting star (HD~46223, O4 V((f)), see
Table~\ref{tab:rosette_OB}) does not lie near the center of the nebula,
but rather near the southeast edge of the ring.  The other early-O star
(HD~46150, O5 V((f)) ) is located near the center of the nebula.  The
main sequence turn-off age of the cluster is about 1.9~Myr
\citep{Park02}, which agrees with the age inferred from the eclipsing
binary V578~Mon \citep{Hensberge00}.  This young age, together with the
absence of nonthermal radio emission, indicates that no supernova
explosion has occurred in the cluster (this point is discussed further
in \S\ref{not_snr.sec}).   The most accurate distance to NGC~2244, $d
= 1.39 \pm 0.1$~kpc, is derived from modeling V578~Mon
\citep{Hensberge00}.  We adopt a distance of 1.4~kpc here.

As mentioned above, an {\it Einstein Observatory} Imaging Proportional
Counter observation of NGC~2244 revealed several individual O stars and
extended X-ray emission from the nebula interior, with $kT \sim 2$~keV
($T \sim 23$~MK), $n_e \sim 0.02$~cm$^{-3}$, and luminosity $L_x = 6
\times 10^{32}$~ergs~s$^{-1}$ in the soft (0.5--3~keV) band
\citep{Leahy85}.  Leahy attributes this to diffuse emission from the OB
winds, but the low resolution provided by this instrument could not
determine the fraction of the emission from the many magnetically
active lower mass pre-main sequence stars expected to accompany the
massive stars.  Indeed, a later higher-resolution {\it ROSAT} study by
\citet{Berghofer02} (see Figure~\ref{fig:rosette_comp}d) attributed all
of the emission to point sources.  This {\it ROSAT} study (see also the
closely related work of \citet{Li02} and \citet{Park02}) identified 128
X-ray emitting pre-main sequence stars in addition to 10 OB stars.  A
separate {\it ROSAT} observation of the RMC revealed dozens of faint,
partially resolved embedded sources attributed to individual Herbig
Ae/Be stars and clusters of low-mass pre-main sequence stars
\citep{GregorioHetem98}\footnote{This study noted strong X-ray emission
at the edge of the {\it ROSAT} PSPC field towards the H{\sc II} region
and speculated that it arose from either unresolved emission from T
Tauri stars or from OB wind emission.  It is now clear that this is
emission from the unresolved O4 V star HD~46223 blurred by the poor
point spread function at that off-axis location in the {\it ROSAT}
image.}.

One aspect of the Rosette star formation region is still unclear: was
the formation of the cluster triggered by an encounter between the
large Monoceros Loop supernova remnant (SNR), seen in radio continuum
and H$\alpha$, and the RMC?  The SNR may have been produced by a
massive member of the Mon~OB2 association; it has a diameter $\sim
80$~pc and is centered $2^\circ$ to the northeast of the Rosette
Nebula.  Older studies placed it at about the same distance as NGC~2264
($\sim$800~pc), but more recent estimates from a CO survey of the
Mon~OB1 region \citep{Oliver96} suggest that it lies at approximately
the same distance as the NGC~2244 stellar cluster; note, however, that
it has an estimated age of only 0.03 to 0.5~Myr \citep{Welsh01}.  Given
a cluster age of 1.9 Myr (see above), the evidence suggests that the
cluster formed earlier than any possible encounter with the SNR.
Extended high-energy $\gamma$-ray emission has been found by the EGRET
instrument around the region of apparent contact between the SNR and
the nebula, interpreted in terms of $\pi^\circ$ decay resulting from
collisions between SNR-accelerated particles and the molecular cloud
\citep{Jaffe97}.  However, no morphological indication of a collision
or compression is evident from optical or molecular maps.
 
In summary, and by analogy with M~17, the major morphological
components of the Rosette star formation region are:  (1) An
edge-brightened, photoionized nebula with a hollow, spherical (or
cylindrical) shape, perhaps with a conical cavity expanding towards the
nearest edge of the surrounding molecular cloud.  The central region
showed extended X-ray emission with {\it Einstein}, with $T_x \sim
23$~MK and $n_e \sim 0.02$~cm$^{-3}$, but {\it ROSAT} studies resolved
this emission into point sources.  Shells of higher density surround
the hollow core, with $T_e \sim 6400$~K, perhaps a clumpy inner shell
with $n_e \sim 10^3$--$10^4$~cm$^{-3}$, and an outer shell or shells
with $n_e \sim 10$~cm$^{-3}$ \citep{Celnik85, Tsivilev02}.  (2) A
photodissociation region between the Rosette blister H{\sc II} region
and its molecular cloud.  (3) Dense, cold molecular gas making up the
Rosette Molecular Cloud and extending far beyond the H{\sc II} region,
where new stellar clusters are forming.  (4) The open cluster NGC~2244
which excites the nebula.

\section{{\it Chandra} Observations and Analysis \label{observations.sec}}

The H{\sc II} regions were observed with the Imaging Array of {\it
Chandra}'s Advanced CCD Imaging Spectrometer (ACIS-I); the observatory
and instrument are described by \citet{Weisskopf02}.  CCDs S2 and S3
from the Spectroscopy Array were also operating, although these chips
are $>15^{\prime}$ off-axis and not positioned to match the ACIS-I
focal surface, so they exhibit large, distorted PSFs that make point
source detection difficult.  They are nonetheless useful for broad
characterization of diffuse emission and for revealing the presence of
bright point sources.  The observing mode was the standard ``Timed
Event, Faint'' mode, with 3.2-sec integration times, 0.041-sec frame
readout times, and $3 \times 3$ pixel event islands.

Table~\ref{tbl:obslog} gives the observing log.  M~17 was observed in a
single 39~ks pointing centered on the heavily obscured O4--O4 binary
known as Kleinmann's Anonymous Star in the NGC~6618
cluster (star \#1 in \citet{Chini80}; \#341 in \citet{Ogura81}; \#189
in \citet{Hanson97}).  The Rosette observation consisted of a
contiguous sequence of four $\sim$20-ks ACIS-I snapshots (Fields 1-4),
resulting in a $\sim 1^{\circ} \times 0.25^{\circ}$ mosaic of the Rosette
Nebula and Rosette Molecular Cloud (RMC).  The aimpoint of Rosette
Field 1 was placed near the center of the Rosette Nebula on the B star
HD~259105, to avoid photon pile-up of the O5 star HD~46150.  An
unfortunate telescope roll angle about this aimpoint led to HD~46150
falling in the gap between CCDs I2 and I3, resulting in substantially
reduced exposure for this important early O star.  Rosette Field 2
yields the best image of the interface between ionized material in the
H{\sc II} region and neutral material in the RMC.  Fields 3 and 4 step
eastward into the molecular cloud, sampling the young, embedded stellar
population there.

Our custom processing, starting with the Level 1 data stream, is
described in Appendix B.  Important steps include removal of the
$0.5^{\prime\prime}$ position randomization and correction of the CCD
charge transfer inefficiency (CTI) to improve spectral resolution
\citep{Townsley02a}.  We generated custom response matrix files (RMFs)
and quantum efficiency uniformity files (QEUs) to facilitate spectral
fitting of CTI-corrected data \citep{Townsley02b}\footnote{Details,
source code, and calibration products for CTI correction using our
method are available at
\url{http://www.astro.psu.edu/users/townsley/cti}.}.  The images are
registered with the USNO A2.0 catalog \citep{Monet98} or Hipparcos
sources \citep{Hog00} in the field to improve the astrometry; this
alignment is accurate to about $\pm 0.3$\arcsec.

\begin{figure}
\centering
\epsscale{0.8}
\plotone{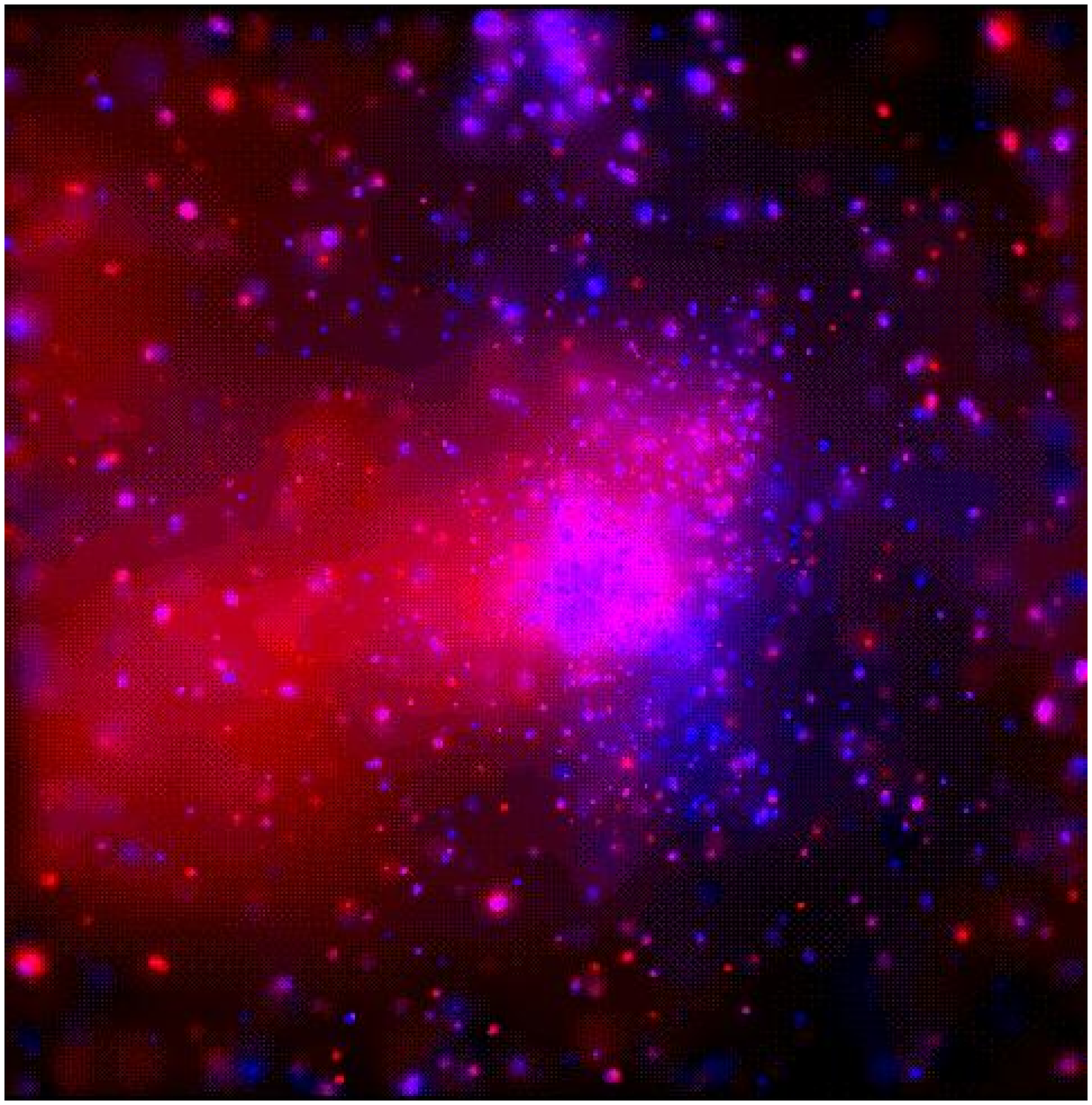}
\plottwo{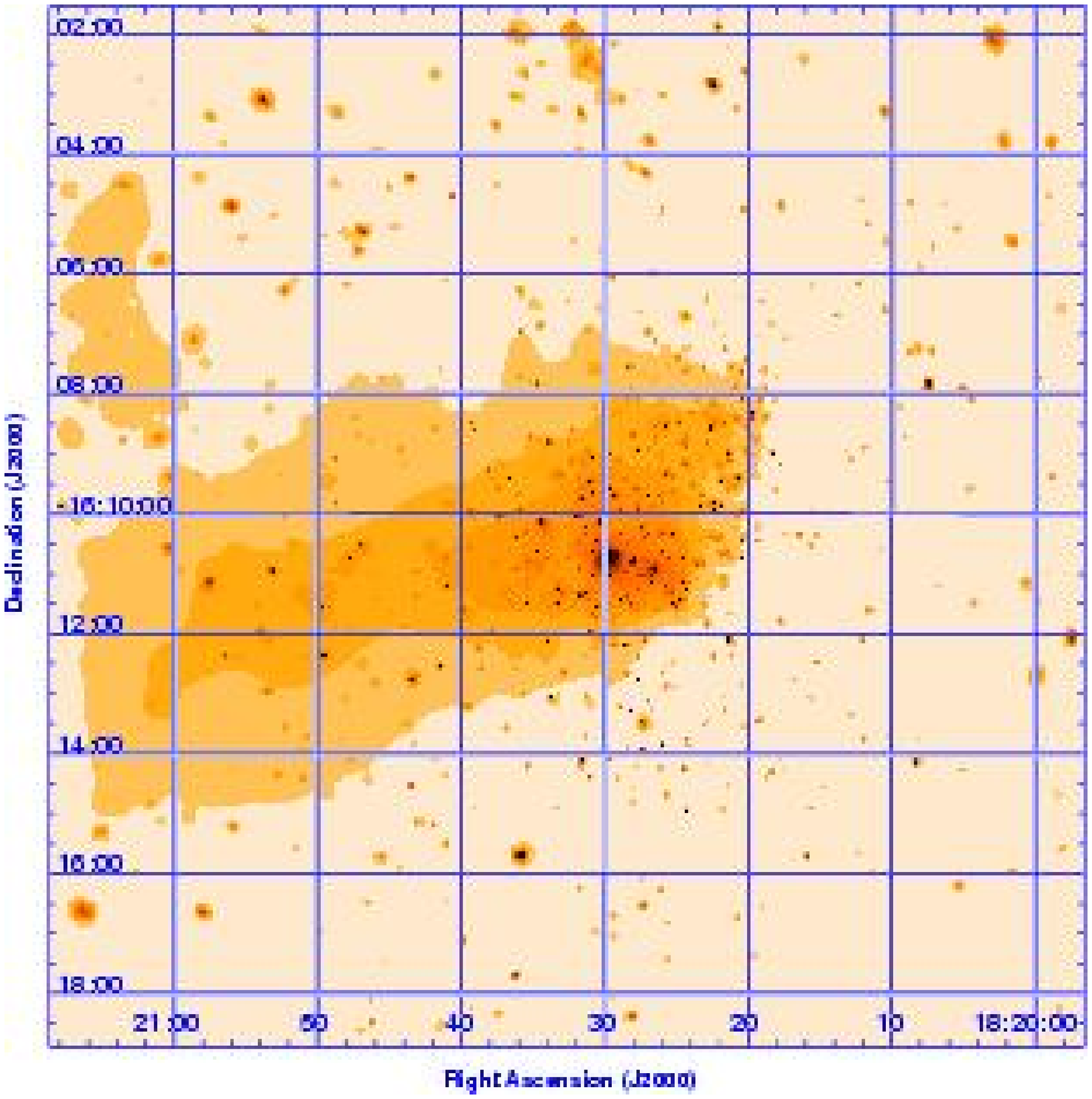}{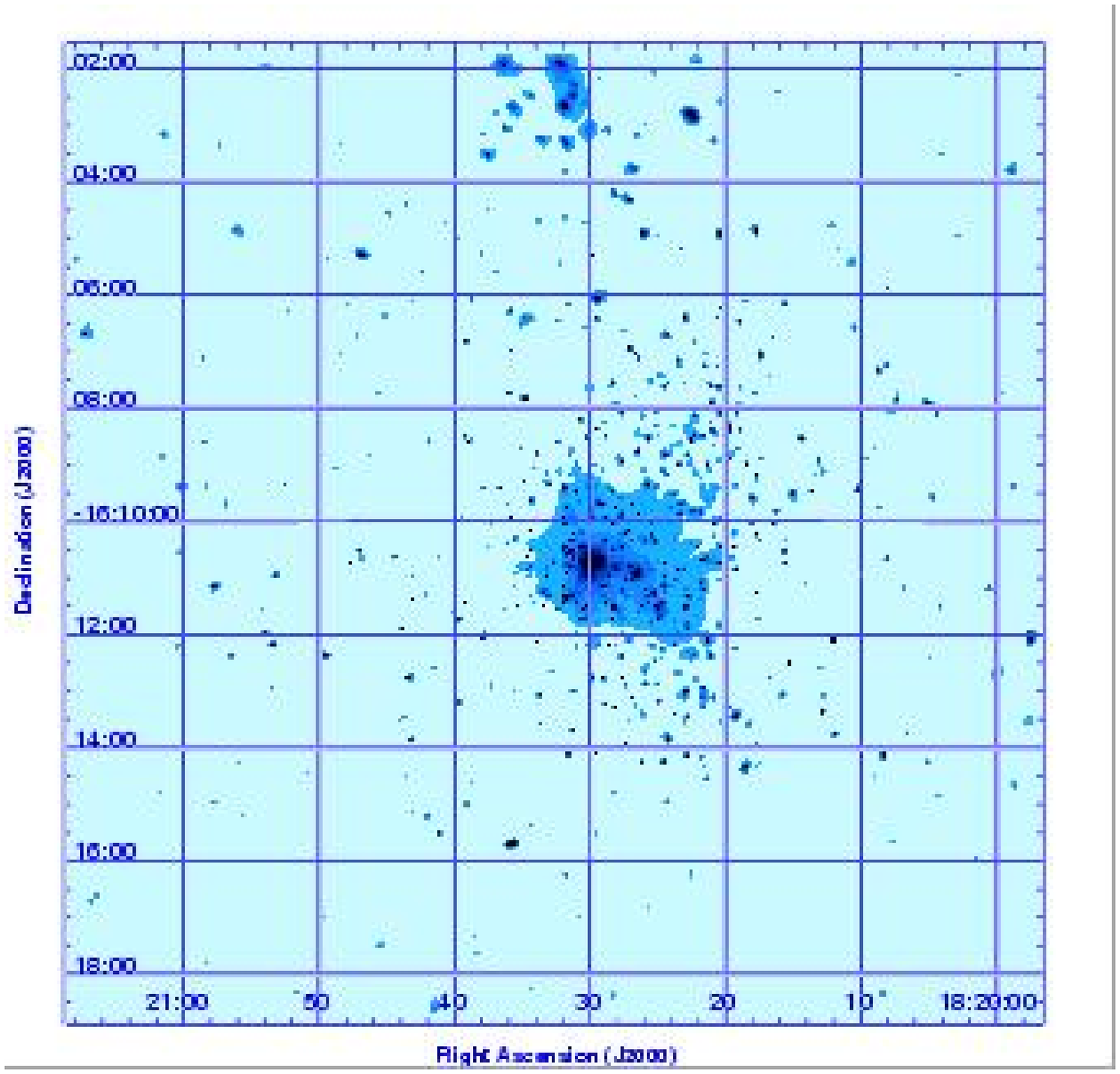}
\caption{
(a) {\it Chandra} ACIS-I image of M~17 with {\it csmooth} adaptive
smoothing and exposure correction (2.07 pixels per bin).  Red intensity
is scaled to the soft (0.5--2~keV) emission and blue intensity is scale
to the hard (2--8~keV) emission.
(b) The smoothed 0.5--2~keV image used for the red intensity in (a),
highlighting the soft diffuse emission.
(c) The smoothed 2--8~keV image used for the blue intensity in (a),
highlighting the hard emission.
\label{fig:m17_color}}
\end{figure}

To search for and quantify any diffuse emission in the regions, we have
performed a variety of smoothing and image processing steps; results
are shown in Figures~\ref{fig:m17_color} through
\ref{fig:m17_soft_patsmooth} for M~17 and
Figures~\ref{fig:rosette_color} and \ref{fig:rosette_patsmooth} for
Rosette.  For display purposes, the photon counting images are smoothed
with an adaptive kernel filter using the {\it csmooth} program (a
translation of the {\it asmooth} code by Harald Ebeling), part of the
CIAO data analysis system supplied by the {\it Chandra} X-ray Center
(CXC).  The ACIS-I exposure maps, which take into account a variety of
effects (e.g., bad CCD columns, gaps between the CCDs, telescope
vignetting, and satellite dithering), are similarly smoothed and the
resulting photon and exposure images are divided to give a view of the
X-ray flux distribution with all known instrumental effects removed. 

Figures~\ref{fig:m17_color}a and \ref{fig:rosette_color}a show the
ACIS-I flux images at this stage of the data processing.  The images
here are shown at reduced resolution after adaptive kernel smoothing
with {\it csmooth}.  In these two-color images, red intensity is scaled
to the soft (0.5--2~keV) brightness and blue intensity is scaled to the
hard (2--8~keV) brightness.  For clarity, the soft and hard smoothed
images are shown separately in Figures~\ref{fig:m17_color}b and c and
\ref{fig:rosette_color}b and c.  Both fields show clear indications of
soft extended emission in addition to hundreds of unresolved sources.

Quantitative study of the extended emission requires removal of the
point source events.  Point sources were found using the wavelet-based
source detection program {\it wavdetect} \citep{Freeman02}.  After
experimentation with the source significance threshold in {\it
wavdetect}, we chose a threshold level $P = 1 \times 10^{-5}$.  While
this results in a number of faint (3-5 counts) and possibly spurious
sources, it is more sensitive to clearly real sources that are missed
(especially when there is underlying diffuse emission) at the usual
threshold of $P = 1 \times 10^{-6}$.  The fields were carefully
examined by eye to identify potential point sources missed by {\it
wavdetect} due to crowding, diffuse emission, or unknown reasons.  The
resulting lists of point sources and identifications with members of the
NGC 6618, NGC 2244, and RMC stellar clusters will be discussed in
separate studies.  

Once this liberal list of potential point sources was established for
each field, we used our custom IDL code {\it acis\_extract} to excise
them from the data\footnote{The {\it acis\_extract} code is available at
\url{http://www.astro.psu.edu/xray/docs/TARA/ae\_users\_guide.html}.}.  
This tool uses the point spread function (PSF) library in the CIAO
Calibration Database to generate a polygon around each source that
represents the 99\% contour of the PSF at a chosen energy (typically
1.5~keV).  A circular exclusion zone is generated with diameter 1.1
times larger than the largest polygon cord.  All events within these
exclusion zones are deleted, resulting in the `sources-removed' (or,
more informally, `swiss-cheese') event list.  The full-band
swiss-cheese images for the M~17 and Rosette data are shown at reduced
resolution (8-pixel binning) in Figures~\ref{fig:m17_patsmooth}a and
\ref{fig:rosette_patsmooth}a, respectively.

\section{Morphology of the Diffuse X-ray Emission \label{morph.sec}}

\subsection{M~17 \label{morph_M17.sec}}

Figure~\ref{fig:m17_patsmooth}b shows the $17' \times 17'$ ACIS image
of the densest parts of M~17, with the $\sim 900$ sources removed by
the `swiss-cheese' procedure explained above, smoothed with our own
adaptive smoothing algorithm.  This algorithm understands the absence
of data (the holes where point sources were) and does not introduce
edge effects into the smoothed image; see Appendix C for details.  Red
areas highlight soft (0.5--2~keV) emission and blue areas highlight hard
(2--7~keV) emission\footnote{For all smoothed swiss-cheese images of
diffuse emission, we clipped the hard band at 7~keV to avoid
contamination from the faint instrumental Ni~K$\alpha$ line at
7.47~keV.}.  The holes from the masked point sources are shown at their
full size in panel b, to give a smoothed representation of the data
that still retains much of the character of the simple binned image
(panel a).  Panel c shows these holes partially smoothed over; in panel
d they are completely smoothed over.  These three versions of the
adaptively-smoothed unresolved emission in M~17 are shown to provide
the reader with an incremental transition from the swiss-cheese image
to the fully-smoothed image; the intent is to illustrate that, although
smoothing can reveal morphological features impossible to see in the
binned data, it can cause the viewer to lose track of the actual amount
of information that went into the smoothed image.  These smoothed
images are reassuringly similar to the {\it csmooth} image in
Figure~\ref{fig:m17_color}a.

Figure~\ref{fig:m17_soft_patsmooth} explores the soft diffuse emission
in more detail.  Panel a shows a three-color image of the soft
emission with the point source masks in place; here red = 600--700~eV,
green = 830--940~eV, and blue = 1280--1540~eV.  The spectrum in panel b
uses 20-eV bins to show the line features that motivated
the bands selected for panel a.  The three-color image allows us to
scale the brightness of the 600--700~eV component up to show faint soft
features; a slight enhancement to the northwest of the core of NGC~6618
is visible in this image.  This image hints at a possible softening of
the X-rays with distance away from the ionizing cluster; although our
spectral analysis to search for changes in the temperature of the gas
as a function of distance from the ionizing cluster showed no such
evolution (see \S\ref{spec_result.sec}), the wider region sampled by
{\it ROSAT} shows a slight spectral softening to the north, consistent
with Figure~\ref{fig:m17_soft_patsmooth}a (Dunne03).
  
The smoothed images in Figures~\ref{fig:m17_color} and
\ref{fig:m17_patsmooth} are clearly closely related to the 2MASS
near-infrared image (Figure~\ref{fig:m17_comp}c): the bright soft X-ray
contours closely follow the hollow structure of the photoionized
region between the northern and southern bars (while fainter emission
is spread over a larger area).   The brightest soft X-ray emission
coincides with the hole in the diffuse H{\sc II} emission in the deep
near-infrared image of \citet{Ando02} immediately around the exciting
stars of the nebula.  The hard emission coincides with the central
region and crosses the dark region to the west of the steep interface
between the photoionized bars and the molecular cloud.  Another patch
of hard emission is associated with the deeply embedded M~17 North
region.  As discussed below, the most likely source of this very slight
excess of hard photons is the unresolved point source population in
these embedded star-forming regions.  The smoothing algorithm enhances
the appearance of this emission but its proximity to regions of large
point source concentrations makes this the simplest explanation.

The overall morphology of the soft diffuse emission in the ACIS image
of M~17 strongly suggests that the hot gas originates in the OB cluster
and flows outwards towards the lower density regions to the east.  This
might be called an `X-ray fountain,' representing the flow of shocked
OB winds into the hot interstellar medium.  The {\it ROSAT} PSPC image
confirms this view:  in the optically thick inner H{\sc II} region, its
contours are elongated identically to the ACIS emission while in the
outer regions, it reveals the X-ray flow extending several parsecs to
the east and north; this is illustrated clearly in Dunne03 (their
Figure 1).  Since $10^4$~K H{\sc II} region gas and $10^7$~K
X-ray gas cannot stably coexist in the same volume, it is likely that
the X-ray gas fills the interior of the cylindrical or conical inner
region defined by the radio/optical H{\sc II} bars.

Farther to the east, the X-ray gas expands into the lower-density,
optically thin region, filling a cavity in the DSS image.  We suspect
that the X-ray filled cavity extends farther to the north and east as
traced by 330 MHz radio emission, but is obscured in both soft X-rays
and H$\alpha$ by intervening cloud material.  The X-ray emitting area
may thus be 2-3 times larger than seen with {\it ROSAT}, and several
times larger than seen in our single $17\arcmin \times 17\arcmin$ {\it
Chandra} pointing.

In summary, the soft X-ray images of M~17 from {\it ROSAT} and {\it
Chandra}, combined with radio, optical, and NIR data, strongly suggest
a dynamic rather than static hot plasma, generated at the OB star
cluster and flowing into (and perhaps even creating) the low-density,
blister-like cavity extending 5--10~pc away from the molecular cloud.
It is then unclear whether the features in the H$\alpha$ maps---wisps,
arcs, and filaments---are manifestations of OB photoionization or
direct thermal contact with the hot, X-ray emitting gas.

\subsection{Rosette \label{morph_rosette.sec}}

Figure~\ref{fig:rosette_patsmooth}b shows the full ACIS image of the
Rosette Nebula, with the $\sim 350$ sources removed by the
`swiss-cheese' procedure explained above and again smoothed with our
own adaptive smoothing algorithm (see Appendix C).  Red areas highlight
soft (0.5--2~keV) emission and blue areas highlight hard (2--7~keV)
emission.  The holes from the masked point sources are shown at their
full size in panel b; panel c shows these holes partially smoothed over
and in panel d they are completely smoothed over.  Again, these images
are similar to the {\it csmooth} versions in
Figure~\ref{fig:rosette_color}.  The brightest areas of diffuse
emission are coincident with the two early O stars in the field; this
is not obvious in the adaptively smoothed images because they have been
scaled to highlight faint diffuse emission.  The `hot spot' to the
southwest of the main region of diffuse emission in
Figures~\ref{fig:rosette_patsmooth}b--d may be due to a point source
that was too faint to be detected so far off-axis.

The adaptively-smoothed images in
Figure~\ref{fig:rosette_patsmooth}b--d retain the off-axis CCDs S2
(upper left) and S3 (upper right).  S3 is a backside-illuminated CCD
and shows higher background, thus the apparent high intensity of soft
emission is dominated by detector effects and little useful information
on the true soft diffuse emission is available from this device.  The
S2 chip may be affected by unresolved point source emission, although
the diffuse emission seen in the smoothed images could be real; its
spectrum is similar to the soft diffuse emission seen on the ACIS-I
CCDs in this pointing.

The diffuse emission found in our {\it Chandra} mosaic of the Rosette
region comes mainly from Field 1, which includes the NGC~2244 exciting
cluster, and to a lesser extent from the northwest part of the adjacent
Field 2 (see Figure~\ref{fig:rosette_color}b).  Diffuse emission may
very well extend outside of our ACIS fields.  The total number of
diffuse photons in our ACIS images of Rosette is three times smaller
than in our M~17 image, thus the spectral and morphological information
is limited.  A longer {\it Chandra} exposure of Field 1 is approved for
more detailed study.

With these limitations in mind, the morphology of the Rosette diffuse
X-ray emission shown in the adaptively-smoothed images in
Figure~\ref{fig:rosette_color} and \ref{fig:rosette_patsmooth}
is similar to that in M~17.  It appears centrally brightened, centered
to the southeast of the axis of the ionized H{\sc II} torus toward the
two earliest stars of the exciting cluster, HD~46223 and HD~46150 (see
Table~\ref{tab:rosette_OB} in Appendix A).  The surface brightness is
highest around these two early O stars and decreases outwards, with
possible leakage beyond the edge of the ionized shell towards the
southeast, where there is a suggestion of a cavity in the optical H{\sc
II} region (Figure~\ref{fig:rosette_comp}b).  In light of the M~17
morphology described in \S\ref{morph_M17.sec}, we can speculate that
the Rosette Nebula diffuse emission might also be an `X-ray fountain'
generated by the exciting cluster NGC~2244 but seen almost along,
rather than perpendicular to, the outflow axis, perhaps flowing out the
conical cavity proposed by \citet{Celnik86}.  The fact that the Rosette
diffuse emission is substantially weaker than that in M~17 is
consistent with production by OB stellar winds, which are fewer and
weaker in Rosette (compare Tables~\ref{tab:m17_OB} and
\ref{tab:rosette_OB}).

\section{Quantitative Study of the Extended Emission \label{diffuse.sec}}

\subsection{Spectral Analysis Methodology \label{spec_meth.sec}}

The ACIS detector gives moderate-resolution X-ray spectra for each of
roughly 100,000 independent spatially resolved locations across its
$17\arcmin \times 17\arcmin$ field of view.  Having spatially
decomposed the images into pixels containing point source photons and
pixels containing diffuse source photons (\S\ref{observations.sec}),
we can readily construct composite spectra for these components.  For
M~17, we selected the part of the ACIS-I image that contained the
brightest diffuse emission (called ``Total Diffuse Region'' in
Figure~\ref{fig:m17_regions}a) and extracted the diffuse-emission
spectrum from only that region; for Rosette, we extracted counts from
the entire ACIS-I array to form the diffuse-emission spectrum.  For
both fields, composite point source spectra were extracted by using all
the point sources in the ACIS-I array.   

The background contribution should be small for the point source
components, since the source events were extracted from regions
tailored to contain only the inner 90\% of each source's PSF.
Background removal for the faint diffuse components, however, must be
performed carefully.  For the Rosette diffuse emission fit, we
subtracted a background spectrum obtained from the S2 chip, which lay
outside the core of the cluster\footnote{We chose to form the
Rosette Field 1 background from the off-axis S2 data, rather than using
the I-array data from one of the other Rosette fields, because there
was a background flare during the Field 1 observation.  This flare was
of modest brightness but lasted for 30\% of the 20-ks observation.  The
hard (2--8~keV) band emission varied erratically up to a factor of 3
higher than the usual background level.  The variations were much less
evident in the soft (0.5--2~keV) band, rising $<50$\% in the 1--2~keV
range and negligibly below 1~keV.  Since flares clearly affect the
spectral shape of the background, we determined that better background
subtraction could be achieved by using the off-axis S2 data (with
appropriate scaling) rather than a different observation.  Our
composite point source spectral analysis for Rosette Field 1 was
restricted to the 0.7--5~keV band to minimize the effect of these
background variations.}.  This background spectrum was scaled by the
effective area (BACKSCAL was converted to an energy-dependent vector)
and by the ratio of the geometric areas.  Further scaling was done to
account for the non-uniform spatial distribution of the particle
background\footnote{Catherine Grant, private comm.; also see
\url{http://space.mit.edu/$\sim$cgrant/cti/cti120.html}.}, obtained by
matching the strength of the instrumental Ni~K$\alpha$ line and the
3--7~keV continuum level.  The prominent Au~M line complex (around 2.1~keV) remaining in the spectrum after background subtraction \citep[see
Table 3 in][]{Townsley02b} may indicate that the spatial distribution
of Au~M fluorescent photons (which come from bombardment of gold-coated
surfaces of the ACIS camera and the telescope by solar energetic
particles) does not follow that of the particle background.  

Accounting for differences in geometric area, ancillary response files
(ARFs), and particle background as described above, and assuming that
all emission on S2 is due to background and not unresolved stars or
diffuse emission, we estimate that the background accounts for about
58\% of the events in our sample of the Rosette diffuse emission.
Subtracting this leaves only 3300 source counts distributed over the
entire ACIS-I array.  Figure~\ref{fig:rosette_patsmooth}b--d implies,
however, that some of the S2 photons are not from the background; if
this is the case then our background subtraction removes legitimate
photons from the Rosette diffuse emission spectrum and a smaller
percentage of unresolved events come from the background.

For M~17, a background region near the southern edge of the I-array
around the molecular cloud was used.  This background spectrum was also
scaled by the effective area and by the ratio of the geometric areas.
Scaling the number of 0.5--2.3~keV counts in the M~17 background area
(lower part of I1 chip) to account for the difference in geometric area
between the diffuse-emission region and the background region, and
ignoring any additional scaling due to ARF differences (the exposure
map shows that this is a small factor), we find that the background
accounts for almost 50\% of the events in the diffuse region.  This
presumes that the region we used for background is free of diffuse
emission, but the {\it ROSAT} image shows that there is faint diffuse
emission in this region.  The spectrum of this region also resembles
that of the diffuse emission.  This shows that again our background
subtraction is almost certainly removing useful diffuse photons and
the actual background contribution to the diffuse emission is likely
less than 50\%.

For both Rosette and M~17, the regions used to sample the background
spectrum are not ideal.  We chose not to use generic background files
supplied by the CXC because our fields are in the Galactic plane, which
is known to contain diffuse X-ray emission from a variety of known
(e.g., the Local Hot Bubble, \citet{Snowden98}) and unknown (e.g., the
Galactic Ridge X-ray Emission, \citet{Ebisawa01}) sources.  Other
nearby regions (Fields 2--4 in the Rosette observation and the western
side of the M~17 observation, which includes the S2 and S3 chips)
sample more obscured parts of the molecular clouds, so backgrounds
generated from these regions would not be representative of the regions
used to sample the diffuse emission.  So rather than choosing
background regions that were clearly incorrect, we decided to sample
more representative regions and suffer the loss of useful photons.

The background-subtracted diffuse emission spectra, the composite point
source spectra, and their associated best-fit plasma models are shown
in Figure~\ref{fig:spectra}, with details given in
Table~\ref{tab:spectra}.  The spectral fits were performed with XSPEC
\citep{Arnaud96} using both the {\it mekal} and {\it vmekal} thermal
plasma models \citep{Mewe86, Kaastra92, Liedahl95} and the {\it apec}
and {\it vapec} models \citep{Smith01}, where the ``v'' designation
provides for non-solar abundances of the elements.  The fits from both
types of models were similar, so only the {\it apec} class of models is
presented here.  One or two absorption components were included via the
{\it wabs} model \citep{Morrison83}.  An additional absorption
component, due presumably to hydrocarbon deposition on the ACIS optical
blocking filters \citep{Chartas02}, was also included in the XSPEC
model\footnote{This instrumental issue is discussed at
\url{http://cxc.harvard.edu/cal/Acis/Cal\_prods/qeDeg/index.html}.  The
correction program and XSPEC model {\it acisabs} was developed by
George Chartas and is available at
\url{http://www.astro.psu.edu/users/chartas/xcontdir/xcont.html}.}.
The spectral fits employed count-weighted ARFs made using the CIAO
utility {\it mkwarf}.  We chose to use the single CTI-corrected
response matrix file (RMF) \citep{Townsley02b} that coincided with the
largest fraction of counts.  For the full-field spectral fits shown in
Figure~\ref{fig:spectra}, the RMFs used were the full-CCD versions, so
the spectral resolution is the average across the entire CCD after CTI
correction (see Appendix B).  These full-CCD CTI-corrected RMFs are
similar to each other, so using a single-CCD RMF for the entire I-array
does not introduce substantial errors.

Table~\ref{tab:spectra} shows that the background-subtracted diffuse
emission in both H{\sc II} regions is reasonably well-represented by a
two-temperature thermal plasma with a single absorbing column.  The
temperatures of both plasma components are comparable between the two
regions.  The Gaussian at $\sim 2.1$~keV was included to model the
instrumental Au~M complex discussed above.  For both regions, the $kT_1$
component was fixed to solar abundances while the $kT_2$ component was
a variable-abundance plasma.

For comparison with the diffuse emission, we performed similar spectral
fits on the emission from the combined point sources detected in the
H{\sc II} regions.  Again the $kT_1$ component was fixed to solar
abundances while the abundances for the $kT_2$ component were allowed
to vary.  The Rosette composite point source spectrum was adequately
fit by plasmas at 0.2 and 2.1~keV with a single absorbing column.
Fitting the combined M~17 point sources required a more complicated
model:  a solar-abundance thermal plasma at 0.19~keV with a relatively
low absorption ($7 \times 10^{21}$~cm$^{-2}$) and a variable-abundance
thermal plasma (3~keV) with higher absorption ($17 \times
10^{21}$~cm$^{-2}$).  This fit was also clearly improved by adding
Gaussians at 1.1, 1.3, and 6.7~keV; lines at these energies are seen in
O stars and in flaring pre-main sequence stars
\citep[e.g.,][]{Schulz00, Tsuboi98}.  No line was needed to model the
instrumental Au~M complex in the combined point source spectra,
indicating that no significant residual instrumental background is
present around 2~keV.

In addition to the introduction of Gaussian lines, we found that
extra-solar abundances of oxygen were needed to give a good fit to all
of the spectra, and extra-solar abundances of neon were needed for the
point source spectra, although the actual abundance values were not
well-constrained.  If both plasma components are forced to have solar
abundances in the M~17 spectra, the fits are statistically worse.  The
reduced $\chi^2$ for the diffuse spectrum increases from 1.2 to 1.4;
the plasma temperatures did not change but $N_H$ increased from $4
\times 10^{21}$~cm$^{-2}$ to $6 \times 10^{21}$~cm$^{-2}$.  For the
composite point source spectrum, the reduced $\chi^2$ went from 1.5 to
3.7; the two absorption values and the first plasma component's
temperature stayed the same, but the second plasma component increased
from $kT_2 = 3.0$~keV to $kT_2 = 3.6$~keV.  This caused the iron line
to be too strong (even with elimination of the Gaussian at 6.7~keV) and
was the major reason for the unacceptable fit.

Similar results were obtained for Rosette.  Requiring both plasma
components to have solar abundances in the fit to the diffuse spectrum
caused the reduced $\chi^2$ to increase from 0.7 to 1.3.  As for M~17,
the plasma temperatures did not change but $N_H$ increased, here from
$2 \times 10^{21}$~cm$^{-2}$ to $5 \times 10^{21}$~cm$^{-2}$.  For the
composite point source spectrum, the reduced $\chi^2$ went from 1.6 to
3.0; the absorption stayed the same, but both plasma components
increased in temperature:  $kT1$ went from 0.20 to 0.29~keV and $kT_2$
went from 2.1 to 2.9~keV.  This caused large residuals below
2~keV---the model appeared to have many more line-like peaks than the
data.

There are several reasons why including non-solar abundances might
improve the fit to these composite spectra.  Excess Ne{\sc IX} and
Ne{\sc X} lines are prominent in the spectra of nearby flaring
late-type stars \citep{Brinkman01} and can be present in pre-main
sequence flaring stars as well.  The origin of excess oxygen is less
clear.  We suspect that it is not astrophysically real, but arises from
inadequate modeling of the absorption.  Each spectrum covers stars or
regions with a wide range of foreground absorptions, leading to a
confused observed spectrum in the $<1$~keV region, where oxygen has
strong lines.  The Gaussians needed to fit the M~17 composite point
source spectrum and other features seen in the (Data$-$Model) residuals
in Figure~\ref{fig:spectra} may also arise from the merging of
different plasma conditions into a single spectrum.  Imperfect
background removal accounts for the residuals at low and high energies.

\subsection{Spectral Results \label{spec_result.sec}}

The resulting plasma spectra are shown in Figure~\ref{fig:spectra} and
the parameters of the best-fit plasma models are given in
Table~\ref{tab:spectra}.  $L_{\rm soft}$ values are the soft-band
(0.5--2~keV) X-ray luminosities assuming distances of 1.6~kpc for M~17
and 1.4~kpc for the Rosette Nebula; absorption-corrected values are
given as $L^c_{\rm soft}$.  There is substantial hard-band (2--7~keV)
flux only for the M~17 combined point sources, where the
absorption-corrected luminosity is $L^c_{\rm hard} \simeq 4.1 \times
10^{33}$ergs~s$^{-1}$.  The temperatures of both diffuse-emission
plasma components are very similar in the two regions and are dominated
by the $T \sim 10$~MK component, with a secondary 1~MK component.  But
the regions are notably different in luminosity and surface brightness,
with the M~17 diffuse emission a factor of $>5$ brighter than the
Rosette diffuse emission and coming from an extraction area $\sim 90$\%
as big as that used to sample the Rosette diffuse emission (roughly
42~pc$^2$ vs.\ 47~pc$^2$).  Equivalently, we see that the diffuse
component comprises $\simeq 36$\% of the total soft luminosity in the
M~17 region sampled but only $\simeq 7$\% in the Rosette Nebula
region.

To within the errors, our $kT_2$ plasma component is consistent with
the Dunne03 results from spectral modeling of the {\it ROSAT} data on
M~17.  This is somewhat remarkable, since the {\it Chandra} data have
shown that point sources contribute substantially to the X-ray emission
in the {\it ROSAT} band.  Our value for $N_H$ is substantially larger
than the Dunne03 value for their Regions A and C but is consistent with
their $N_H$ for Region D.  The large section of the ACIS field that was
used to sample the diffuse emission encompasses most of Dunne03 Regions
C and D and part of their Region A, but it also extends farther to the
north and west than those regions, into parts of the nebula that are
more highly absorbed.  This may account for our higher $N_H$ estimate.  

No hard diffuse component was detected in either ACIS field, when all
diffuse counts were combined into a single spectrum.  The smoothed
X-ray images in Figures~\ref{fig:m17_color}c and
\ref{fig:m17_patsmooth}b--d suggest that a diffuse hard component might
be present in M~17, towards the M~17SW molecular cloud and M~17 North
protostellar region.  However, this emission is spatially coincident
with high concentrations of point sources and, once those point sources
are removed from the data, few hard counts are left to contribute to
the global diffuse spectrum (see the swiss-cheese image,
Figure~\ref{fig:m17_patsmooth}a).

There are enough diffuse counts in M~17 to warrant subdividing them
into smaller regions to search for spatial gradients in temperature.
Two such schemes are shown in Figure~\ref{fig:m17_regions}, one
sensitive to temperature effects associated with proximity to the
molecular boundaries of the H{\sc II} region cavity, and the other
sensitive to cooling as the gas flows eastward from the central O
stars.  In almost all cases, no changes in plasma components were
found.  The only discrepant case is the westernmost box in (b) which
showed 2.3 and 23~MK (0.2 and 2~keV) components (plus stronger
absorption, with $N_H = 14 \times 10^{21}$~cm$^{-2}$).  These properties
may reflect either a change in plasma conditions or contamination by an
unresolved population of lower-mass stars (see \S\ref{not_lowmass.sec}).

Finally, we note that the composite spectra of the point sources are
complex in different ways (Figure~\ref{fig:spectra}).  Even complicated
spectral models do not give statistically acceptable fits.  This is
understandable given that we have combined hundreds of different stars,
each of which suffers different absorption and may have spatially and
temporally variable plasma emission, to generate these composite
spectra.  The presence of excess neon and the iron line at 6.7~keV
supports the interpretation that these are magnetically active flaring
stellar sources.  A soft ($kT \sim 0.2$~keV) spectral component is seen
in the composite point source spectra of both M~17 and Rosette; this
may be simply a component of the stellar plasma, although
\citet{Wolk02} see a similar soft component in extended emission in
RCW~38 and note that it may be partially caused by X-ray halos around
the point sources, due to dust scattering.  In any case, we have no
evidence that the physical mechanism generating this soft plasma
component in the stellar spectra is related to the soft component seen
in the diffuse emission.  The most important result for our purposes
here is that in both star-forming regions the point sources are
dominated by a substantially harder plasma component (2--3~keV, or
23--35~MK) than seen in the diffuse emission ($\sim 0.7$~keV, or 8~MK).

\subsection{Physical Properties of the Plasma \label{phys.sec}}
 
It is clear from the above analysis that the diffuse X-ray emission in
both the Rosette Nebula and M~17 is due predominantly to a hot plasma.
The physical parameters can readily be estimated, with results summarized
in Table~\ref{tab:phys}.

For both nebulae, our {\it Chandra} results show relaxed center-filled
morphologies (\S\ref{morph.sec}).  M~17 lacks any measurable
temperature gradients (\S\ref{diffuse.sec}), although
Figure~\ref{fig:m17_soft_patsmooth} hints that such gradients may be
present; too few diffuse counts are present in our Rosette observation
to search for such gradients.  We therefore consider the simple
situation of a uniform, optically thin, isothermal plasma in a simple
geometry.  For simplicity, we will assume that the diffuse X-ray
luminosities given in Table~\ref{tab:spectra} come from the idealized
emitting regions defined below; in reality of course the emission has a
more complex structure.  

For M~17, we adopt a cylindrical geometry viewed perpendicular to its
axis with radius $3.5\arcmin$ ($r=1.6$~pc) and length $8\arcmin$ ($l =
3.7$~pc), giving a plasma volume $V_x = \pi r^2 l = 9 \times
10^{56}$~cm$^3$.  For the plasma volume $V_x$ in Rosette, the emitting
region can be approximated by a circle of radius $4.7\arcmin$ ($r =
1.9$~pc) and we adopt a spherical geometry with volume $V_x =
\frac{4}{3} \pi r^3 = 9 \times 10^{56}$~cm$^3$.  We introduce a
geometric correction factor $\eta$ of order unity to account for the
possibility that the emitting regions are flattened or elongated along
the line of sight.  This is simultaneously used as a filling factor
($\eta < 1$) if the plasma density is inhomogeneous due to shocks or
the penetration of cooler gas.  We thus retain $\eta$ as a free scaling
parameter of the inferred physical parameters of the X-ray gas, but
find that most quantities depend only weakly on $\eta$.  Recall that
the true X-ray luminosity and emitting volume is undoubtedly larger
than used here (probably by a factor $>2$ for M~17 and $<2$ for
Rosette), as we restrict our diffuse photon extraction to regions with
high surface brightness.

The plasma luminosities $L_{x,bol} = \beta L_x$ considered here scale
the absorption-corrected X-ray luminosities given in
Table~\ref{tab:spectra} by a bolometric correction factor $\beta \simeq
1.3$ for $kT = 0.6$--0.8~keV to account for emission lying outside of
the Chandra spectral band.  Although the best fit spectra are
multitemperature, for simplicity we consider here the plasmas to be
isothermal at the dominant energy around $kT = 0.7$~keV.

From the three measured quantities $L_{x,bol}$, $V_x$, and $T_x$, the
basic physical parameters of the plasma can be derived. The plasma
electron density $n_{e,x}$ is obtained from $L_{x,bol} = \Lambda
n_{e,x}^2 \eta V_x$~ergs~s$^{-1}$, where $\Lambda = 5 \times
10^{-23}$~ergs~cm$^{3}$~s$^{-1}$ is the bolometric X-ray emissivity for
a cosmic abundance plasma at $kT_x = 0.7$~keV based on the CHIANTI
emission line database \citep{Landi99}.  $\Lambda$ is only weakly
dependent on temperature close to $kT_x \approx 1$~keV.  For a fully
ionized cosmic abundance plasma, the mean molecular weight per electron
is $\mu_x = 0.62$.  Knowing $n_{e,x}$ and $T_x$, the plasma pressure is
then $P_x/k = \rho T_x/(\mu_x m_p) = 2.2 n_{e,x}T_x$, the thermal
energy $E_x = \frac{3}{2}P_x V_x$, and the cooling timescale
$\tau_{cool} = E_x/L_{x,bol}$.  The total mass of the X-ray gas is $M_x
= \mu_x m_p n_{e,x} V_x$.  The timescale for dynamical equilibration
across the X-ray regions is $\tau_{equil} = r/v_s \simeq 1 \times
10^4$~yr where $v_s = 0.1 T_x^{1/2}$~km~s$^{-1}$ is the plasma sound
speed.  This short equilibration timescale indicates that the X-ray
plasma should quickly fill most of the H{\sc II} region volume,
supporting the assumption of a constant temperature over the X-ray
emitting region except where shocks are present.  The X-ray
morphologies of both regions are center-filled and roughly consistent
with the uniform plasma assumed here, although deeper exposures may
reveal inhomogeneities due to shocks.

In addition to providing these quantities for the M~17 and Rosette
X-ray emitting regions in Table \ref{tab:phys}, we compare them with
properties of the H{\sc II} photodissociation regions that surround the
X-ray gas based on published radio and optical studies.  Note that at
these lower temperatures helium is only partially ionized:
\citet{Celnik85} gives N(He$^+$)/N(H$^+$) = 0.12 for the Rosette.  Then
the mean molecular weight per electron in the photodissociation region
is $\mu_{II} = 0.95$, and $P_{II}/k = 1.4 n_{e,II}T_{II}$.  We also assume
$\eta = 1$ for this H{\sc II} gas, which may not be correct if the 
warm gas is filamentary as suggested by recent high-resolution
optical and infrared images \citep{Jiang02}.  We use the values
of $T_{II}$ and $n_{e,II}$ given in \citet{Tsivilev02} for Rosette.

We can first compare the gas pressures of the X-ray and H{\sc II}
regions.  In M~17, the X-ray plasma and photodissociation region
pressures are approximately equal, so that the plasma probably does not
expand radially, perpendicular to the axis of the H{\sc II} cylinder.
Unfortunately, the H{\sc II} pressure in the low-density side east of
this cylinder is unmeasured, but it is possible that the X-ray plasma
is unconfined and expands in that direction.  The similarity between
the radio, optical, and {\it ROSAT} images (Figure~\ref{fig:m17_comp})
would however suggest that it is at least partially confined.  
We can compare our estimate of $T_x$ and $n_{e,x}$ to the {\it ROSAT}
results for M~17:  Dunne03 find $T_x = 8.5 \times 10^6$~K
and $n_{e,x} = 0.09$~cm$^{-3}$.  The X-ray temperatures are comparable to
within the errors but the density is somewhat lower; this is expected
because the {\it ROSAT} data sample the fainter, more extended emission
eastward of the ionizing cluster that is missed in the smaller ACIS-I
field of view.

Rosette is similar: the nominal pressure of the X-ray plasma is the
same as the photodissociation region pressure, if we assume that
the model of \citet{Tsivilev02} is correct and the inner shell contains
high-density clumps ($n_{e,II} = 10^3$~cm$^{-3}$).  However, there is
still a difficulty in explaining why the Rosette cavity is much smaller
than it should be according to standard bubble models
(\S\ref{winds_elab.sec}).  There are two possible explanations here:
either the plasma volume is greatly elongated along the line of sight
so that $\eta >> 1$ and the density is considerably lower than we
estimate here, or the plasma escapes and flows into the surrounding
interstellar medium.  As mentioned above, this flow might be an
unimpeded `X-ray fountain' roughly along the line of sight, as we
clearly see in M~17 nearly in the plane of the sky, and/or a leakage of
X-ray gas through fissures in the H{\sc II} and molecular gas.

\section{Possible Origins of the Extended Emission \label{origins.sec}}

\subsection{Instrumental Effects \label{not_instrum.sec}}

We are quite confident that the extended emission does not arise from
some instrumental effect.  For each source, the {\it acis\_extract}
procedure (\S\ref{observations.sec}) calculates the polygon enclosing
99\% of the PSF, then masks out a circular region 1.1 times larger than
the largest cord of this polygon to create the swiss-cheese image.
Thus we have been very careful to remove all of the emission from
detected point sources, accounting for the size and shape of the point
spread function at different locations on the detector.  Residual
contamination from known point sources should therefore contribute
negligibly to the measured diffuse emission (Table~\ref{tab:spectra}).

There are no known spatial variations in the background events in the
ACIS-I array which mimic either the M~17 or Rosette diffuse emission.
As discussed in \S\ref{spec_meth.sec}, a temporal flare occurred during
the Rosette observation but had little effect on the soft band where
the diffuse emission appears; flare events also should be distributed
roughly uniformly across the detector.

\subsection{Low Mass Pre-Main Sequence Stars \label{not_lowmass.sec}}

There is a more serious concern that a large number of cluster members
which are too faint to be individually detected as point sources may
mimic diffuse emission.  This indeed was a significant problem with the
earlier {\it Einstein} and {\it ROSAT} studies of high-mass star
formation regions (\S\ref{sec:intro}).  We can quantitatively evaluate
this effect by comparing the M~17 and Rosette clusters to the Orion
Nebula Cluster (ONC) and its environs, for which there is both a more
complete stellar census and deeper {\it Chandra} X-ray coverage than
for the clusters considered here \citep[][see their Tables 2, 3, and
5]{Feigelson02}.  We assume here that the ONC, M~17, and Rosette
stellar clusters have identical IMFs with identical X-ray luminosity
functions, differing only in their total population.  Since M~17
(Rosette) has 51 (up to 31) members with spectral type B3 or earlier
(if we assume that most stars listed in Table~\ref{tab:rosette_OB} are
members of NGC~2244) compared to 8 high mass stars in the ONC, we infer
that they have $\sim 6$ ($\sim 4$) times the low mass population of the
ONC.  The ONC has 1075 {\it Chandra} detections above
0.10~cts~ks$^{-1}$ in the ACIS-I detector.  Our sensitivity to point
sources in the M~17 (Rosette) fields, scaling by both exposure time and
distance, is equivalent to sensitivities of 2.7 (4.0)~cts~ks$^{-1}$ in
the ONC.  Examining the ONC source count distribution and scaling
upward by a factor of 6 (4) to match the overall cluster populations,
we predict the detection of 1960 sources in M~17 and 900 sources in
Rosette.  This is about a factor of 2 higher than the number of sources
detected in M~17 and  in Rosette Field 1.  However, since many of the
ONC sources are from an embedded population behind the H{\sc II}
region, while in M~17 and Rosette these embedded stars are mainly
displaced from the H{\sc II} region, we believe this difference is not
significant.

Having established that the detected ONC, M~17, and Rosette source
counts are at least roughly compatible, we now use the deeper ONC
observation to obtain the coarse estimate that 4500 (6100) undetected
sources emitting $3 \times 10^{32}$ ($5 \times 10^{32}$)~ergs~s$^{-1}$
should be present in the M~17 (Rosette) fields (more Rosette sources
remain undetected mainly because our observation of Rosette Field 1 is
half as long as our M~17 observation).  This is about 8\% (80\%) of the
absorption-corrected diffuse X-ray luminosity $L^c_{\rm soft}$ shown in
Table~\ref{tab:spectra}.  For M~17, this result clearly supports the
already convincing morphological evidence that very little of the
`diffuse' emission can arise from the undetected population of
X-ray-faint cluster members.  The situation is not quite so clear for
Rosette, where our calculation permits that a significant fraction,
even most, of the `diffuse' emission could come from undetected
stars---note that this calculation is very rough, though, and could
easily be in error by a factor of 2.  The difference in spectral shape
between the `diffuse' and the detected point source components (compare
panels b and d in Figure \ref{fig:spectra}), however, independently
indicates that the point source contribution to the diffuse emission is
small.  The issue will be easily settled with the upcoming deeper {\it
Chandra} observation of Rosette:  from the ONC source count
distribution, we estimate that a 100~ks exposure should resolve about
half of the emission from stars that are currently undetected.

\subsection{Supernova Remnants \label{not_snr.sec}}

In M~17, there are many arguments against a hypothesis that the X-rays
are independent of the H{\sc II} gas and instead arise from the remnant
of a recent supernova explosion.  Foremost, the X-ray morphology
follows very closely the morphology of the thermal radio and optical
emitting H{\sc II} region (\S\ref{morph_M17.sec}) which does not show
kinematic disturbances greater than $\simeq 100$~km~s$^{-1}$.  In
addition, the stellar cluster currently has no stars evolved off of the
main sequence, indicating a young age, and is insufficiently rich to
host a population of extremely massive stars which could evolve
extremely rapidly into a supernova.  Even in a cluster with 100 O
stars, the first supernova explosion does not occur for $\simeq 4$~Myr
\citep{Knodlseder02}.  There is no report of polarized nonthermal radio
emission in the region despite many studies (although conceivably
nonthermal emission from a small supernova remnant could be absorbed by
free-free processes in the H{\sc II} gas).  The only hints of energetic
activity that could be associated with a supernova remnant are several
arc-like structures seen in the optical and low-frequency radio images
and some high-excitation optical emission lines
(\S\ref{review_M17.sec}).  But the latter cannot be supernova remnants
because there are no corresponding X-ray enhancements in the {\it
ROSAT} image (see the simulations of \citet{Velazquez03}).

The issue is somewhat more complicated for the Rosette Nebula because
of the presence of the large Monoceros Loop, a well-known old supernova
remnant, in its vicinity (\S\ref{review_Rosette.sec}). However, the
Monoceros Loop is far larger than the Rosette and is displaced, in the
plane of the sky and probably also along the line of sight, from the
region we study here.  Furthermore, the X-ray morphology of the nebula
is center-brightened and nicely fills the hole within the toroidal
H{\sc II} emission.  Most supernova remnants are either edge-brightened
with thermal spectra or plerionic with nonthermal spectra; only a
handful are known which are center-filled with thermal gas. Finally, we
recall that, like M~17, the NGC~2244 cluster is too young (1.9 Myr;
\S\ref{review_Rosette.sec}) and too poor to plausibly have produced a
supernova remnant.

We conclude that in both M~17 and the Rosette nebula the interpretation
of the diffuse emission as arising from the interior or shock of a
supernova remnant can be ruled out with considerable confidence.

\subsection{O Star Winds \label{yes_winds.sec}}
 
To investigate whether the properties of the plasma derived in
\S\ref{phys.sec} are consistent with an origin in the O star winds, we
apply the stellar wind properties obtained from ultraviolet P Cygni
lines for large samples of O stars \citep{Howarth89} to the O stellar
populations of the two clusters under study (see Appendix A).
Specifically, we apply the empirical relation $\log \dot{M} = 1.69 \log
(L_\star/L_\odot) - 15.41$, converting spectral subtypes to bolometric
$L_\star$ following \citet{Panagia73}.  This gives an estimated total
mass loss rate of $\Sigma(\dot{M}_w) \sim 3 \times 10^{-5}$ ($1 \times
10^{-5}$) M$_\odot$~yr$^{-1}$ for the M~17 (Rosette) ionizing clusters.

The derived X-ray emitting plasma masses (Table \ref{tab:phys}) can
thus be replenished by winds in $t_w \sim 5000$~yr for M~17, $t_w \sim
4000$~yr for Rosette.  As this is much shorter than the lifetime of the
OB stars, we infer that only a tiny fraction of the wind material
generated over the OB star lifetimes is now in X-ray emitting plasma.
A typical early O star wind with terminal velocity $v_w =
2500$~km~s$^{-1}$ would travel $r_w = v_w t_w = 12.8$~pc in M~17
(10.2~pc in Rosette) during the replenishment timescale.  This wind
kinetic timescale $t_w$ is also much shorter than the plasma cooling
timescale $\tau_{cool}$, consistent with the absence of any strong
cooling in the plasma as a function of distance from the O stars.  Most
importantly, we find that the wind kinetic energy $E_w \sim
\frac{1}{2}\Sigma(\dot{M}_w) v_w^2 t_w$ is 10 (7) times the plasma
thermal energy estimated above for M~17 (Rosette).  Thus, only $\sim
10$\% of the wind energy need be converted via shocks into hot thermal
plasma to account for the observed X-ray emission.

\section{Elaboration of the OB Wind Bubble Model \label{winds_elab.sec}}

From the analysis above, we find that the only plausible origin for the
$\sim 0.7$~keV (8~MK) diffuse emission suffusing the M~17 and Rosette
H{\sc II} regions is the thermalization of O star wind kinetic energy.
Our findings thus support the basic concept of shocked wind-swept
bubbles \citep[e.g.][]{Pikelner68, Dyson72, Weaver77} and provide some
of the first clear detections of X-ray temperature plasma in H{\sc II}
regions.

The calculations in \S\ref{yes_winds.sec} with respect to a simple wind
shock model show that only about 10\% of the expected wind kinetic
energy is converted into observed thermal energy.  The shocked wind
bubbles in M~17 and Rosette are $<10$~pc in radius.  For comparison,
the hydrodynamical calculation of \citet{Weaver77} predicts, for a {\bf
single} $t_\star = 1$~Myr O7 star expanding into a uniform cold medium,
that 30\% of the wind energy appears in the X-ray temperature gas,
which has a typical radius of 25~pc.  Similar discrepancies are found
between the relatively low displacement velocities in H$\alpha$ found
in recently studied LMC H{\sc II} regions \citep{Naze01}.  Thus, most
of the wind kinetic energy is not seen in the X-ray gas and does not
expand the bubble.

Several ideas have been raised to account for the missing energy and
momentum of OB winds.  The most obvious possibility, for blister H{\sc
II} regions which are only partially confined by cold gas, is that the
hot gas flows freely into the low density interstellar medium, away from
the cloud.  Hydrodynamical models of the expansion of stellar wind
bubbles in the presence of strong density gradients are presented by
\citet{Rozyczka85} and \citet{Comeron97}.  A flow of this type is
directly seen in our X-ray image of M~17 where the X-rays extend
asymmetrically several parsecs to the east, and may also be present in
the Rosette Nebula, where gas may flow nearly along the line of sight
or through fissures in the inhomogeneous cold cloud.  In the latter
case, the hot medium probably quickly mixes with cloud material and
becomes subject to rapid radiative cooling \citep{McKee84,
Capriotti01}.  

Another possible sink for wind kinetic energy is its conversion into a
turbulent cascade in a thick boundary layer where the wind approaches
the cold medium \citep{Kahn90}.  In this case, a high velocity shock
producing X-ray plasma is weak and most of the wind energy is
eventually radiated at longer wavelengths.  The soft $kT \simeq 0.1$~keV 
(1~MK) component seen in the Rosette and M~17 diffuse emission (Table
\ref{tab:spectra}) may be direct evidence for rapid cooling in a
post-shock or turbulent boundary layer.

A related possibility is that the wind loses most of its kinetic energy
through the entrainment of cooler material close to the star
\citep[e.g.][] {Arthur93, Dyson95, Pittard01}.  While mass loading
close to the star is unlikely in these mature H{\sc II} regions which
have evacuated molecular material within the stellar association, this
process may be occurring at the boundaries between the X-ray and cooler
gases.  In the Rosette, optical images show elephant trunk structures
which might be good sites for ablation and entrainment.  In M~17, the
stellar winds may be encountering a fissured neutral medium in the
highly fragmented southern H{\sc II} bar \citep[e.g.][]{Clayton85,
Meixner92}.

Another possibility is that the X-ray gas is produced, and the kinetic
energy spent, primarily in collisions between the OB winds rather than
with the exterior cold cloud.  \citet{Canto00} present a 3-dimensional
hydrodynamical calculation of colliding OB winds which are otherwise
unconfined by dense cold gas.  This is an attractive model for our data,
as it immediately accounts for the center-filled rather than
edge-brightened morphology of the X-ray emission and for the bulk flow
of the merged, shocked winds into the diffuse medium seen in M~17.
They find that the X-ray emission measure integrated along the line of
sight produces rather uniform emission filling the center of the
cluster, even though strong pressure gradients are present at the
wind-wind collision shocks.  The gas temperatures over the inner parsec
of the cluster are in the range $5-10$~MK, essentially identical to the
temperatures found here for the M~17 and Rosette diffuse emission.

We finally note that we do not have any clear evidence for very hard
diffuse X-ray emission that might emerge from non-Maxwellian particle
acceleration at a wind-cloud shock \citep{Dorland87}.  The smoothed
ACIS image of M~17 suggests that a hard (blue) diffuse component may be
present where the hot bubble encounters the M~17SW molecular cloud
(Figure \ref{fig:m17_color} panels a and c), but the emission is not
evident in the swiss-cheese image after point sources have been removed
(Figure \ref{fig:m17_patsmooth}a).  Our adaptive smoothing tool also
shows hints of diffuse hard emission near the cluster core and the
young embedded star-forming region M~17 North
(Figure~\ref{fig:m17_patsmooth}b-d) but again this is more likely to be
unresolved emission from embedded point sources in these regions.

\section{X-rays from High-mass Star Forming Regions (HMSFRs) \label{hmsfr.sec}}

\subsection{Summary of HMSFR X-ray results to date}

We place the M~17 and Rosette results in the broader context of diffuse
X-ray characteristics of other Galactic star forming regions to seek an
understanding of the conditions necessary for large-scale X-ray
emitting OB wind shocks.  Results from recent {\it Chandra} and {\it
XMM-Newton} observations (and one older {\it Einstein Observatory}
result) are summarized in Table \ref{tab:hmsfrs}, listed in order of
increasing reported diffuse X-ray luminosity.  We caution that this
should be considered a progress report rather than an authoritative
tabulation\footnote{Earlier versions of this table were presented at
conferences by \citet{Feigelson01} and \citet{Townsley03}.}.  The
regions are subject to different levels of absorption, were observed
with different sensitivities (in ergs~s$^{-1}$) and resolution (in pc),
and were subject to different treatments of the T Tauri X-ray source
population and diffuse emission.  In most regions, the underlying T
Tauri population is poorly characterized, and in some regions even the
OB stars are not fully identified. 

{\bf Nearby low-mass star forming regions (LMSFRs)}~~  Here we bundle
together recent high-resolution studies of the Taurus-Auriga, Ophiuchi,
and Perseus molecular clouds \citep[e.g.][]{Imanishi01, Preibisch01,
Getman02, Favata02}, along with a region of low-density star
formation in the Orion cloud \citep{Pravdo01}.  Two cases of faint
$10^{28}$~ergs~s$^{-1}$ diffuse (i.e. from interstellar gas rather than
the immediate vicinity of young stars) X-ray emission with extent
($<10^{-3}$~pc) have been found in these regions, associated with
small-scale shocks in high-velocity Herbig-Haro outflows
\citep{Pravdo01, Favata02, Bally03}.  No stars earlier than late-B are
present in these images, and none of these studies report diffuse X-ray
emission on parsec scales.  The sensitivity limit to such emission is
of order $10^{29}$--$10^{30}$~ergs~s$^{-1}$~pc$^{-2}$ or
$10^{-3}$--$10^{-2}$ of that seen here in M~17 and the Rosette.

{\bf Orion Nebula (M~42)}~~  With its `Trapezium' of the O6 star
$\theta^1$C Ori surrounded by a group of O7 V to B0 V stars, a full
initial mass function down to the brown dwarf regime, and an embedded
cluster associated with the OMC-1 core, the Orion Nebula is the nearest
HMSFR and a bright H{\sc II} region.  $\theta^1$C Ori is the principal
ionizing source of the H{\sc II} region and has a wind with mass loss
rate $\dot{M} = 8 \times 10^{-7}$~M$_\odot$~yr$^{-1}$ and terminal velocity
$v_\infty = 1650$~km~s$^{-1}$ \citep{Leitherer88}.   In extensive
studies with $Chandra$, no diffuse emission has been reported
\citep[e.g.][]{Schulz01, Feigelson02, Flaccomio03}. The upper limit to
parsec-scale emission is around $10^{30}$~ergs~s$^{-1}$~pc$^{-2}$.   Note,
however, that the O6 star is extremely bright in X-rays and populates
the wings of its point spread function, so that faint diffuse emission
could go unnoticed in the inner $\simeq$0.4~pc radius.

{\bf Eagle Nebula (M~16)}~~ A 77~ks {\it Chandra} image of this region
reveals over 1000 point sources, most associated with the ionizing
cluster NGC 6611 \citep{Mytyk01}.  The earliest stars in the cluster
\citep{Hillenbrand93} are the O5.5 V ((f)) star BD -13$^\circ$4923 and
HD~168076 (O5 V ((f)), $v_\infty = 3305$~km~s$^{-1}$, $\dot{M} = 3
\times 10^{-6}$~M$_\odot$~yr$^{-1}$).  (For this and following O stars,
we have taken terminal wind velocities $v_\infty$ from \citet{Prinja90}
and mass loss rates $\dot{M}$ from \citet{Howarth89}, both based on
studies of ultraviolet line profiles.)  Our inspection of the M~16 ACIS
image reveals no obvious diffuse emission at levels greater than $\sim
10^{30}$~ergs~s$^{-1}$, although detecting any faint diffuse emission
spatially coincident with the large number of point sources in the
image would require more careful analysis.

{\bf Lagoon and Hourglass Nebulae (M~8)}~~  The bright bipolar H{\sc
II} region in M~8 known as the Hourglass Nebula was studied with {\it
XMM-Newton} by \citet{Rauw02}.  Although limited by its small extent
and confusion with embedded stellar sources, soft diffuse emission with
luminosity $L_x \simeq 7 \times 10^{32}$~ergs~s$^{-1}$ and energy $kT =
0.6$~keV ($T = 7$~MK) was ``probably'' detected from the southern lobe of the
nebula.  The diffuse plasma luminosity lies below this value as some
contribution from the exciting star, Herschel~36 (O7 V), and lower mass
stars is undoubtedly present.  As we find here with M~17 and Rosette,
the authors note that this X-ray luminosity is far below the $10^{35}$~ergs~s$^{-1}$ kinetic power of Her~36.  An existing 60 ks
{\it Chandra} observation of M~8 did not include the Hourglass Nebula,
but a 180~ks observation that includes the Hourglass is scheduled.  The
{\it XMM-Newton} image of the larger M~8 region shows no diffuse
emission associated with the main ionizing star, 9~Sgr (O4 V, $v_\infty
= 2750$~km~s$^{-1}$, $\dot{M} = 5 \times 10^{-6}$~M$_\odot$~yr$^{-1}$),
and none is seen in our qualitative examination of the 60~ks {\it
Chandra} image. Note that the other major ionizing stars of M~8,
HD~165052 (O6.5 V + O6.5 V binary, $v_\infty = 2295$~km~s$^{-1}$,
$\dot{M} = 3 \times 10^{-7}$~M$_\odot$~yr$^{-1}$) lie just outside the
fields of view of both the {\it XMM-Newton} and {\it Chandra}
observations.

{\bf Rosette Nebula}  This HMSFR is discussed in detail above, with OB
stars listed in Appendix A.  The area of diffuse emission listed here
is the size of the extraction region used to obtain the diffuse
spectrum (the full ACIS-I array).  The diffuse emission is soft, with
spectral components at 0.1 and 0.8~keV (1 and 9~MK) and with an
absorption-corrected X-ray luminosity of $L_c \leq 6 \times
10^{32}$~ergs~s$^{-1}$ (Table~\ref{tab:spectra}).  The true plasma
emission is less than the observed value because of the uncertain, but
possibly substantial, contribution of unresolved T Tauri stars
(\S\ref{not_lowmass.sec}).  The H{\sc II} region is mainly ionized by
two stars with similar wind powers:  HD~46223 (O4 V ((f)), $v_\infty =
2910$~km~s$^{-1}$, $\dot{M} = 2 \times 10^{-6}$~M$_\odot$~yr$^{-1}$)
and HD~46150 (O5 V ((f)), $v_\infty = 2925$~km~s$^{-1}$, $\dot{M} = 2
\times 10^{-6}$~M$_\odot$~yr$^{-1}$).  Unlike the Trapezium and M~17,
where the O stars are concentrated within the inner $\simeq 0.5$~pc,
the two dominant stars in Rosette are widely separated by at least
3~pc.

{\bf RCW 38}~~ Both soft and hard diffuse X-ray emission is reported
from a 97-ks {\it Chandra} observation of this bright southern H{\sc
II} region \citep{Wolk02}.  Its ionization is dominated by a single O5
star, IRS~2.  The authors fit the diffuse X-rays with a thermal
bremsstrahlung plus power law model, attributing the soft
bremsstrahlung component to dust-scattered X-rays and the hard power
law component to some non-thermal magnetic process, perhaps synchrotron
emission from an embedded supernova remnant (see our discussion in
\S\ref{not_snr.sec}).  We note, however, that the diffuse spectrum in
RCW~38 is similar to the integrated point source spectrum we find in
M~17 (Figure~\ref{fig:spectra}, lower left).  We fit the latter sources
with a thermal plasma ($kT = 3.0$~keV, $N_H = 1.7 \times
10^{22}$~cm$^{-2}$) similar to one which fits the RCW 38 diffuse
emission ($kT = 2.23$~keV, $N_H = 1.15 \times 10^{22}$~cm$^{-2}$ with
low metal abundances, \citet{Wolk02}).  The authors use a deep $K$-band
image to argue that the diffuse emission does not arise from the
low-mass pre-main sequence population; it would be useful to make a
comparison between the diffuse spectrum and the combined point source
spectrum to elucidate the level of T~Tauri contamination.  We thus list
in Table \ref{tab:hmsfrs} the diffuse luminosity $L_x = 1.6 \times
10^{33}$~ergs~s$^{-1}$ (which we estimated based on the count rates and
spectral parameters reported in \citet{Wolk02}) as an upper limit to
the diffuse emission generated by OB winds.

{\bf Omega Nebula (M~17)}~~  These results are also discussed in detail
above, with OB stars listed in Appendix A.  The area of diffuse
emission listed here is the size of the extraction region used to
obtain the diffuse ACIS spectrum (see Figure~\ref{fig:m17_regions}a);
the {\it ROSAT} data show that fainter, soft diffuse emission extends
eastward for at least another 7~pc (see Figure~\ref{fig:m17_comp}d and
Dunne03 Figure 2).  The ACIS diffuse emission is soft like Rosette's,
with absorption-corrected $L_c = 3.4 \times 10^{33}$~ergs~s$^{-1}$.
Emission outside the ACIS field of view contributes another $\sim 1
\times 10^{33}$~ergs~s$^{-1}$ (Dunne03).  Here, spatial structure
combined with spectral comparison shows negligible contamination by low
mass stars.  The O stars in M~17 lack direct measurements of $v_\infty$
and $\dot{M}$, but Dunne03 calculate these quantities using the method
of \citet{Schaerer97} and find $v_\infty = 2500$--$3000$~km~s$^{-1}$,
$\dot{M} = 0.3$--$4 \times 10^{-6}$~M$_\odot$~yr$^{-1}$ (see their
Table 2).  Ionization of the H{\sc II} region is probably dominated by
the obscured O4--O4 binary called Kleinmann's Anonymous Star.
 
{\bf Arches Cluster}~~ The Arches is one of three very massive and
heavily obscured young stellar clusters in the inner 100~pc of the
Galactic Center.  A 51-ks {\it Chandra} image shows a complex
morphology:  most of the emission comes from two structures in the
interior of the core, which may arise from either localized wind-wind
collisions from close binaries or from more extended wind shocks
\citep{YusefZadeh02}.  The third component (A3), with $L_x \sim 1.6
\times 10^{34}$~ergs~s$^{-1}$, is clearly extended several parsecs away
from the stellar association, similar to the flow we see in M~17.  It
is this component which we list in Table \ref{tab:hmsfrs} as a lower
limit to the diffuse X-ray emission, for comparison with other regions.
Note that the emission is considerably harder than that seen in Rosette
or M~17 (although the fit parameters are uncertain due to the small
number of counts detected) and shows a strong 6.4~keV iron fluorescent
line, suggesting close interaction between the winds and ambient
molecular gas.  The heavy obscuration along this line of sight
precludes detection of a soft component analogous to the M~17 and
Rosette diffuse emission.

{\bf NGC~3603}~~ This very dense and luminous massive stellar cluster, often
considered to be the Galactic analog of the R~136 cluster at the center
of 30~Doradus in the Large Magellanic Cloud, lies at a distance similar 
to the Arches Cluster but is less obscured.  A 50-ks {\it Chandra}
study shows hard diffuse emission with an extent $\simeq 8$~pc with
$L_x \simeq 2 \times 10^{34}$~ergs~s$^{-1}$ and energy ranging from $kT
= 2$ to 3~keV ($T = 23$ to 35~MK), in addition to nearly 400 point sources
\citep{Moffat02}.  The diffuse component is spatially coincident with
the cluster core, but the authors calculate that the underlying
pre-main sequence population should be too faint to contribute
substantially to the emission.  The diffuse emission is attributed to
the collisions of multiple stellar winds from a large population of
massive (including O3-5 V and Wolf-Rayet) stars in the region.

{\bf Carina Nebula}~~ Carina is a giant H{\sc II} region extending over
$\sim 50$~pc, with an extremely rich young stellar population.  It is a
complicated mix of several very rich star clusters of varying ages and
evolutionary states, including many very massive O3 V stars,  post-main
sequence Wolf-Rayet stars, and the unique eruptive $\eta$~Car system.
Early study with the {\it Einstein} satellite revealed pointlike
emission from several early-type stars and a soft diffuse X-ray
component, with $L_x \simeq 10^{35}$~ergs~s$^{-1}$ extending over tens
of parsecs \citep{Seward82}.  \citet{Dorland87} concluded that a
substantial fraction of the {\it Einstein} emission at low energies
($<3$~keV) could be explained by T~Tauri stars similar to those seen in
$\rho$~Ophiuchus, while the hard X-ray contribution ($>3$~keV) could be
truly diffuse emission.  Portions of the complex have been observed
with {\it XMM-Newton} and {\it Chandra}; our examination of these
images confirms the presence of diffuse emission with large-scale
($>5$~pc) gradients that can not be attributed to an unresolved
low-mass stellar population.  \citet{Evans03} examined early {\it Chandra}
data and conclude that over half of the X-ray emission in Carina might
come from a diffuse component.  However, given the presence of many
evolved massive stars, it seems likely that one or more supernovae have
occurred in the region.  The relative contribution of supernova
remnants and OB wind shocks to the diffuse X-ray component is, in our
opinion, unknown.

{\bf Other regions}~~ Closely related observations have been made
which, for reasons of insufficient knowledge, we omit from
Table~\ref{tab:hmsfrs}.  A {\it Chandra} observation of the heavily
obscured group of ultracompact H{\sc II} regions in W~3 showed over 200
point sources but no diffuse emission \citep{Hofner02}.  However, any
soft emission with $kT \leq 1.5$~keV (17~MK) would be absorbed by intervening
material.  A faint X-ray structure with $kT \simeq 6$~keV (70~MK) and $L_x
\simeq 10^{33}$~ergs~s$^{-1}$ has been detected with {\it Chandra} in a
$1 \times 1$~pc region within the Sgr B2-North H{\sc II} region
\citep{Takagi02}. The absorption here is extremely high, with $N_H = 6
\times 10^{23}$~cm$^{-2}$ corresponding to over 100 magnitudes
of visual absorption.  It is not possible here to discriminate between
emission arising from individual massive stars, many low-mass stars,
and wind shocks.  X-ray emission has also been found with {\it ROSAT}
and {\it ASCA} in regions such as W~51, NGC~6334, and the Trifid Nebula;
{\it Chandra} data have been recently obtained for these regions.  Detailed
analyses of the contribution from diffuse hot wind bubbles based on
these new high-resolution images have not yet been reported, however.

\subsection{Implications for X-ray Emission from OB Wind Shocks
and the Structure of H{\sc II} Regions \label{implications.sec}}

Some tentative conclusions regarding the generation of parsec-scale
X-ray gas in star forming regions can be reached from these recent
findings.  Except for faint ($L_x \sim 10^{28}$~ergs~s$^{-1}$) and
localized ($<10^{-3}$~pc) emission from some shocks in Herbig-Haro
outflows \citep{Pravdo01, Favata02}, low-mass star forming regions show
no diffuse X-rays.  The terminal shocks of T Tauri winds do not produce
diffuse X-rays detectable with the current generation of X-ray
observatories.

\citet{Weaver77} and similar models of wind-blown bubbles predict
measurable X-ray emission associated with a single O star, even a
relatively late O7 star.  However the {\it Chandra} data show that the
powerful 2000--3000~km~s$^{-1}$ winds of {\bf early} O stars are clearly a
necessary condition to produce the shocks responsible for diffuse soft
X-ray emission.  \citet{Abbott82} showed that stellar wind power is a
strong function of spectral type; late-O stars have wind power reduced
by a factor of five from that of early-O stars, so HMSFRs without stars
earlier than O6 may be unlikely to exhibit diffuse soft X-rays.  Also,
the absence of such diffuse emission in the Orion, Eagle, and Lagoon
nebulae, where only one or two stars earlier than $\sim$O6 are present,
suggests another discrepancy with the \citet{Weaver77} prediction:  the
O stars must be sufficiently numerous to produce detectable diffuse
X-rays.  The proximity of the O stars may also be important:  it is
possible that colliding winds, as discussed by \citet{Canto00}, may be
more efficient at creating the kind of `X-ray fountain' we see in M~17,
with plasma flowing away from the exciting O star cluster.

The principal exception to these ideas is the Hourglass Nebula, where a
single O7 star may have associated soft diffuse X-rays.  Perhaps this
star's youth and proximity to molecular material enhances the
production of soft X-rays.  For parsec-scale emission, the Rosette
Nebula, with two widely-separated O4--O5 stars and $L_x {\rm
(diffuse)} \sim 10^{32}$~ergs~s$^{-1}$, may represent the minimal wind
environment necessary to produce detectable diffuse soft X-rays.  Even
in this minimal case, the X-ray plasma appears to have a sufficiently
strong dynamical effect to evacuate cooler material from the interior
of the nebula, resulting in the annular structure of its H{\sc II}
gas.

The Omega Nebula, Arches Cluster, and NGC~3603 cluster have richer O
star populations, producing $L_x {\rm (diffuse)} \sim
10^{33}$--$10^{34}$~ergs~s$^{-1}$.  Here the gas fills the interior of the
photoionized H{\sc II} region, pushing against the cold cloud material
where it is present and flowing away from the stellar cluster where it
is not.  In \S\ref{yes_winds.sec} and Table~\ref{tab:phys}, we give a
quantitative analysis of this type of nebula.  Less than 1\% of the
wind material produced over the history of the O stars is present in
the nebula, suggesting that most of the wind has flowed outward,
contributing to the Galactic hot interstellar medium.  Some of the gas
may cool, as evidenced by the 0.1~keV (1~MK) spectral component we see here.
The X-ray gas in M~17 is roughly in pressure equilibrium with the
cooler H{\sc II} gas.  The Rosette data imply that soft X-rays are
present without multiple O stars in close proximity.  The diffuse
emission of M~17 is not concentrated around the O4--O4 binary system,
implying that the collision of its component winds is not the principal
source of diffuse X-rays.  The $\geq 6$-fold difference in X-ray
luminosity we see between Rosette and M~17 is roughly consistent with
the difference in their O star populations (Rosette has 2 stars of type
O6 or earlier; M~17 has 7) and their wind kinetic energies ($2 \times
10^{48}$~ergs in Rosette; $9 \times 10^{48}$~ergs in M~17).

The Carina Nebula, where the diffuse soft X-ray luminosity is of order
$10^{35}$~ergs~s$^{-1}$, represents the first stages of starburst
phenomena, where rich stellar clusters of a variety of ages have an
enormous effect on their environment.  Many powerful winds are
colliding with each other, encountering the cloud edges, and quite
possibly interacting with supernova remnants.  The X-ray plasma is
probably not relatively uniform with a simple laminar outward flow as
we see in M~17.  Rather, motions are likely in multiple directions, with
shocks, turbulence, and strong density inhomogeneities.  Such a
superheated, shock-dominated hot interstellar environment is seen in
starburst superbubbles such as 30 Doradus \citep{Chu90, Dennerl01}, the
mild starburst nucleus in our Galactic Center region \citep{Wang02,
Muno03}, and in true nuclear starbursts such as NGC~253
\citep{Strickland02} and M~82 \citep{Griffiths00}.  In the latter case,
the overpressure of the supernova- and wind-heated interstellar medium
drives  kiloparsec-scale winds of enriched material into the
intergalactic medium.

A critical issue raised by these studies is the likely presence of a
hard ($kT \sim 5$~keV, or 58~MK) as well as soft ($kT \simeq 0.8$~keV,
or 9~MK) plasma in some portions of some H{\sc II} regions.  It is
reported in RCW~38, NGC~3603, and the Arches Cluster but not in W~3 or
the less rich stellar associations like the ONC or Rosette.  In the
intermediate case of M~17, unresolved hard X-rays are spatially
coincident with the highest concentrations of stars; a diffuse
component cannot be distinguished from an unresolved component due to
the large population of faint, embedded pre-main sequence stars.  In
all these examples, we must be aware of the possible contamination by
the population of low-mass embedded stars masquerading as diffuse
gaseous emission; in regions with evolved stars or a range of cluster
ages, contamination by supernova remnants must also be considered.  If
real, this diffuse hard component must represent a more efficient
conversion of wind kinetic energy into plasma thermal energy.  It may
require nonlinear transport processes at the interface of the wind and
molecular cloud as described by \citet{Dorland87}.

One possibility for the eventual dissipation of the wind energy is that
the plasma pressure overcomes the ambient pressure at a weak point,
such as a low-density region at the periphery of a molecular cloud;
this would be reminiscent of the `champagne flow' invoked for H{\sc
II} regions like Orion.   This is probably the case for M~17, which
displays a clear flow of gas toward the boundary with the interstellar
medium.  Alternatively, the plasma may `leak' through fissures, simply
flowing past denser regions; this might be the case for Rosette,
although a flow nearly along the line of sight is also possible.  In
both cases, the gas then escapes into the surrounding low-pressure
interstellar medium, adding to the network of worms and chimneys
created in the galactic disk by supernova explosions \citep{deAvillez01}.  

If the surrounding medium is dense enough to hold the plasma, the wind
energy must be efficiently dissipated at or near the periphery of the
plasma bubble, since there appears to be little or no expansion of the
H{\sc II} shell.  In the case of a steep density gradient (such that
the slope is smaller than the mean free path of the plasma electrons in
the H{\sc II} region), \citet{Dorland87} have shown that no Maxwellian
electron velocity distribution is possible, and that a high-energy
electron tail efficiently takes away the wind energy via non-linear
conduction.  These non-thermal electrons, which would radiate hard
X-rays, end up losing their energy unnoticed in the surrounding
photodissociation region.  This may be occurring around the embedded
RCW~38 cluster, if the hard diffuse X-ray emission there is not due to
a supernova remnant and/or the embedded stellar population.  Another
possibility for wind energy dissipation is that the hot plasma directly
evaporates the surrounding medium by photoevaporation resulting from
the complete absorption of the (soft) X-ray photons and/or from heating
by the bremsstrahlung keV electrons.

\section{Concluding Comments \label{conclusions.sec}}

\subsection{Summary of Findings}

{\it Chandra} ACIS-I images of the Rosette and M~17 H{\sc II} regions
in the 0.5--8~keV X-ray band reveal diffuse emission extending over
several parsecs.  Unlike studies with previous satellites, the high
resolution of the {\it Chandra} mirrors and detectors and low detector
noise allow faint diffuse emission to be studied in quantitative detail,
even in the midst of hundreds of stellar X-ray sources.  Strong
arguments are presented that the emission is not attributable to
supernova remnants or low-luminosity stellar sources.  It is produced
by plasma at $T \sim 1$~MK and $\sim 10$~MK with luminosities of
$\leq 6 \times 10^{32}$~ergs~s$^{-1}$ in Rosette and $3.4 \times
10^{33}$~ergs~s$^{-1}$ in M~17.  The emission has a roughly uniform,
center-brightened appearance.  In Rosette, the X-ray gas surrounds the
ionizing cluster, filling the hole within the well-known ring of
H$\alpha$ and radio continuum emission.  In M~17, the gas is roughly
coincident (in projection) with the H$\alpha$ emission, brightest
around the cluster core and flowing eastward for several parsecs.

\subsection{Astrophysical Implications}

We establish that only a small portion of the wind energy and a tiny
fraction of the mass appears in the observed diffuse X-ray plasma.
Although cooling into the H{\sc II} component is possible, we suspect
that most of the wind energy and mass of these blister H{\sc II}
regions flow without cooling into the low-density interstellar medium.
In this case, the fate of deeply embedded winds where the stars are
completely surrounded by dense molecular gas will be different than in
these blister H{\sc II} regions.

Together with other recent reports from {\it Chandra} and {\it
XMM-Newton}, this is the first unambiguous detection of the `wind-swept
bubble' which has long been predicted to arise from the thermalization
of OB stellar winds.  We estimate that $\sim 10$\% of the wind kinetic
energy is converted to X-ray emitting plasma, although it is still not
clear whether most of the dissipation occurs in wind-wind or wind-cloud
shocks.  We suspect that both types of shocks are present.  
The low-temperature spectral component suggests that some post-shock
gas is cooling, but most of the gas mass and most of the kinetic energy
of the winds probably stay in the form of bulk flow.  This is clearly
seen in M~17, where the X-ray emission flows from the core of the
stellar cluster, away from the dense molecular cloud and towards low
density interstellar environments.  The OB wind energy is not, at least
in these two cases, principally dissipated in turbulent layers or mass
loading close to the O stars.

For some regions (such as Rosette), the physical conditions of the
$10^4$K gas may not be well-established, if the derived pressure can
change by a factor of 100 with different reasonable assumptions for
geometry, clumpiness, and filling factor; compare the models of
\citet{Celnik85} and \citet{Tsivilev02}.  Derived physical parameters
can also change substantially if the stellar mass loss rates change by
factors of 3 or more when clumping due to turbulence is considered
\citep{Moffat94}; this was illustrated by Dunne03 for the case of
M~17.  Dunne03 also showed that the presence of strong magnetic fields
may inhibit heat conduction, thus reducing the X-ray luminosity in M~17
below the values predicted by the standard wind-blown bubble models.

New high-resolution optical and near-infrared images of the hydrogen
line structures in H{\sc II} regions are showing a rich, sub-pc
filamentary structure in the photodissociation regions
\citep{Jiang02}.  This may change our ideas about the interface between
the inner ten million degree plasma and the outer ten thousand degree
photodissociated gas, hence about their possible exchanges of energy.
The two media may well coexist in some fashion over a certain distance,
a situation reminiscent of supernova remnants expanding into the
interstellar medium.  These considerations require that the classical
`Str\"omgren sphere' paradigm be seriously revised to better account
for the existence of stellar winds.

In fact, one of the potential implications of this finding may be the
need to reformulate our language and overall conception of H{\sc II}
regions.  In M~17, the apparent outflow of X-ray gas from the NGC~6618
cluster away from the molecular cloud might be termed an `X-ray
fountain' or `X-ray champagne flow,' in analogy to the champagne model
of photoevaporation of dense gas by OB ultraviolet radiation
\citep{TenorioTagle79}.  For H{\sc II} regions where most of the
interior volume is filled with an X-ray emitting plasma, as in M~17 and
Rosette, the cooler ionized gas responsible for H$\alpha$ and radio
continuum emission is confined to a relatively thin layer near the
boundary with the molecular cloud.  Thus the phrase `Str\"omgren
sphere' might be replaced with `Str\"omgren shell' or `Str\"omgren
surface.'  For M~17, this idea is supported by a recent high-resolution
$JHK$ image where the hydrogen line emission is mostly concentrated in
filaments and a thin `silver lining' around the dark
cloud\footnote{\url{http://www.pparc.ac.uk/Nw/Press/uist.asp}}. 

The current sample of HMSFRs studied with modern X-ray instrumentation
shows a large range of diffuse emission properties, with reports of
both hard and soft, bright and faint emission and spatial scales of
fractions of a pc$^2$ to thousands of pc$^2$.  Generally, nearby,
less-obscured regions show soft, faint diffuse emission filling
moderate-sized, presumably wind-blown bubbles.  More distant, obscured
regions ionized by more early stars may sometimes exhibit hard diffuse
emission, but the contribution of low-mass stellar sources is difficult
to untangle.  The soft component may be present but is probably
absorbed.  The specific character of diffuse X-rays in HMSFRs is
governed by the details of star formation and the distribution of
molecular material associated with each region. 

In addition to clarifying H{\sc II} region astrophysics, establishing
the effects of OB winds on their environment is relevant for
understanding the energetics of the interstellar medium on a Galactic
scale.  Stellar wind bubbles within an H{\sc II} region can affect the
evolution of later supernova remnants and superbubbles \citep{Oey01}.
The integrated mechanical input of stellar winds generally exceeds that
of supernova remnants over the full lifetime of a starburst event and,
if efficient thermalization occurs, winds and supernovae together can
power galactic superwinds \citep{Leitherer92}.

\subsection{Prospects for Further Study}

{\it Chandra} study of H{\sc II} regions containing a variety of
ionizing stellar clusters, with different wind sources and ages, and in
different gaseous environments, is likely to give additional insights
into the fate of OB winds.  We suspect that X-ray plasma will be
commonly found in H{\sc II} regions when several O stars with
sufficiently strong winds are present.  A deeper {\it Chandra}
observation of the Rosette Nebula is planned, both to improve the
discrimination between diffuse gas and the low-mass stellar population
and to search for subtle gradients in the gas properties.  For example,
spectral or brightness changes in the X-ray emission around the
molecular gas would show the wind termination shock (see
\S\ref{winds_elab.sec}), while bright X-ray arcs between the principal
O stars would reveal a wind-wind collision shock.  The wind-cloud
vs.\ wind-wind heating mechanisms can also be tested by comparing the
diffuse X-ray emission in embedded clusters to those which have largely
evacuated cold gas from their environment.

Study of the optical line emission from these H{\sc II} regions
specifically to elucidate the wind-swept bubble model would also be
helpful.  High-resolution imaging may confirm that the line emission is
confined to a shell around the photodissociation region and does not
permeate the entire volume around the ionizing cluster.  Optical line
spectroscopy can investigate high-velocity components in the H{\sc II}
region, such as knots with $\simeq 100$~km~s$^{-1}$ velocities reported
by \citet{Meaburn87} in M~17 and by \citet{Clayton98} in Rosette, to
study the dynamical influence of the winds on the cooler gas.

While hot wind-swept bubbles in H{\sc II} regions have been extensively
modeled for several decades, most of these studies do not make specific
predictions of X-ray luminosities, temperatures, and morphologies.
Notable exceptions are the studies by \citet{Dorland87},
\citet{Comeron97}, and \citet{Canto00}.  These models are beginning to
be revisited in light of the new {\it Chandra} findings
\citep{Stevens03}.  We suspect, for example, that winds with heavy mass
loading or thick turbulent mixing layers are incompatible with our
results while winds with strong shocks are favored, but \citet{Stevens03}
favor mass-loading from the disks of pre-main sequence stars to explain
the luminosity and temperature of Rosette.  It would also be
desirable to construct numerical three-dimensional hydrodynamical
calculations with the specific configuration of wind sources and cold
gaseous boundaries present in Rosette and M~17.  Comparing the
predicted X-ray properties to our observations should give new insights
into the structure (e.g., wind-wind vs.\ wind-cloud) and physics of the
shocks that heat H{\sc II} regions to X-ray temperatures.

\acknowledgments{Support for this effort was provided by the Chandra
X-ray Observatory grant G01-2008X and by NASA contract NAS8-38252.  EDF
thanks CEA/Saclay for hospitality during the study period.  This
research has made use of NASA's Astrophysics Data System and the SIMBAD
database, operated at CDS, Strasbourg, France.  We appreciate the
insight gained through conversations with A.\ Dalgarno, K.\ Getman,
C.\ Grant, M.\ Mac~Low, C.\ McKee, J.\ Raymond, and C.\ Sarazin.  We thank
R.\ Chini for supplying data and our referee, N.\ Miller, for his time
and helpful comments, which improved the quality of this paper.  This
study was a true collaboration, with equal contributions made by the
first three authors in the development of the paper and substantial
supporting roles played by the other authors.  LKT is most grateful for
this ensemble effort.}


\section{Appendix A: Massive Stars in M~17 and Rosette}

Tables \ref{tab:m17_OB} and \ref{tab:rosette_OB} give properties of the
O and early B stars (through B3) in NGC~6618, the cluster exciting
M~17, and NGC~2244, the cluster exciting the Rosette Nebula,
respectively.  The tables are ordered by spectral type.  All values are
obtained from the literature except for the final column, which
indicates whether the star is detected in our {\it Chandra} ACIS
images.  There may be additional early-type members of NGC~6618; visual
studies of this cluster are complicated by heavy obscuration and
confusion with the bright H{\sc II} region and near-infrared studies
have not covered the entire cluster, which may extend up to $25\arcmin$
($\sim 11.6$~pc) in diameter \citep{Perez91}.  In Rosette, the census
of early stars is probably more complete because the cluster core is
less absorbed.  Some of the stars listed in Table~\ref{tab:rosette_OB},
however, lie well outside the cluster coronal radius of 7.95~pc ($\sim
19.5\arcmin$) measured by \citet{Nilakshi02} and thus may not be
members of NGC~2244.  They may be associated with other star-forming
regions in the Mon~OB2 association or the RMC, since they possess
other properties that are consistent with NGC~2244 cluster membership
\citep{Kuznetsov00}.

\section{Appendix B: ACIS Data Reduction Procedure \label{cti.app}}

Our ACIS data reduction begins with the Level 1 event list (`evt1
file') and associated files (e.g., the telescope aspect solution,
contained in the `asol1' file).  We require Level 1 data products
because they contain information that is unavailable in the more highly
processed Level 2 files.  Many of the diagnostic plots described below
are easily generated by {\it Event Browser}, part of the `Tools for
ACIS Review and Analysis' (TARA) package developed by Patrick Broos and
other members of the ACIS team and available to all interested
users\footnote{\url{http://www.astro.psu.edu/xray/docs/TARA/}}.

The first step is to improve the absolute astrometry of the field by
registering X-ray point sources with (typically) visual catalogs of
high astrometric precision.  If enough matches are found in the field,
we use the Tycho 2 catalog of Hipparcos positions \citep{Hog00}.  If
the field contains few Tycho matches (as is the case for M~17) we can
usually obtain enough matches with the USNO A2.0 catalog
\citep{Monet98} to improve the positions.  Preliminary {\it Chandra}
source positions are obtained by running {\it wavdetect}
\citep{Freeman02} on a Level 2 event list, including only the central
part of the ACIS-I array for speed and because the small PSFs in this
region will yield better X-ray positions.  We then compute the median
offset between X-ray and visual positions in RA and Dec separately,
discarding outliers that are likely due to mismatched sources.  The
astrometry of the observation is adjusted by offsetting the RA and Dec
columns in the aspect file, offsetting the nominal pointing
coordinates, and reprojecting the event data.  

We remove the pixel randomization when the events are reprojected; in
our experience, any observation longer than a few ksec samples enough
pixels due to spacecraft dither that any pixel-to-pixel gain variations
are already averaged out, and the best possible positions and sharpest
PSFs are critical here, to enable source matching with catalogs from
other wavebands.  Note that the sky coordinates (x,y) of the events
will not change (except for the removal of the pixel randomization);
reprojection simply moves the x/y system on the sky (in both asol1 and
evt1 files).  A comparison of the old and new positions is made by
examining the distribution of the distance each event moved.  The pixel
randomization consists of uniform random numbers [-0.5,0.5] added to
the CCD chip coordinates.  With a non-zero roll angle, the
randomization in x/y coordinates is not a uniform distribution but it
should have a characteristic shape, with a maximum value of
$\sqrt{0.5^2 + 0.5^2} = 0.7$~pixels.

The data are then examined for possible processing errors.  The set of
flight event grades [24, 66, 107, 214, 255] are removed in the onboard
processing (they are not telemetered to the ground).  Thus, the
appearance of many events with these grades in the Level 1 data
strongly suggests a processing error, such as the use of an incorrect
bias.  If the field contains a very bright source then we have the
opportunity to look for a variety of problems in the aspect solution.
By plotting event x and y positions versus event arrival time for a
small region around the brightest source we may look for ``movement''
of the source across the sky.  Incomplete de-dithering might produce
periodic movement; aspect camera problems might produce jumps in the
position; motion of the Science Instrument Module at the beginning or
end of the observation will make the source appear to run off the field
completely.

If the data were obtained in ``Timed Event, Very Faint'' mode, each
event consists of a $5 \times 5$ pixel event island.  Code was
developed by Alexey Vikhlinin
(CXC)\footnote{\url{http://hea-www.harvard.edu/\-~alexey/\-vf\_bg/vfbg.html}}
to use the outer pixels in this island as a way to reduce background.
This routine sets a status bit that flags events with high outer
pixels.  It will flag good events from bright sources as well as bad
events, so it must be used with caution.  For appropriate data, this
routine is applied at this stage of the processing and the event list
is filtered to remove flagged events.  As with all filtering steps, we
carefully examine the events that are discarded to make sure that the
filtering was appropriate and was applied correctly.

The data are then corrected for radiation-induced charge transfer
inefficiency (CTI) using the method developed at Penn State
\citep{Townsley02a}.  Correcting for CTI improves spectral resolution
of extended sources by $\sim$20--40\%.  This is illustrated in
Figure~\ref{fig:cti-res}, which gives the energy resolution of ACIS-I
CCDs as a function of CCD row number both before and after CTI
correction, shown using the FWHM of the Al~K$\alpha$ and Mn~K$\alpha$
calibration lines in the ACIS External Calibration Source (ECS)
\citep{Townsley02b}.

The latest release of CIAO (November 2002, Version 2.3) now includes a
CTI corrector for front-illuminated (FI) CCDs, developed by the MIT
ACIS group and the CXC.  It employs the same forward-modeling approach
used in the Penn State corrector, although the implementation and
details of the model are different.  This code was not available when
the M~17 and Rosette data were analyzed and does not apply to
back-illuminated (BI) devices, so the Penn State corrector was used;
which code we use in the future will depend on performance, as assessed
by CTI correction of the ACIS ECS data.  These calibration observations
are made every orbit, as a way to monitor CTI and other aspects of ACIS
performance.  They can be used as a check on the uniformity of the
energy calibration across the field and are available via the
Provisional Data Retrieval
Interface\footnote{\url{http://cxc.harvard.edu/cgi-gen/cda/retrieve5.pl}}.
The CXC can provide users with the observation identification number
(``obsid'') of the calibration observations most appropriate for a
given celestial observation.

To check the Penn State CTI corrector performance on the M~17 data, we
ran the corrector on the closest ECS data, observation ID 61277.  A
plot of chipy vs.\ energy showed that the corrector still worked well
for improving spectral resolution; it was tuned using data from January
2000 through January 2001, so it was prudent to check its performance
for the M~17 data which were obtained in March 2002.  This test did
reveal a problem, however:  the corrector was not doing a good job of
regularizing the gain across the CCD, so the event energies showed a
clear decline of $\sim 1$\% from the bottom of the CCD to the top.
This test motivated us to include the time-dependence of the CTI as
measured by Catherine
Grant\footnote{\url{http://space.mit.edu/$\sim$cgrant/cti/cti120.html}}
in the Penn State CTI corrector.  This new code restored the flat gain
across the CCDs for this observation; in fact, this time-dependent
corrector surpasses the original corrector for {\it all} ACIS
observations made at a focal plane temperature of -120C, for both FI
and BI devices.  Analysis products developed for use with the original
Penn State CTI corrector (RMF and QEU files) are still valid for the
time-dependent corrector; these products and the corrector code are
available
online\footnote{\url{http://www.astro.psu.edu/users/townsley/cti}}.

The events are then filtered in various ways.  The standard grade
filter (retaining only {\it ASCA}-type grades 0, 2, 3, 4, and 6) is
applied and events with bad grades are examined to make sure that they
are truly bad (for example, legitimate X-ray point sources might show
up in an image made of the bad grades---this would be an indication of
photon pile-up in these sources and a warning to the user that these
sources must be analyzed more carefully).  The Level 1 event lists
include a ``status'' column containing 32 single-bit flags representing
various reasons why an event may be deemed undesirable\footnote{See
\url{http://space.mit.edu/$\sim$gea/docs/acis\_event\_status\_bits.html}
for the definitions of these flags.}.  Most observers will want to
respect most of these flags, removing events which have a flag value of
1.  There is one notable exception:  bits 16--19 are set by the
pipeline tool {\it acis\_detect\_afterglow} in an attempt to identify
spurious events due to cosmic ray afterglows.  This tool suffers the
problem of flagging real events from even moderately bright
sources\footnote{See
\url{http://www.astro.psu.edu/users/tsuboi/FlaringPixel/} for an
explanation of this problem.}.  By simply applying a [status=0] filter
or by using the Level 2 event list, significant numbers of legitimate
events from point sources will be discarded.  Again, this behavior is
not restricted to bright point sources.  To avoid this problem, at this
stage of the processing we follow the uncontroversial status flags but
ignore flags 16--19 by applying this filter:
[status=000000000000xxxx0000000000000000].

The CXC distributes a `flight timeline' (flt1.fits) file containing
Good Time Interval (GTI) tables which designate certain time intervals
deemed ``bad'' by the CXC pipelines.  We do not know exactly what
criteria are used by the CXC to determine these Good Time Intervals,
but many observations contain small gaps in the aspect solution which
are recorded in the ``flight timeline.''  Filtering with these GTI
tables is applied as part of the Level 2 processing, and we apply it
here.  We search for additional aspect glitches by examining the
acis.asol1 file, a timeseries of the absolute pointing (RA, Dec, Roll
angle).  

Due to solar activity, the particle background can increase
substantially over short timescales.  We search for these background
flares, which are especially troublesome on BI CCDs,  by examining
lightcurves from each CCD using the CXC's {\it lightcurve} tool.  We then
modify the GTI files and apply them to the event list to exclude times
containing flares.  We always verify that the cleaned data no longer
show flares by re-computing the lightcurves.

The data are then searched for hot columns or hot pixels that are not
removed by the standard processing.  These can be found by careful
examination of each CCD (in chip coordinates); the image must retain
1-pixel binning in order to see the hot columns.  Another good way to
find hot columns is to make a histogram of events as a function of the
CHIPX coordinate; columns with too many events are then easily
noticed.  It is important to check the energies at which these bad
columns go away.  We have found that most are gone above 0.7~keV,
virtually all are gone above 1~keV, and absolutely all are gone above
2~keV.  The worst CCD by far is the noisy CCD8 (S4).  We find that we
almost always have to filter CCD0 (I0), column 3, to remove events
below 700~eV.  Also, there is a hot pixel in CCD6 (S2), at chip
coordinates (766,501).  Note that any spatial filtering done here is
{\it not} reflected in the bad pixel file and is thus not accounted for
in ARF or exposure map calculations.  For making smoothed images, it is
better to remove the bad columns and suffer an imperfect exposure map
than to leave them in, especially if soft extended emission might be
present or there are other reasons to work in very soft bands
($<1$~keV).

CIAO provides the command {\it acisreadcorr} to remove readout streaks
due to bright
sources\footnote{\url{http://asc.harvard.edu/ciao/ahelp/acisreadcorr.html}}.
We do not use it in this recipe, however, because we are mainly
concerned with large fields full of point sources.  In these fields, it
is highly likely that other point sources fall under the readout streak
of the bright one.  The {\it acisreadcorr} operation removes {\it all}
the events in the streak and replaces them with artificial data
according to some spectral shape of the nominal background that the
user provides.

This filtered event list is then examined to make sure that it still
contains a GTI table for each CCD and that appropriate FITS header
keywords are present and set to the right values.  A very coarse energy
filter is then applied to remove events that we know are background.
These energy limits (0--10.5~keV) are well outside the range that will
be used for spectral fitting.  This needs to be done here even if it
was done before CTI correction, because the corrector moves some events
up to very high energies when it corrects cosmic rays.

Note that these data still have flaring pixels in them, because
applying the flaring pixel filter removes good events.  These data have
not been band-limited (filtered strongly by energy or PI).  Further
filtering by energy is useful when making images and for timing
analysis, but must not be done for spectral analysis because the low
and high energy ranges specified will, in general, fall in the middle
of PI or PHA bins, which are used to construct spectra.  Generally
speaking, the user should specify the energy range over which spectra
will be fit directly in the fitting package, rather than by
band-limiting the data.  The flaring pixel filter and an energy filter
to minimize background are applied before source-searching is done.

\section{Appendix C:  An Alternative Method for Adaptive-kernel Image Smoothing}

\begin{figure}
\centering
\includegraphics[width=0.45\textwidth]{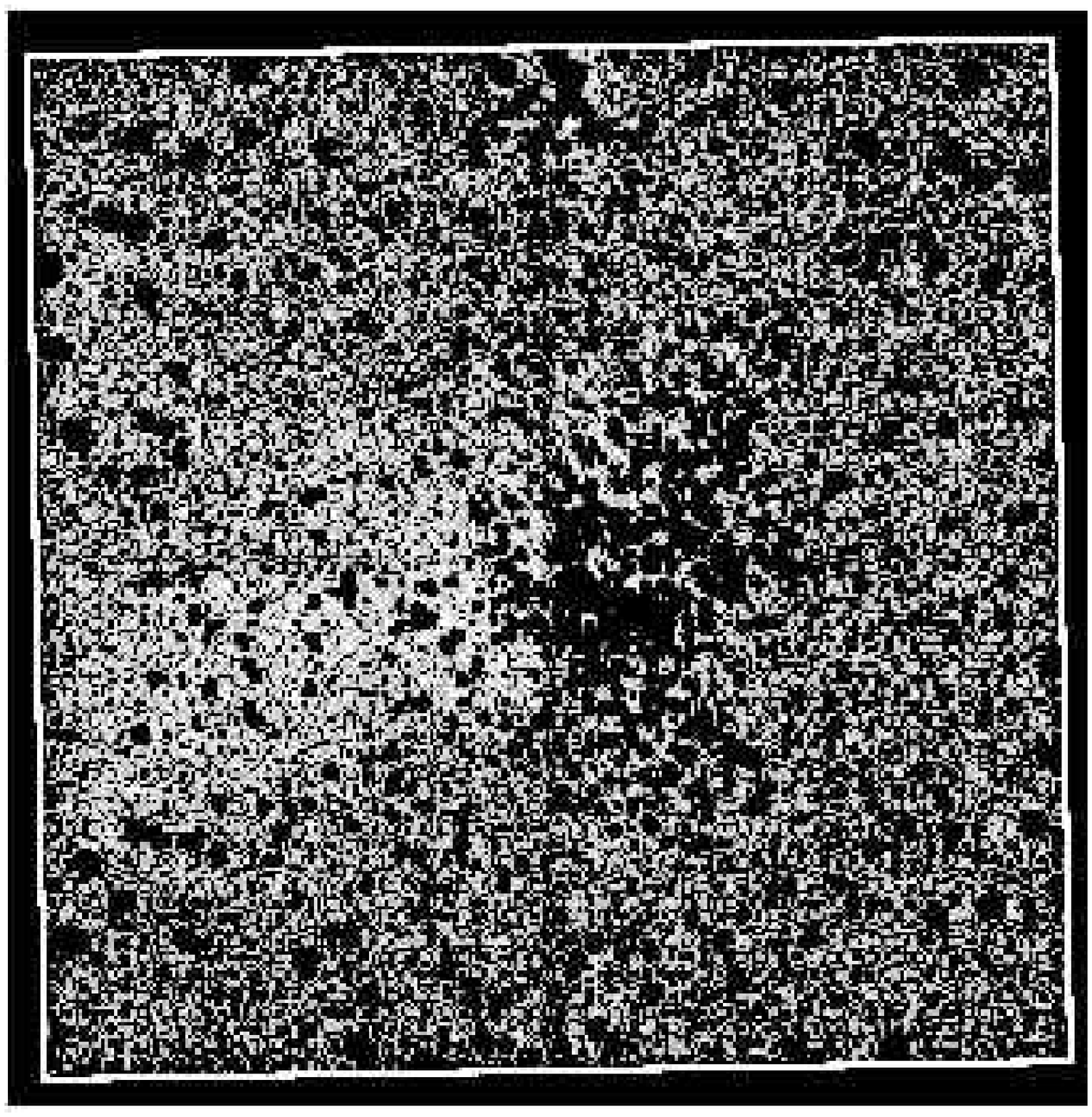}
\hspace{0.05in}
\includegraphics[width=0.45\textwidth]{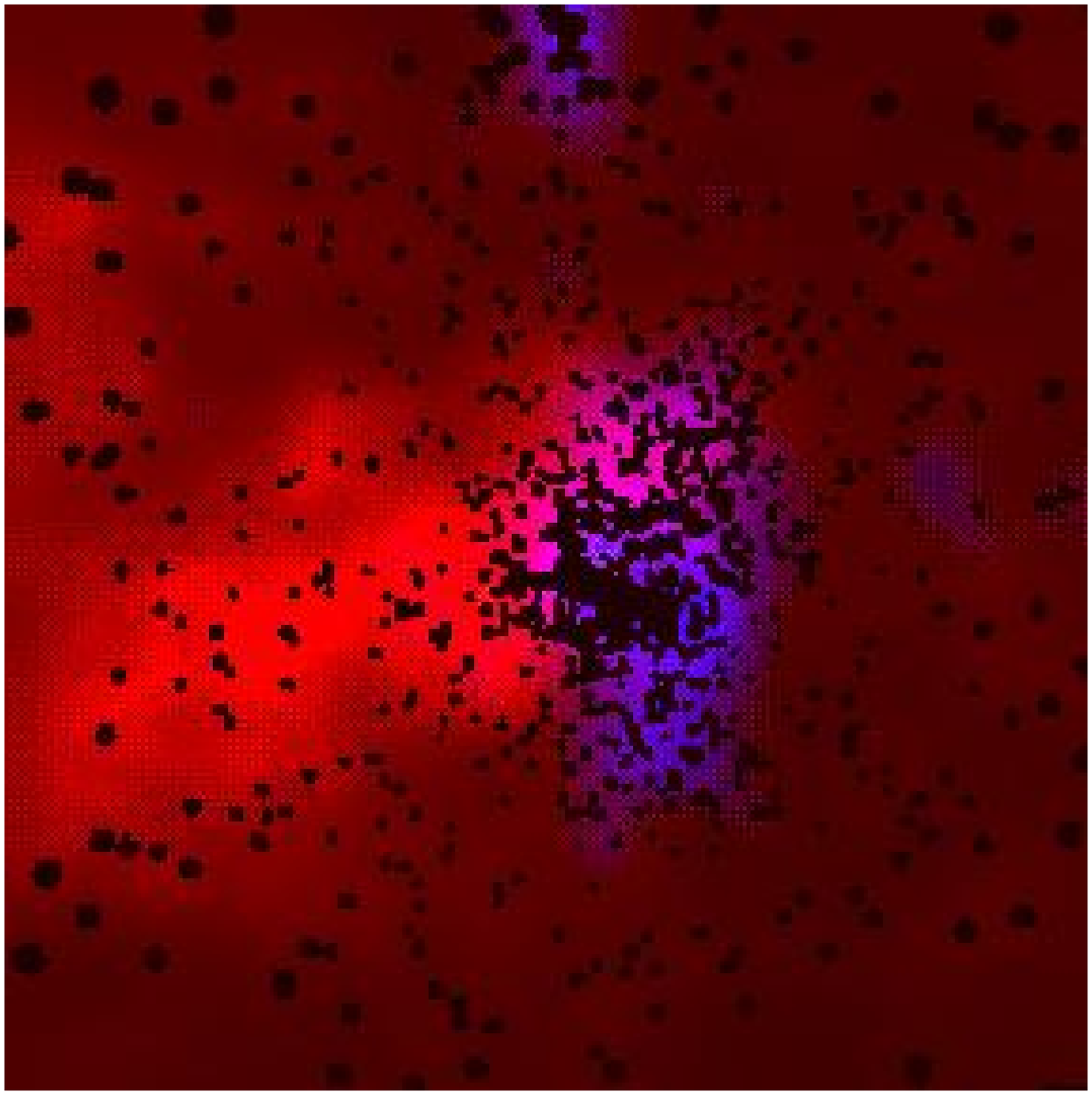}
\includegraphics[width=0.45\textwidth]{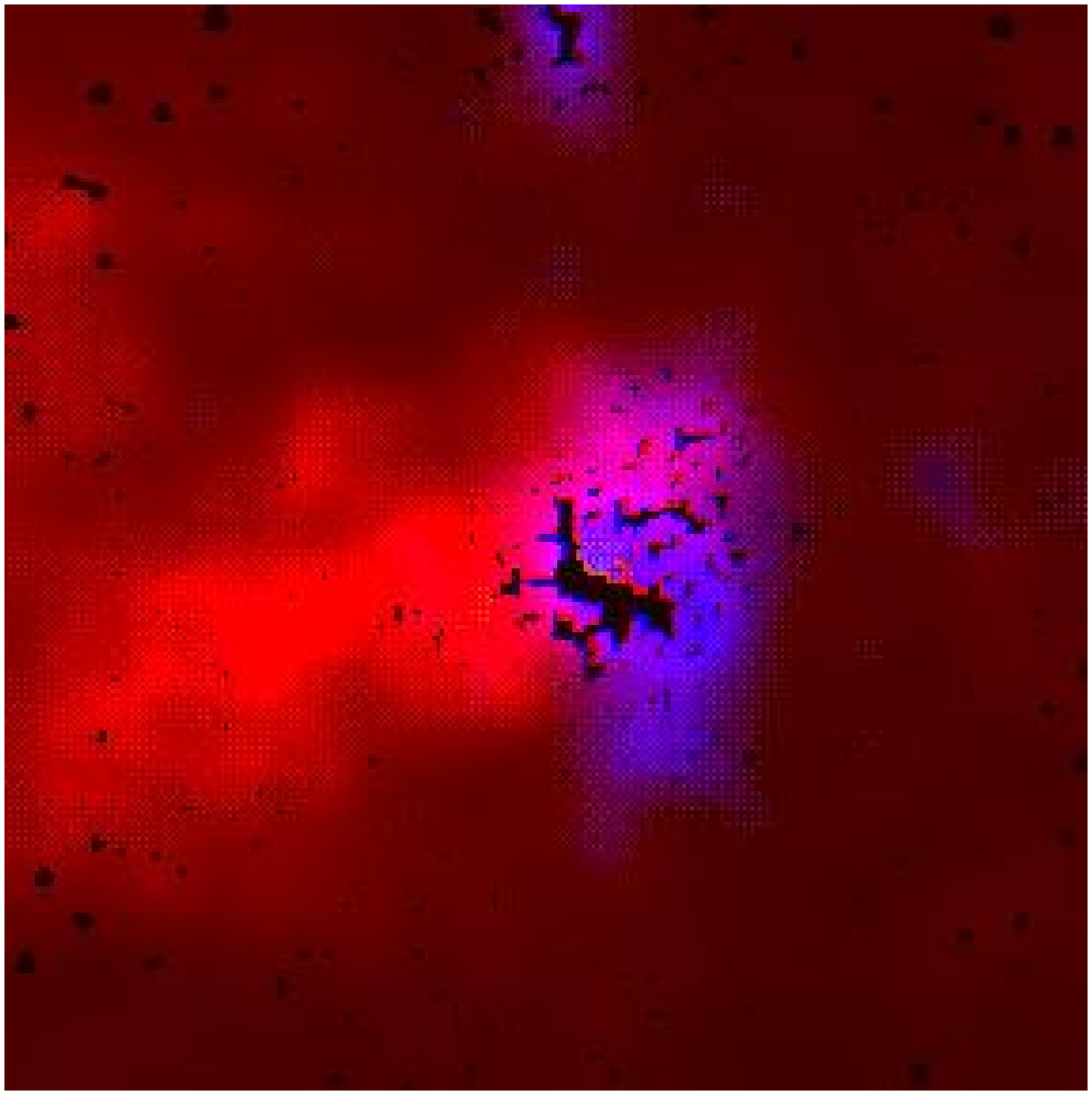}
\hspace{0.05in}
\includegraphics[width=0.45\textwidth]{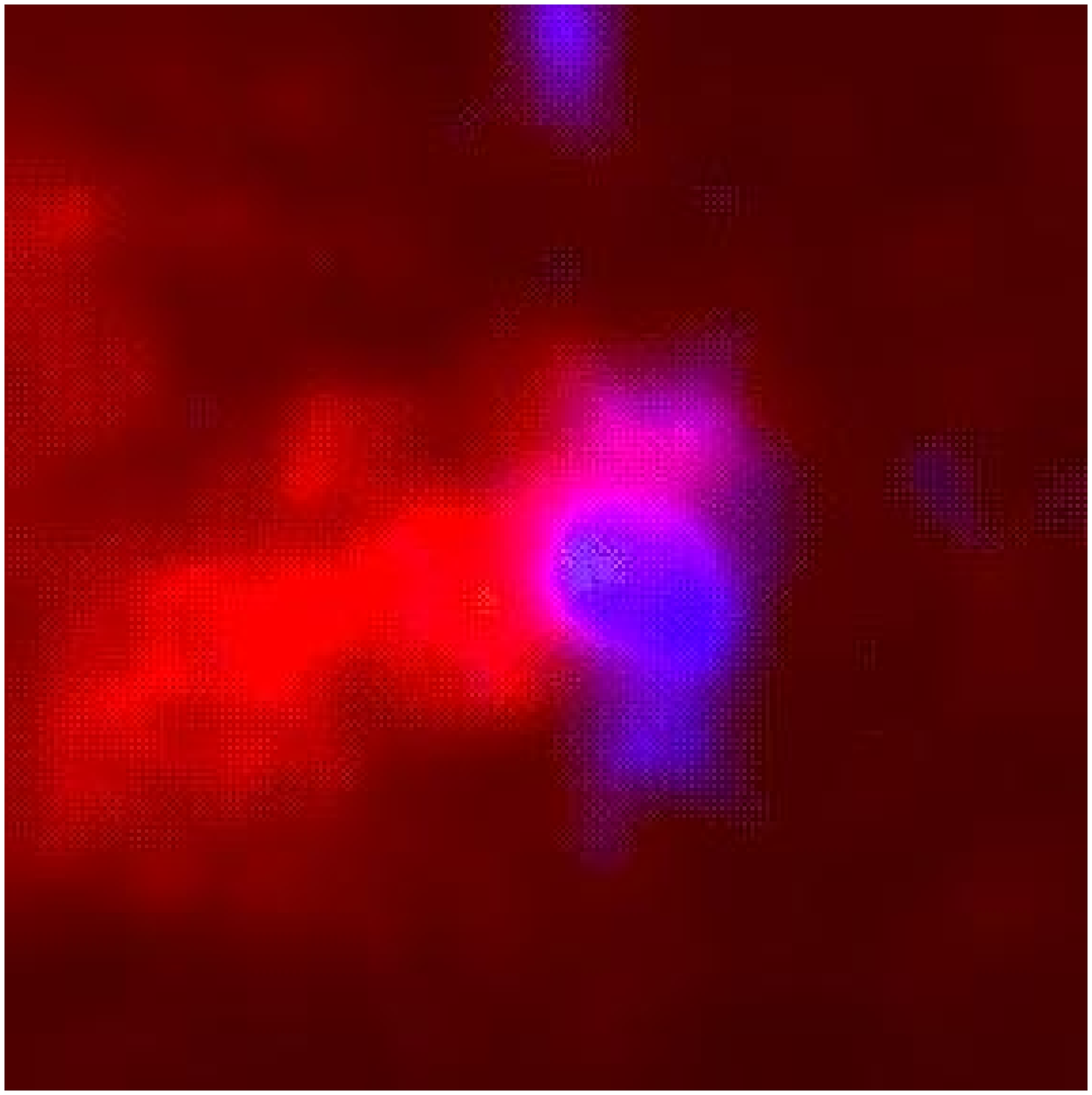}
\caption{
(a) {\it Chandra} ACIS-I binned image ($8 \times 8$ pixels) of M~17,
0.5--7~keV, with the $>900$ detected point sources removed for the
study of diffuse emission.  See \S\ref{observations.sec} for details.
The gaps between the 4 CCDs in the ACIS-I array are faintly visible.
(b) {\it Chandra} ACIS-I swiss-cheese (sources-removed) image of M~17
using our own adaptive smoothing algorithm and exposure correction (8
pixels per bin).  Red intensity is scaled to the soft (0.5--2~keV)
emission and blue intensity is scale to the hard (2--7~keV) emission.  
(c) The same datasets as in (b) with the holes partially smoothed over.
(d) The same datasets as in (b) with the holes completely smoothed over.
The three smoothed images use identical intensity scaling.
\label{fig:m17_patsmooth}}
\end{figure}

\begin{figure}
\centering
\includegraphics[width=0.45\textwidth]{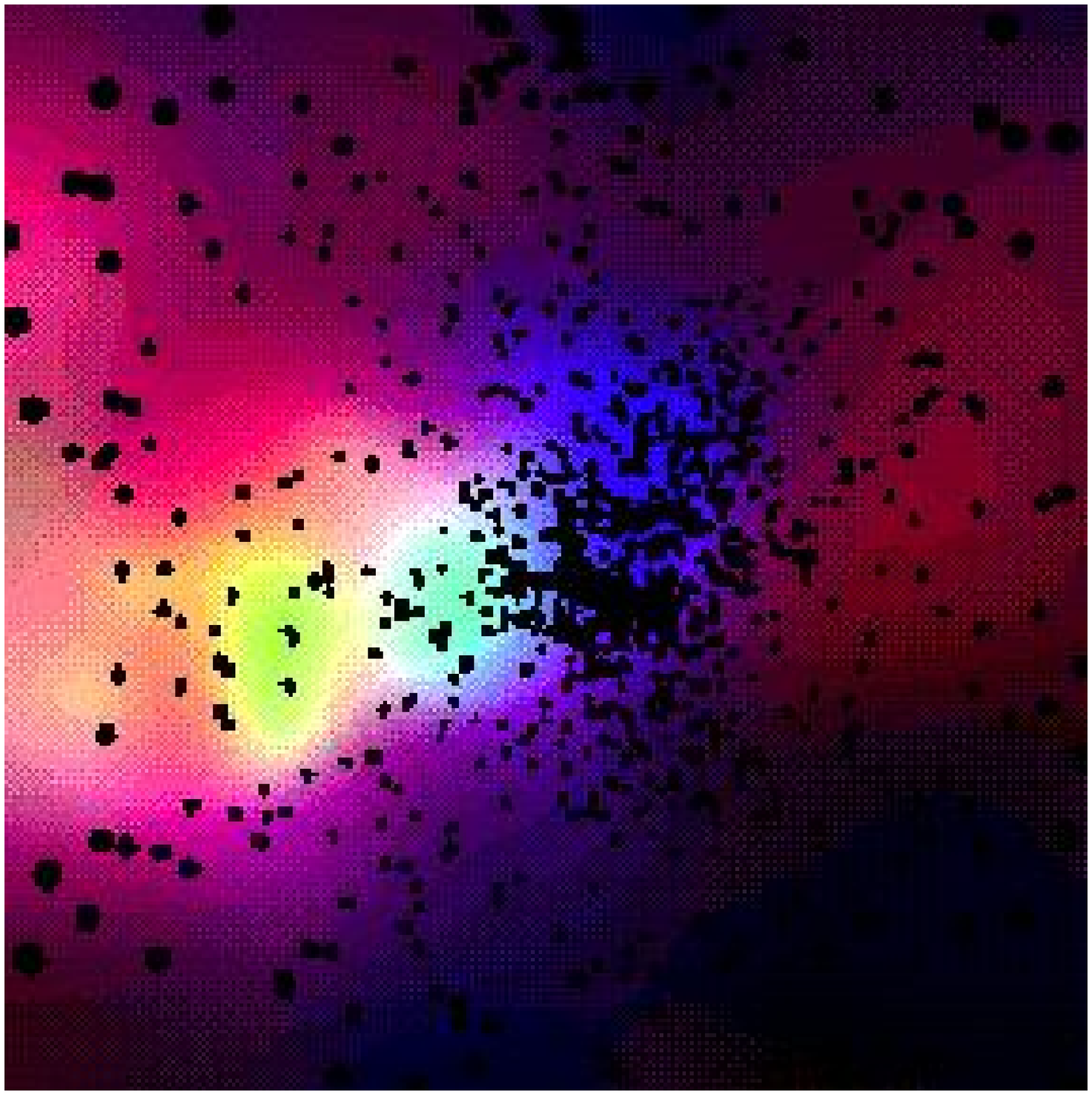}
\includegraphics[width=0.45\textwidth]{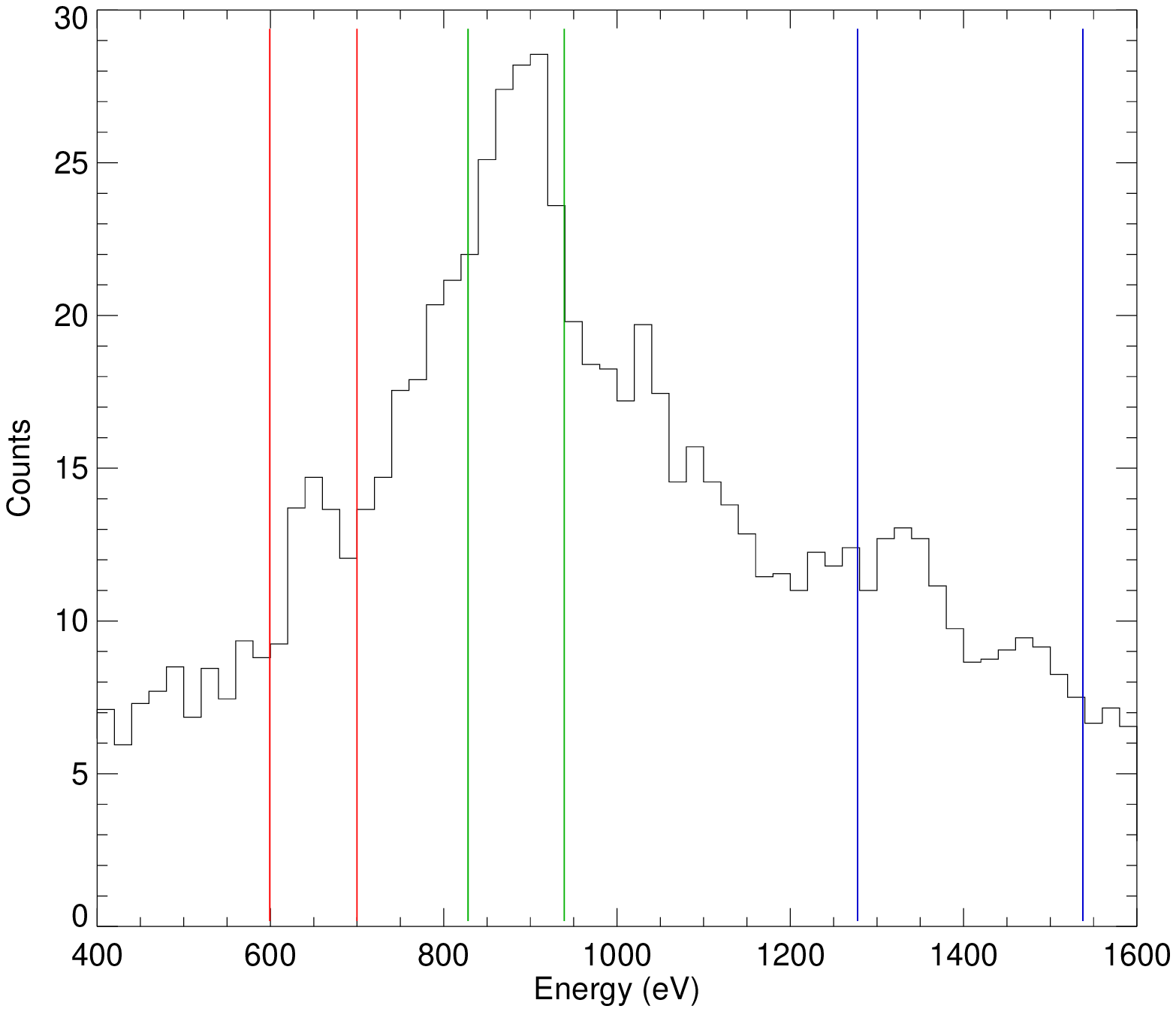}
\caption{
(a) A three-color image of the smoothed soft diffuse emission in M~17:
red = 600--700~eV, 1267 events; green = 830--940~eV, 2871 events; blue = 1280--1540~eV, 2623 events.
(b) A spectrum of the soft diffuse emission, binned at a constant 20~eV
per bin to show the peaks in the emission that were used to define the
three soft bands for (a).  Vertical bars delineate the bands.
\label{fig:m17_soft_patsmooth}}
\end{figure}

\begin{figure}
\centering
\raisebox{-1.5in}[0.0in][2.5in]{
\includegraphics[width=0.9\textwidth, angle=-25]{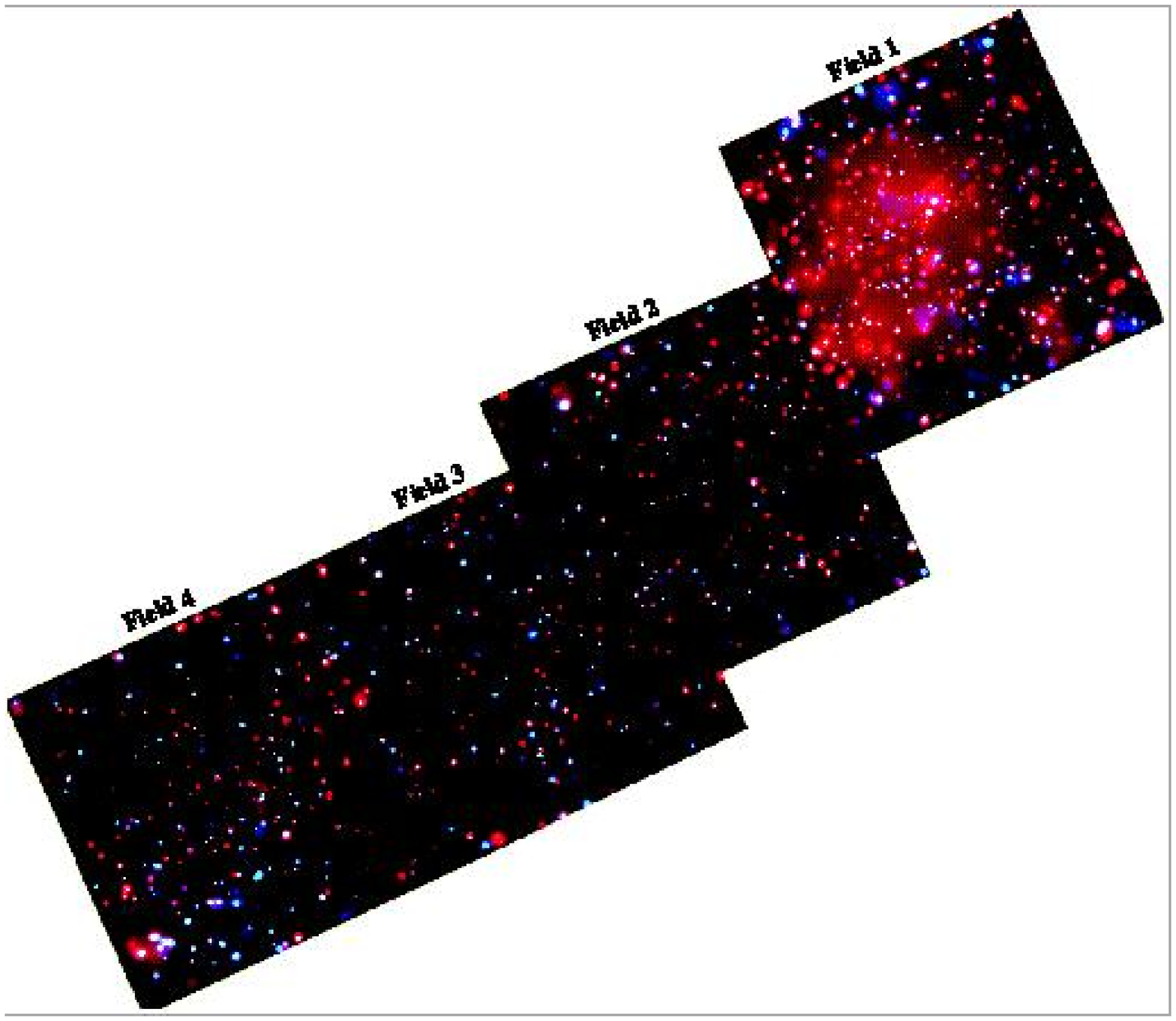}}
\mbox{\includegraphics[width=0.45\textwidth]{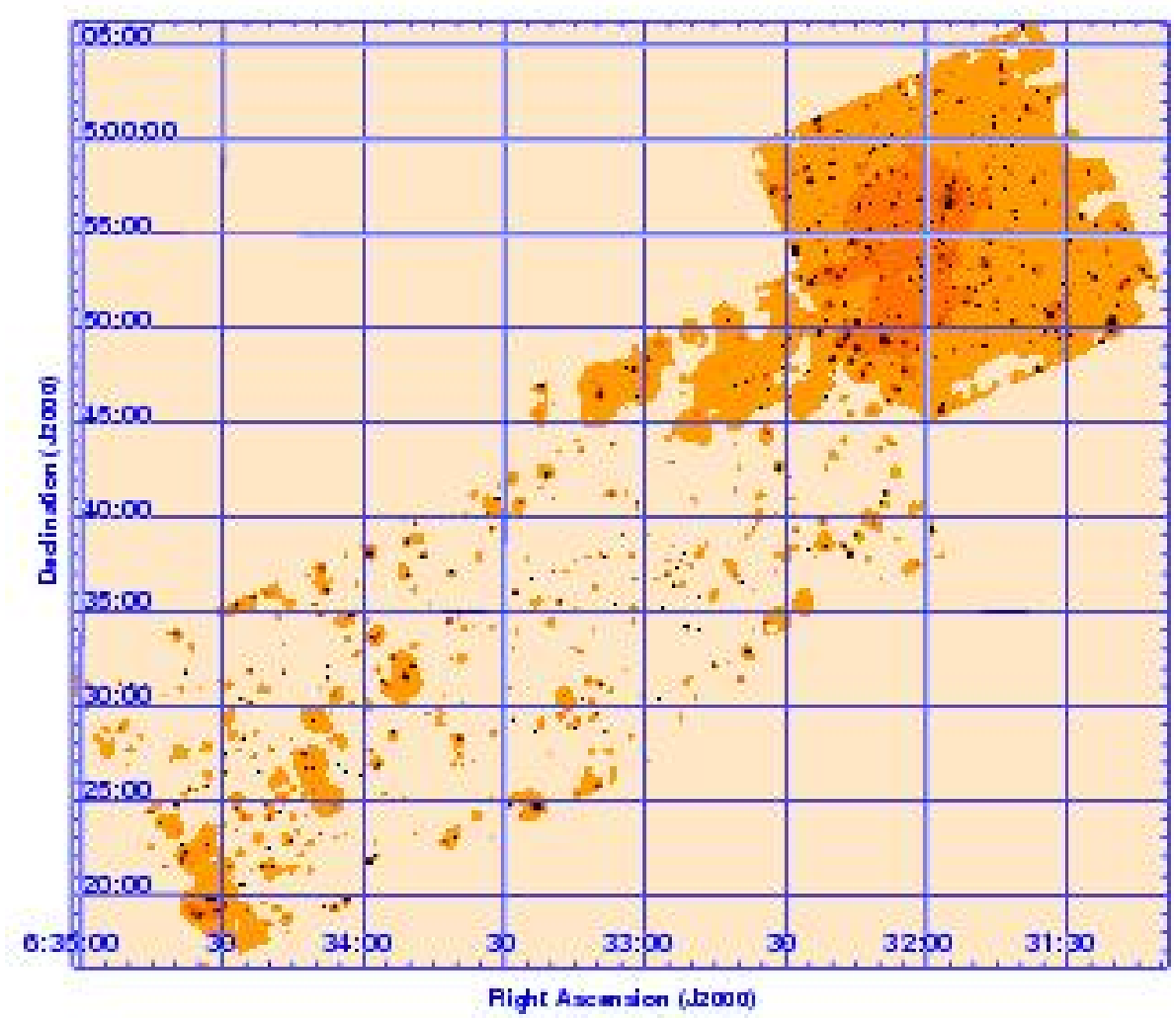}
\includegraphics[width=0.45\textwidth]{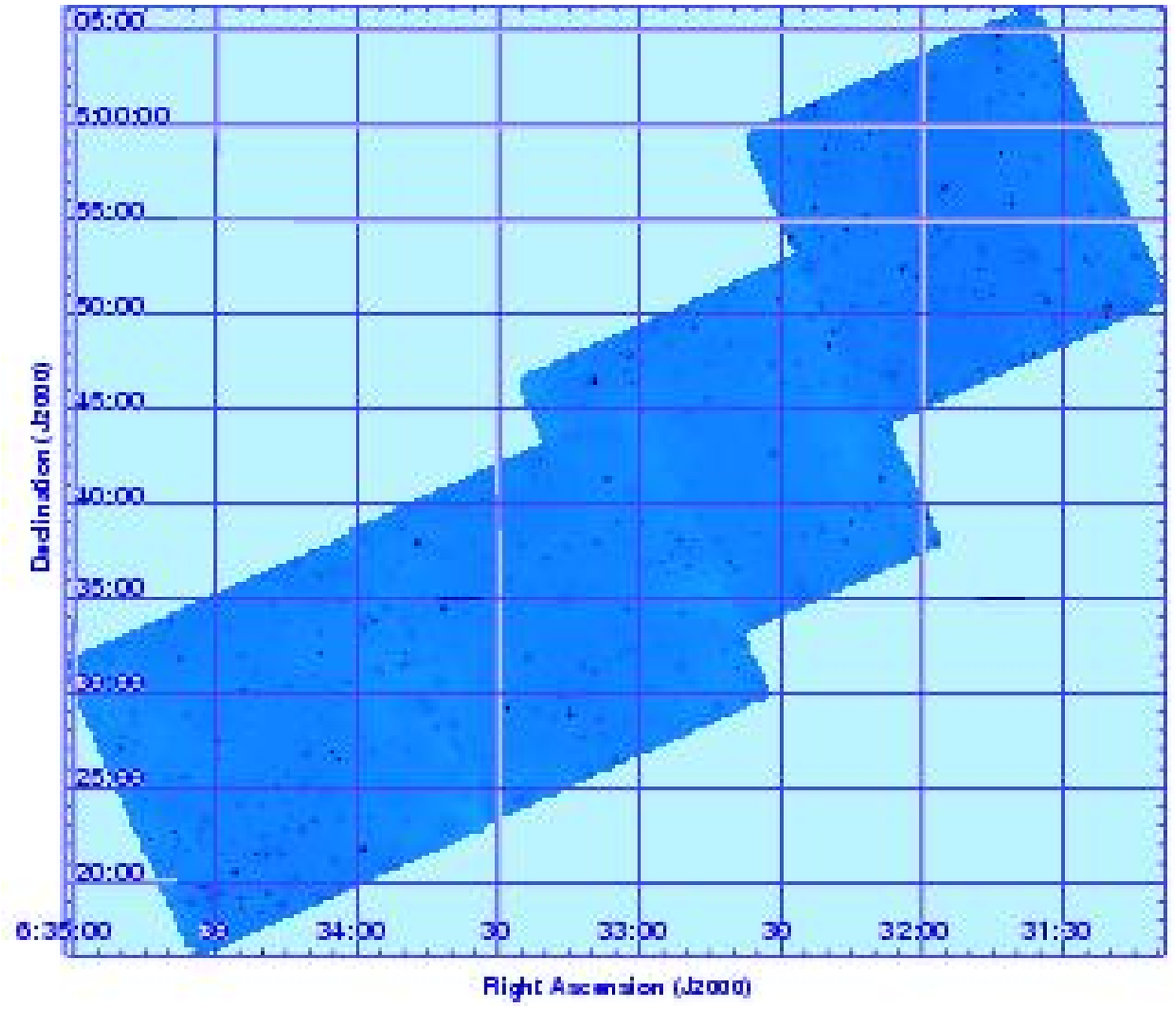}
}
\caption{
(a) {\it Chandra} ACIS-I $1^{\circ} \times 0.25^{\circ}$ mosaic of the
Rosette Nebula and RMC, after {\it csmooth} adaptive smoothing and exposure
correction (2.65 pixels per bin).  Red intensity is scaled to the soft
(0.5--2~keV) emission and blue intensity is scaled to the hard
(2--8~keV) emission.  This image has been rotated to fit better on the
page; the actual orientation of the field on the sky is given in the
lower panels.
(b) The smoothed 0.5--2~keV image used for the red intensity in (a),
highlighting the soft diffuse emission.
(c) The smoothed 2--8~keV image used for the blue intensity in (a), showing
that there is no significant hard diffuse emission in this field.
\label{fig:rosette_color}}
\end{figure}

\begin{figure}
\centering
\includegraphics[width=0.45\textwidth]{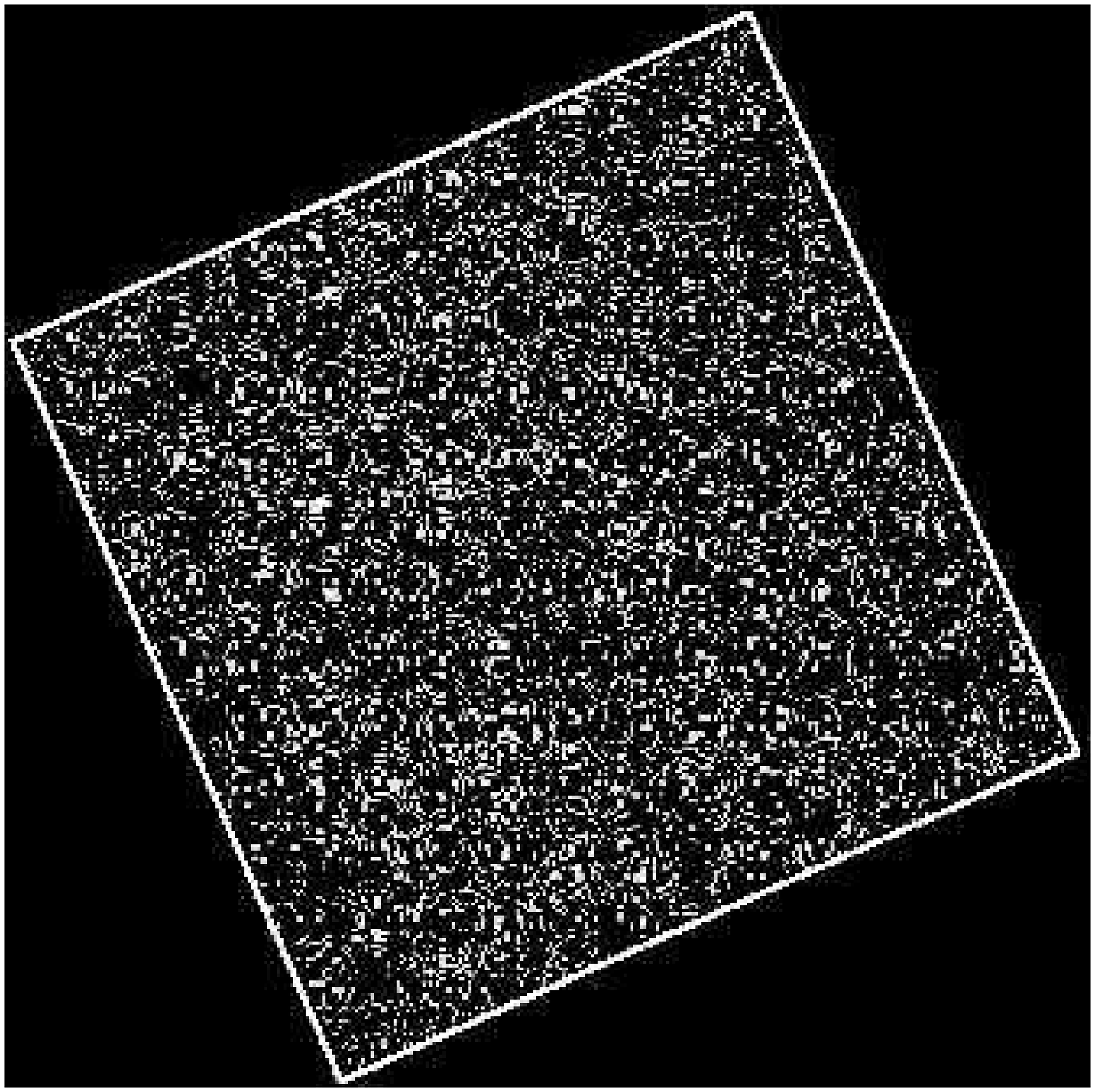}
\hspace{0.05in}
\includegraphics[width=0.45\textwidth]{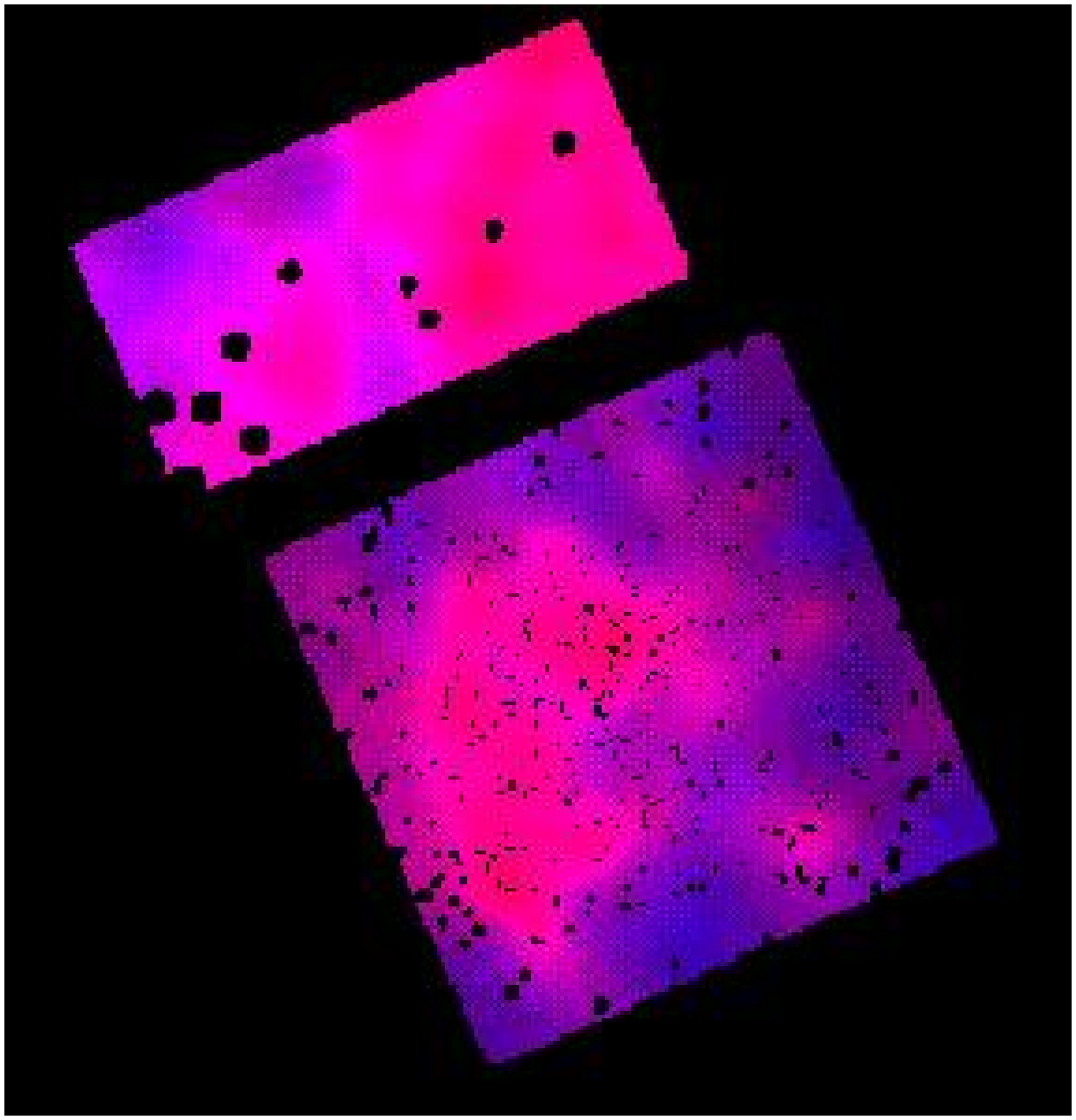}
\includegraphics[width=0.45\textwidth]{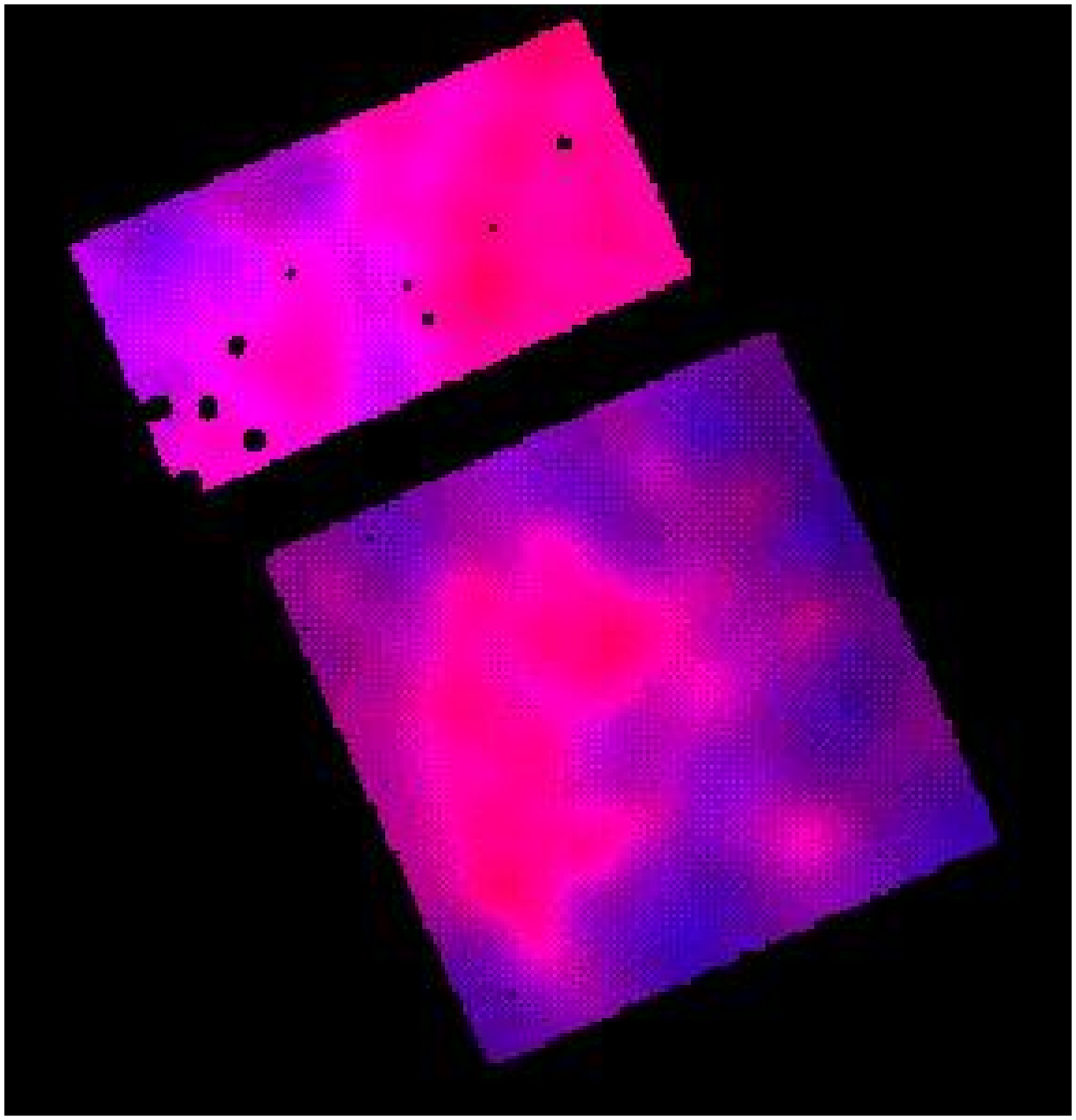}
\hspace{0.05in}
\includegraphics[width=0.45\textwidth]{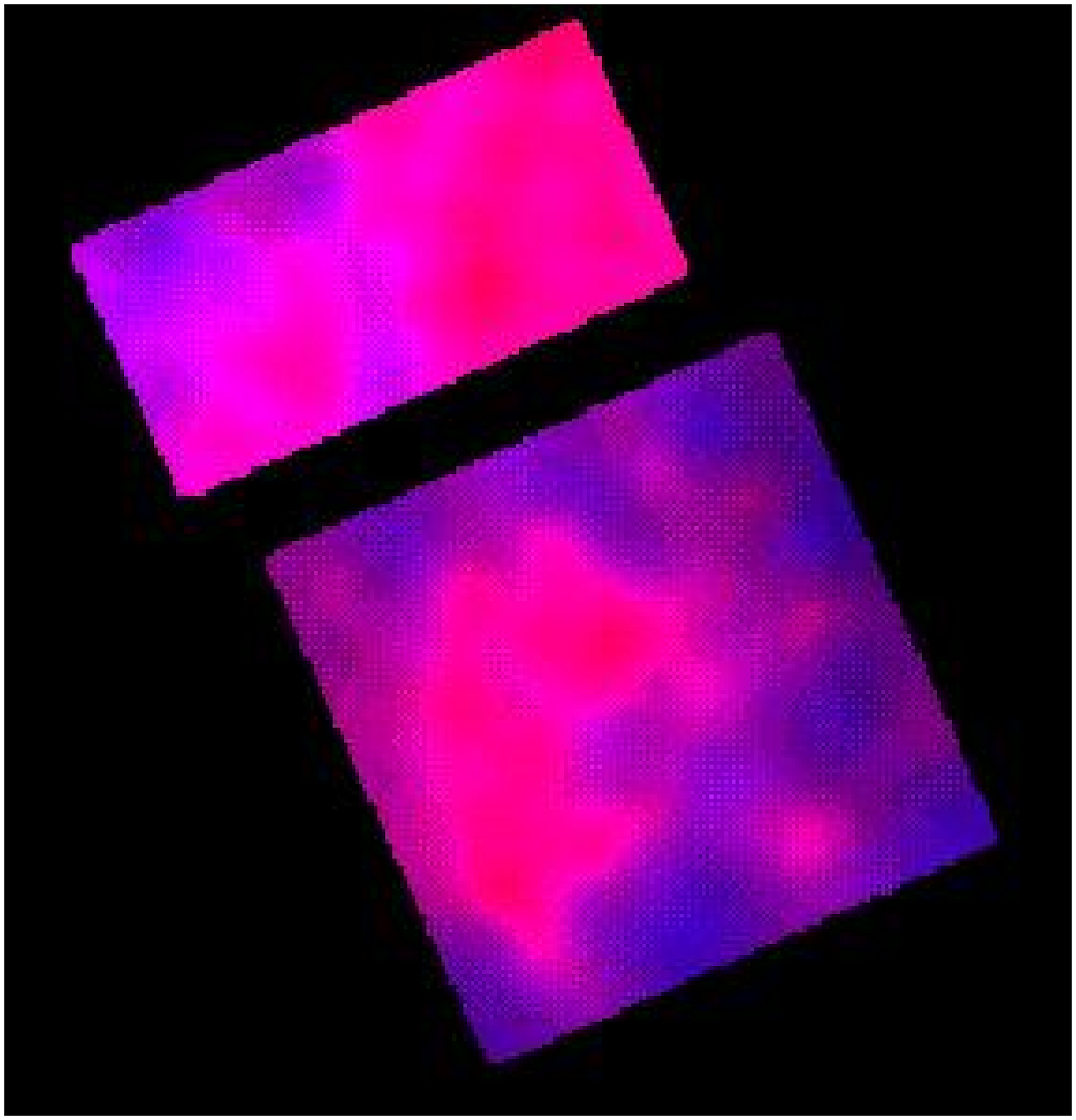}
\caption{
(a) {\it Chandra} ACIS-I binned image ($8 \times 8$ pixels) of the
Rosette Nebula pointing (Field 1, the westernmost of the four pointings
in the mosaic in Figure~\ref{fig:rosette_color}a), 0.5--7~keV, with the
$>350$ detected point sources removed for the study of diffuse
emission.  See \S\ref{observations.sec} for details.  The gaps between
the 4 CCDs in the ACIS-I array are faintly visible.
(b) {\it Chandra} ACIS swiss-cheese (sources-removed) image of Rosette
using our own adaptive smoothing algorithm and exposure correction (8
pixels per bin).  Red intensity is scaled to the soft (0.5--2~keV)
emission and blue intensity is scaled to the hard (2--7~keV) emission.
The off-axis S2 and S3 chips are included at the top of this image; the
background on S3 (a back-illuminated CCD, top right in the image) is
higher than the other CCDs, so no information on diffuse emission there
can be inferred from this image.
(c) The same datasets as in (b) with the holes partially smoothed over.
(d) The same datasets as in (b) with the holes completely smoothed over.
The three smoothed images use identical intensity scaling.
\label{fig:rosette_patsmooth}}
\end{figure}

\begin{figure}
\centering
  \includegraphics[width=0.45\textwidth]{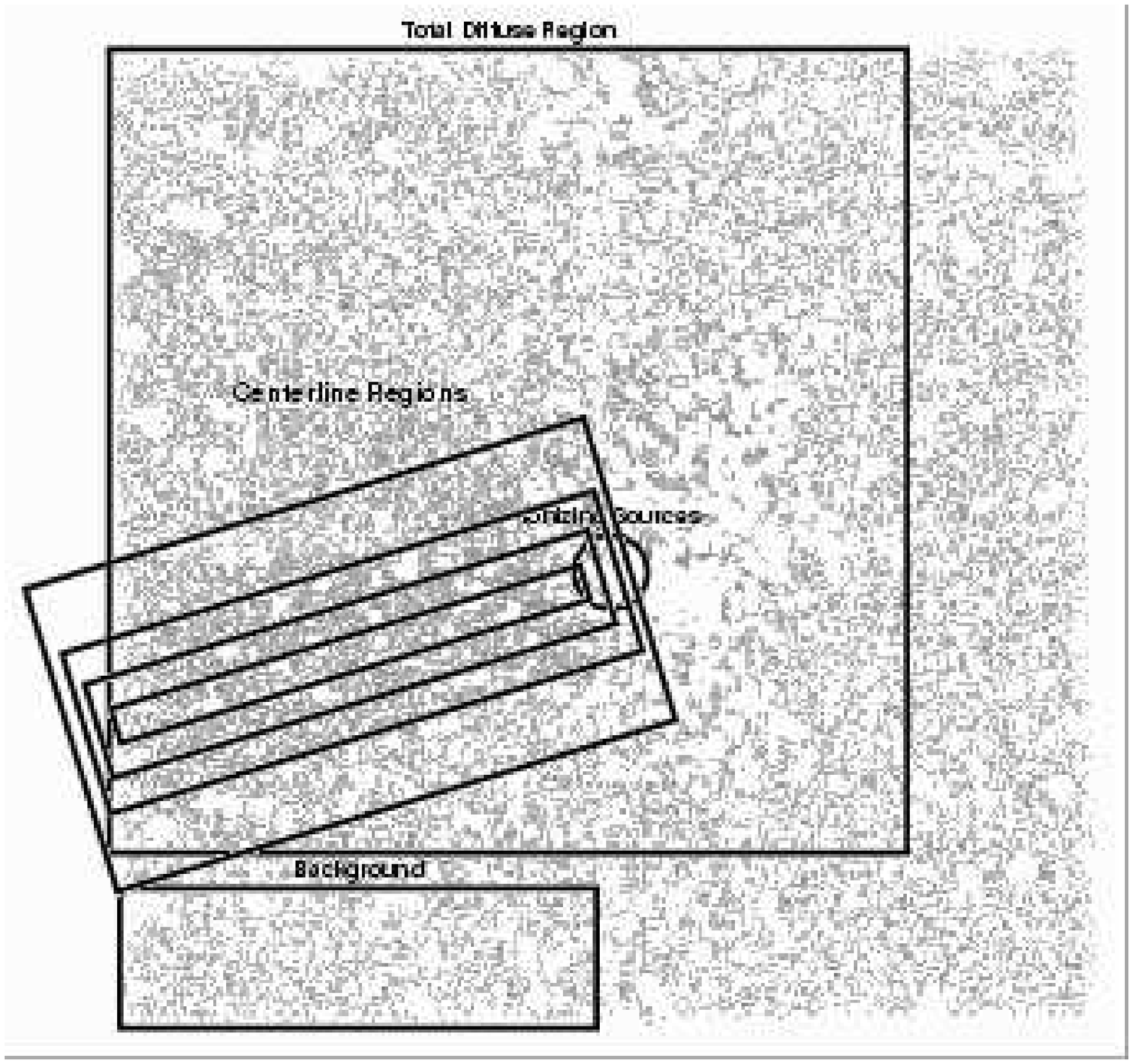}
  \hspace{0.1in}
  \includegraphics[width=0.45\textwidth]{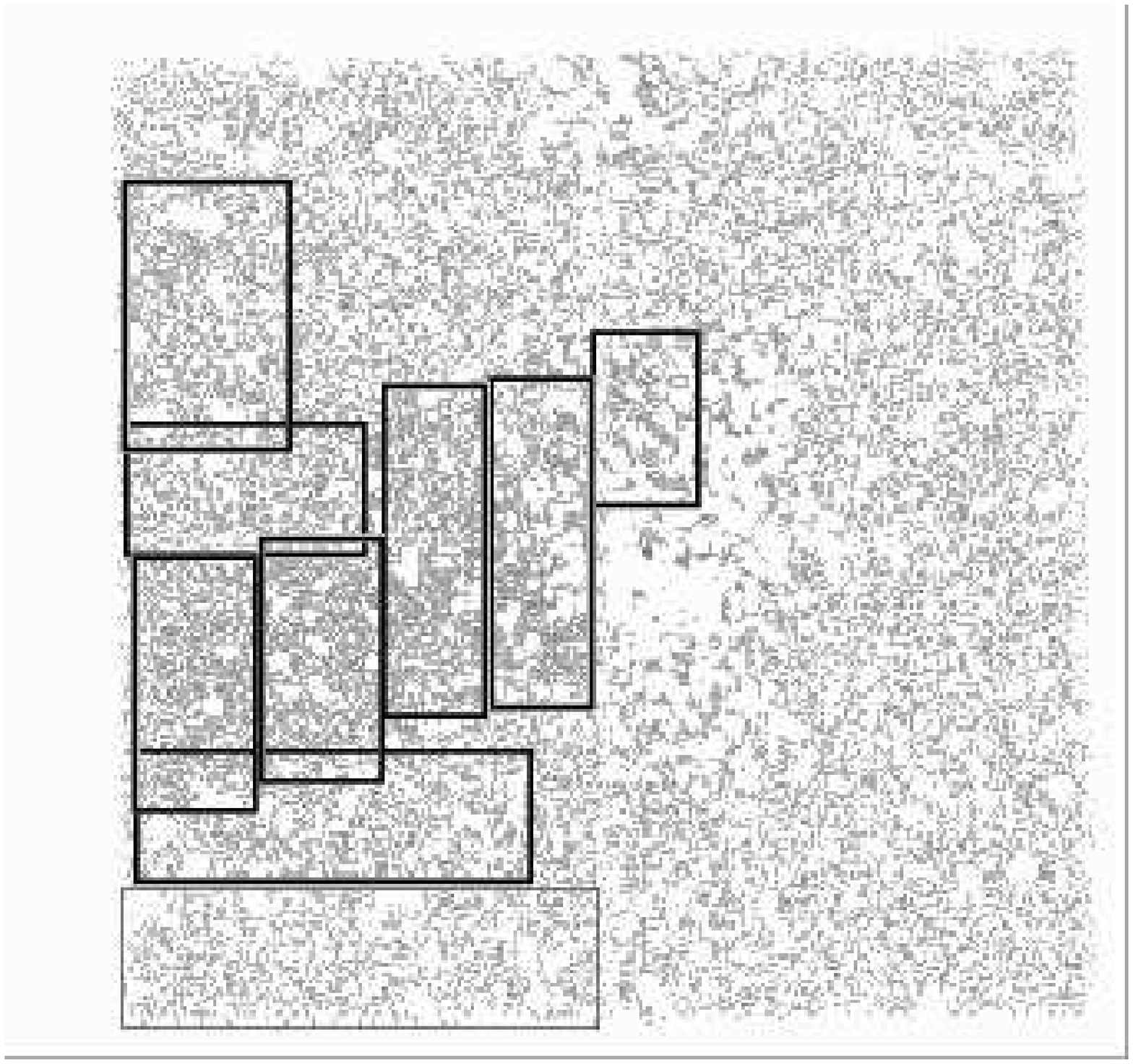}
\caption{Regions used for spectral fitting of diffuse emission in M~17,
overlaid on a full-band, binned, sources-removed image of the diffuse
emission.  (a) Outlined are the large region used for our spectral
analysis of the diffuse emission seen in M~17, the region at the bottom
of the I1 CCD used as background for that spectral fitting, and the
rectangular annuli following the centerline of the soft diffuse
emission, used to search for temperature gradients away from the
centerline.  The circle represents the approximate location of the ring
of O stars at the center of NGC~6618.  (b) Other regions used to look
for temperature variations in the diffuse emission, particularly as a
function of distance from the ionizing O stars.  The I1 background
region is again shown in light outline at the bottom left.  
\label{fig:m17_regions}} 
\end{figure}

\begin{figure}
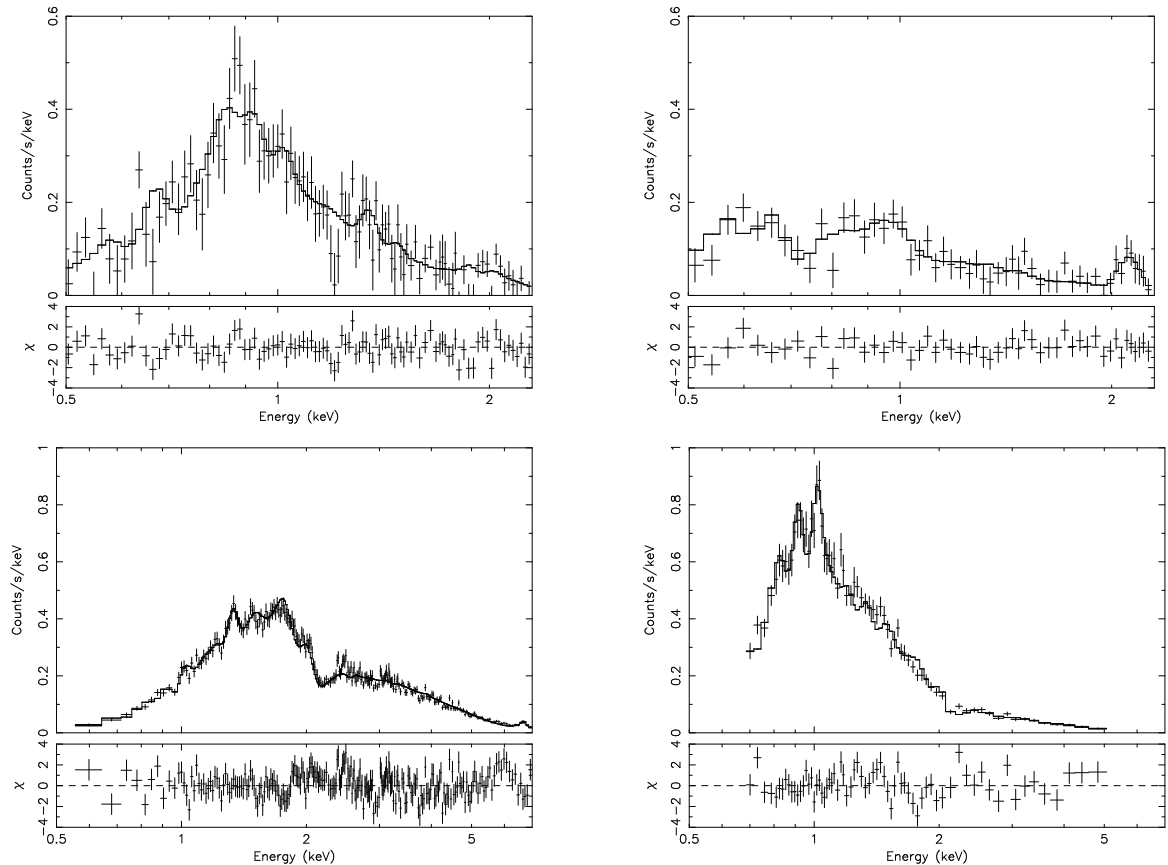


\centering
  \begin{minipage}[t]{1.0\textwidth}
  \centering
  \includegraphics[width=0.3\textwidth,angle=-90]{f9a.eps}
  \hspace{0.5in}
  \includegraphics[width=0.3\textwidth,angle=-90]{f9b.eps}
  \end{minipage} \\ [0.1in]
  \begin{minipage}[t]{1.0\textwidth}
  \centering
  \includegraphics[width=0.3\textwidth,angle=-90]{f9c.eps}
  \hspace{0.5in}
  \includegraphics[width=0.3\textwidth,angle=-90]{f9d.eps}
\caption{ACIS spectra of the following composite components using the
spectral models given in Table~\ref{tab:spectra}:  
(a) M~17 diffuse emission; 
(b) Rosette diffuse emission; 
(c) M~17 point sources; 
(d) Rosette point sources.
The upper panels show the data and best-fit models; lower panels give
the (Data$-$Model) residuals.
\label{fig:spectra}}
  \end{minipage}
\end{figure}

\begin{figure}
\centering
\includegraphics[width=0.5\textwidth]{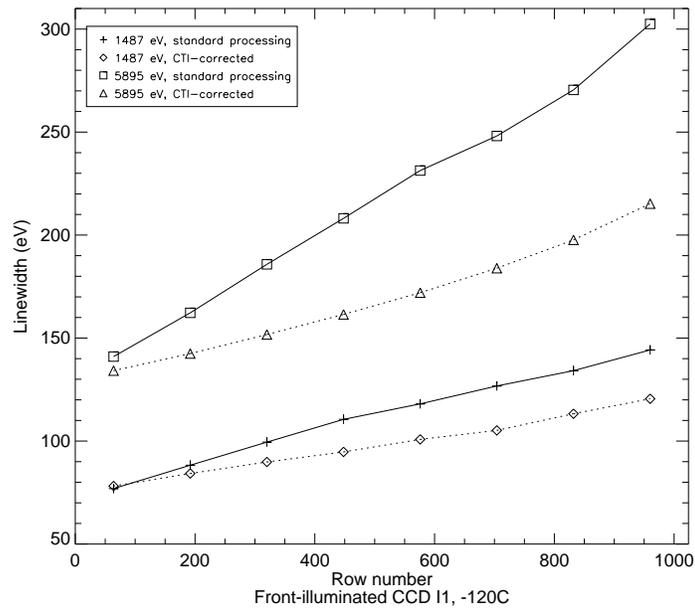}
\caption{ACIS spectral resolution on a front-side illuminated CCD as a
function of row number, before and after charge transfer inefficiency
(CTI) correction using the PSU algorithm.  By reducing the spectral
resolution variation across the CCDs, CTI correction substantially
improves spectral analysis of diffuse structures on ACIS-I.
\label{fig:cti-res}}
\end{figure}

The smoothed images shown in Figures~\ref{fig:m17_patsmooth} and
\ref{fig:rosette_patsmooth} were constructed using a simple adaptive
kernel smoothing method.  The smoothing kernels consist of the null
kernel \[ k(0,i,j) = \delta(i,j) = 1 \mbox{ when } i=0,j=0 \mbox{ (and 0 otherwise)}\] which
produces no smoothing, plus a family of 2-D Gaussians with integer
sigma values $r = \{1,2,3,\ldots\}$, 
\[ k(r,i,j) = \exp^{-\frac{i^2+j^2}{2r^2}} \hspace{0.1in}. \] 
For a particular kernel size, $r=R$, the flux at position $(X,Y)$ is
estimated as
\begin{equation} 
flux(X,Y) = \frac{counts}{exposure}
\label{eqn:flux}
\end{equation}
where 
\[ counts = \sum_{x \in Obs} \sum_{y \in Obs} k(R,x-X,y-Y) image(x,y) \hspace{0.1in}, \]
\[ exposure = \sum_{x \in Obs} \sum_{y \in Obs} k(R,x-X,y-Y) emap(x,y) \hspace{0.1in}, \]
$Obs$ is the set of pixels in the `observed' region
(positive exposure) of the scene, $image$ is the integer-valued image
of observed counts, and $emap$ is the exposure map.  

To derive an estimate of the error on the flux value, we first note
that each pixel in our observed counts image, $image(x,y)$, is of
course a sample from an unknown underlying Poisson distribution
associated with that pixel.  We make the simplifying assumption that
all the pixels under the kernel have the {\bf same} true flux (which we
just estimated).  By multiplying that flux by the pixel's exposure
value, we then compute the mean number of counts that should be
observed in each pixel if the image was repeatedly observed:
\[mean(x,y) = flux(X,Y) emap(x,y) \hspace{0.1in}. \]
Each pixel's (Poisson) variance is assumed to be that mean value.  We
then propagate the variances of the distributions through the linear
equation (\ref{eqn:flux}) in the usual way to obtain an error for our
flux estimate:
\[ error = \frac{\sqrt{\sum_{x \in Obs} \sum_{y \in Obs} [k(R,x-X,y-Y)]^2 flux(X,Y) emap(x,y)}}{exposure} \hspace{0.1in}. \]

The flux and flux error are used to define a simple `significance' value, 
\begin{eqnarray*} 
significance & = & \frac{flux}{error} \\
             & = & \frac{\sum_{x \in Obs} \sum_{y \in Obs} k(R,x-X,y-Y) image(x,y)} {\sqrt{\sum_{x \in Obs} \sum_{y \in Obs} [k(R,x-X,y-Y)]^2 flux(X,Y) emap(x,y)}} 
\end{eqnarray*}
which increases as $r$ increases since more counts are encompassed
under the kernel.  The observer supplies a significance threshold, and
at each position $(X,Y)$ the flux is estimated using the smallest
kernel size, $r$, that produces a significance exceeding the
threshold.  Thus, regions of low flux are more heavily smoothed than
regions of higher flux.  The significance threshold controls the
overall level of smoothing.  A map of the kernel sizes chosen is
returned by the smoothing program.

\begin{figure}
\centering
\includegraphics[width=0.5\textwidth]{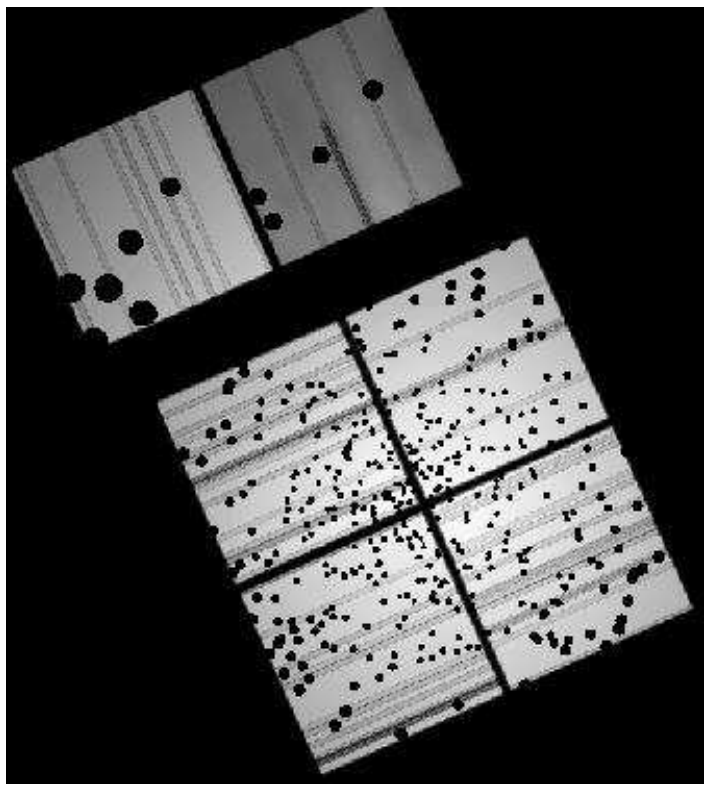}
\caption{An example ACIS-I exposure map, including the `swiss-cheese' masks.
\label{fig:ad2d_exposure}}
\end{figure}

Application of the kernel to the exposure map in Equation
(\ref{eqn:flux}) is required for accurate smoothing of any wide-field
ACIS image because the exposure varies significantly due to chip gaps
and bad regions of the detector, as shown in the example swiss-cheese
exposure map in Figure~\ref{fig:ad2d_exposure}.  The algorithm
naturally handles unobserved regions, such as areas explicitly masked
by the observer and areas outside the field of view of ACIS, because
those zero-count and zero-exposure pixels do not affect either the flux
or flux error computations.  As you would expect, an unobserved region
tends to cause neighboring observed regions to be more heavily smoothed
since a larger kernel must be used to find an adequate number of
counts.  Note that the algorithm will estimate a flux for unobserved
pixels (zero exposure) using the nearby observed regions, as in
Figures~\ref{fig:m17_patsmooth}d and \ref{fig:rosette_patsmooth}d.
This is of course a form of interpolation.

The smoothing program recognizes a negative value in the exposure map
as a flag specifying that the position is `off-field,' meaning that no
flux estimate should be computed there.  We flag unobserved pixels in
this way for several purposes:  (1) Pixels outside the field of view of
ACIS represent regions of the sky totally unobserved.  Estimates of the
sky flux on those regions are inappropriate since they would be
extrapolations.  It is desirable to include in this `off-field'
category pixels at the edge of the detector which have very small
exposure (due to dithering) because normalization by very small
exposure values leads to noisy results.  The off-field category can be
most easily defined by simply thresholding the original exposure map
(prior to application of the swiss-cheese mask regions).  (2) In
Figures~\ref{fig:m17_patsmooth}b and \ref{fig:rosette_patsmooth}b, we
flagged as off-field all the pixels that were masked by the observer,
preventing any interpolation over the holes.  (3) In
Figures~\ref{fig:m17_patsmooth}c and \ref{fig:rosette_patsmooth}c, we
flagged as off-field those masked pixels that were more than a certain
distance from any on-field pixel.  This compromise approach fills in
many small holes that are distracting to the eye, while avoiding making
gross interpolations across large masked regions.

This algorithm has been implemented as the IDL program {\it
adaptive\_density\_2d.pro} and is part of the publicly-available TARA
software package mentioned in Appendix B.

\clearpage

\newpage

\newpage

\newpage

\newpage

\newpage

\newpage

\newpage

\newpage

\newpage

\newpage

\clearpage


\begin{deluxetable}{ccccc} 
\tabletypesize{\normalsize}
\tablewidth{0pt}
\tablecolumns{5}
\tablecaption{Log of {\it Chandra} Observations \label{tbl:obslog}}

\tablehead{
\colhead{Target} &
\colhead{Date} &
\colhead{R.A.\ (J2000)} &
\colhead{Dec (J2000)} &
\colhead{Integration (s)}
}
\startdata
M~17		& 2 Mar 2002 & 18 20 30.39 & $-$16 10 29.7 & 39440 \\
Rosette Field 1 & 5 Jan 2001 & 06 31 52.85 &   +04 55 42.0 & 19700 \\
Rosette Field 2 & 5 Jan 2001 & 06 32 40.84 &   +04 42 45.0 & 19500 \\
Rosette Field 3 & 5 Jan 2001 & 06 33 17.15 &   +04 34 42.0 & 19410 \\
Rosette Field 4 & 6 Jan 2001 & 06 34 17.34 &   +04 27 45.9 & 19510 \\
\enddata

\tablecomments{Positions are for the ACIS Imaging Array aimpoint.  The
exposure time is the net usable time after various filtering steps are
applied in the data reduction process.}
\end{deluxetable}

\newpage

\begin{deluxetable}{lrcccccccccccc} 
\tabletypesize{\tiny}
\tablewidth{0pt}
\tablecolumns{14}
\tablecaption{X-ray Spectra and Luminosities \label{tab:spectra}}

\tablehead{
\colhead{Target} &
\colhead{Counts} &
\colhead{Scaled Bkgd} &
\multicolumn{2}{c}{Absorption\tablenotemark{a}} &&
\multicolumn{2}{c}{Plasma Temp (keV)} &
\colhead{Gaussians} &
\multicolumn{4}{c}{X-ray Luminosities\tablenotemark{b}} &
\colhead{reduced} \\ \cline{4-5} \cline{7-8} \cline{10-13}
\colhead{} &
\colhead{} &
\colhead{Counts} &
\colhead{$N_{H,1}$} &
\colhead{$N_{H,2}$} &&
\colhead{$kT_1$} &
\colhead{$kT_2$} &
\colhead{(keV)} &
\colhead{$L_{\rm soft}$} &
\colhead{$L^c_{\rm soft}$} &
\colhead{$L_{\rm hard}$} &
\colhead{$L^c_{\rm hard}$} &
\colhead{$\chi^2$}
}

\startdata
M~17 diffuse    & 19,700 & $\sim 9700$ & 4 $\pm$ 1 & \nodata && $0.13\pm0.02$ & $0.6\pm0.1$ & 2.1           & 0.4 & 3.4 & \nodata & \nodata & 1.2\\
Rosette diffuse &   7800 & $\sim 4500$ & 2 $\pm$ 1 & \nodata && $0.06\pm0.02$ & $0.8\pm0.1$ & 2.1           & 0.2 & 0.6 & \nodata & \nodata & 0.7\\
M~17 pt srcs    & 36,600 & \nodata & 7 $\pm$ 1 & 17 $\pm$ 1 && $0.19\pm0.02$ & $3.0\pm0.1$ & 1.1, 1.3, 6.7 & 0.6 & 6.1 & 3.5     & 4.1     & 1.5\\
Rosette pt srcs &   9500 & \nodata & 6 $\pm$ 1 & \nodata && $0.20\pm0.02$ & $2.1\pm0.2$ & \nodata       & 0.6 & 7.4 & 0.6     & 0.7     & 1.6\\
\enddata

\tablenotetext{a}{in units of $10^{21}$ cm$^{-2}$}
\tablenotetext{b}{in units of $10^{33}$ ergs~s$^{-1}$}
\end{deluxetable}

\newpage

\begin{deluxetable}{cccc} 
\tabletypesize{\normalsize}
\tablewidth{0pt}
\tablecolumns{4}
\tablecaption{Physical Properties of the H{\sc II} Region Plasmas \label{tab:phys}}

\tablehead{
\colhead{Parameter} &
\colhead{Scale factor}  &
\colhead{M~17} &
\colhead{Rosette}
}

\startdata
\multicolumn{4}{l}{\it Observed X-ray properties} \\
$kT_x$ (keV)           	& \nodata     &  0.6               &  0.8               \\
$L_{x,bol}$ (ergs~s$^{-1}$)&\nodata   & $4 \times 10^{33}$ & $8 \times 10^{32}$ \\
$V_x$ (cm$^3$)       	& $\eta$      & $9 \times 10^{56}$ & $9 \times 10^{56}$ \\
                        &             &                    &                    \\
\multicolumn{4}{l}{\it Derived X-ray plasma properties} \\
$T_x$ (K)		& \nodata     & $7 \times 10^6$    & $9 \times 10^6$    \\
$n_{e,x}$ (cm$^{-3}$)  	&$\eta^{-1/2}$&  0.3               &  0.1               \\
$P_x/k$ (K~cm$^{-3}$)  	&$\eta^{-1/2}$& $5 \times 10^6$    &  $2 \times 10^6$   \\
$E_x$ (ergs)            &$\eta^{1/2}$ & $9 \times 10^{47}$ & $3 \times 10^{47}$ \\
$\tau_{cool}$ (yr)      & $\eta^{1/2}$& $7 \times 10^6$    & $1 \times 10^7$    \\
$M_x$ (M$_\odot$)       & $\eta^{1/2}$&  0.15              &  0.04              \\
		        &             &                    &                    \\
\multicolumn{4}{l}{\it Derived photodissociation region properties} \\
$n_{e,II}$ (cm$^{-3}$)  & \nodata     & $10^3$                 & $10^3$             \\
$T_{II}$  (K)           & \nodata     &  8000              &   6400             \\
$P_{II}/k$ (K~cm$^{-3}$)& \nodata     & $1 \times 10^7$    & $9 \times 10^6$    \\
                        &             &                    &                    \\
\multicolumn{4}{l}{\it O star wind properties} \\
Stellar age $t_\star$ (yr)&\nodata    & $1 \times 10^6$    & $2 \times 10^6$   \\
$\Sigma(\dot{M}_w)$ (M$_\odot$ yr$^{-1}$)&\nodata&$3 \times 10^{-5}$ & $1 \times 10^{-5}$ \\
$\Sigma(\dot{M}_w) t_\star$ (M$_\odot$)&\nodata & 30       &   20               \\
$E_w$ (ergs)            & \nodata     & $9 \times 10^{48}$ & $2 \times 10^{48}$ \\
\enddata

\tablecomments{$\eta$ is a multiplicative factor for the volume of the emitting 
plasma, correcting the observed volume to the actual volume, taking 
extinction, possible inhomogeneities, etc.\ into account. Note that all 
derived plasma properties, being $\propto \eta^{1/2}$, are not very sensitive 
to this correction. See \S\ref{phys.sec} for details.}
\end{deluxetable}

\newpage

\begin{deluxetable}{lrcccccccc}
\centering
\tabletypesize{\scriptsize}
\tablewidth{0pt}
\tablecolumns{10}
\tablecaption{Diffuse X-rays from High Mass Star Forming Regions (HMSFRs) \label{tab:hmsfrs}}

\tablehead{
\colhead{Region} &
\colhead{Distance} &
\colhead{Earliest} &
\colhead{\# O6} &
\colhead{Diffuse} &
\colhead{Diffuse} &
\colhead{$N_H$} &
\colhead{kT} &
\colhead{$L_x$\tablenotemark{b}} &
\colhead{Ref.} \\ 
\colhead{} &
\colhead{(pc)} &
\colhead{Star\tablenotemark{a}} &
\colhead{or earlier} &
\colhead{X-rays?} &
\colhead{Area (pc$^2$)} &
\colhead{$10^{21}$~cm$^{-2}$} &
\colhead{(keV)} &
\colhead{$10^{33}$~ergs~s$^{-1}$} &
\colhead{}
}
\startdata
LMSFRs\tablenotemark{c} &150-350& late B &     0  &  no    &\nodata&\nodata&  \nodata  & $\leq 10^{-5}$ & \nodata \\
Orion Nebula            &   450 & O6     &     1  &  no    &\nodata&\nodata&  \nodata  &  $< 10^{-3}$   &   1     \\
Eagle Nebula            &  2000 & O5     &     2: & \nodata&\nodata&\nodata&  \nodata  &  $< 10^{-3}$   &   2     \\
Lagoon -- NGC 6530      &  1800 & O4     &     3: &  no    &\nodata&\nodata&  \nodata  &  $<10^{-2}$    &   3     \\
Lagoon -- Hourglass     &  1800 & O7     &     0  &probably&  0.04 & 11.1  &    0.63   & $\leq 0.7$\tablenotemark{d} &  3 \\
Rosette Nebula          &  1400 & O4     &     2  &  yes   &  47   &   2   & 0.06, 0.8 & $\leq 0.6$\tablenotemark{d} &   4     \\
RCW 38                  &  1700 & O5     &     1: &  yes   &   2   &  11.5\tablenotemark{e} & 2.2\tablenotemark{e} & 1.6\tablenotemark{e} & 5 \\
Omega Nebula\tablenotemark{f} &  1600 & O4     &     7  &  yes   &  42   &   4   & 0.13, 0.6 &   3.4   &   4     \\
Arches Cluster          &  8500 & O3/W-R & $>$30  & yes    &  14   & 100   &    5.7    &    16   &   6     \\
NGC 3603                &  7000 & O3/W-R & $>$20  &  yes   &  50   &   7   &    3.1    &    20   &   7     \\
Carina Nebula           &  2300 & O3/W-R & $>$30  &  yes   & 1270  & 3--40 &   0.8:    &   200:  &   8     \\
\enddata

\tablenotetext{a}{Some of these regions contain heavily obscured
ultra-compact H{\sc II} regions or high-mass protostars, so the
``earliest star'' is approximate.  We list here the earliest spectral
type given in the literature.}

\tablenotetext{b}{Luminosities are obtained from the listed reference.
They are corrected for absorption and usually cover the 0.5--2~keV band for
soft emission, the 2--8~keV band for hard emission.  For diffuse
emission that extends beyond the ACIS-I FoV, these are lower limits.
The X-ray luminosity in LMSFRs refers to two cases of diffuse
X-rays from Herbig-Haro outflow shocks.}

\tablenotetext{c}{Low-mass star formation regions near the Sun such as
the Taurus-Auriga, Ophiuchi, Chamaeleon, Lupus and Perseus clouds.  Very
faint and localized regions of diffuse X-rays have been found in two
young stellar outflows; see text.}

\tablenotetext{d}{The upper limits here refer to likely contributions
from unresolved stellar X-ray sources.}

\tablenotetext{e}{Thermal plasma fit to the core of the diffuse emission;
the authors prefer a power law plus thermal plasma fit, with $N_H = 9.5
\times 10^{21}$~cm$^{-2}$, $\Gamma = -1.6$, and $kT = 0.2$~keV for
the total diffuse emission.  We estimated $L_x$ from these parameters using
PIMMS (\url{http://xte.gsfc.nasa.gov/Tools/w3pimms.html}).}

\tablenotetext{f}{Values for area, $N_H$, $kT$, and $L_x$ are
determined from the ACIS data.  Additional diffuse soft X-rays are seen
in the {\it ROSAT} data, outside the ACIS field of view; including
this emission would roughly double the area and increase $L_x$ by $\sim 1
\times 10^{33}$~ergs~s$^{-1}$ (Dunne03).}

\tablerefs{1: \citet{Feigelson02};~~2: \citet{Mytyk01};~~3:
\citet{Rauw02};~~4: this work;~~5: \citet{Wolk02};~~6:
\citet{YusefZadeh02};~~7: \citet{Moffat02};~~8: \citet{Seward82}.}

\end{deluxetable}   

\newpage

\begin{deluxetable}{rrrllcc}
\centering
\tabletypesize{\scriptsize}
\tablewidth{0pt}
\tablecolumns{7}
\tablecaption{O and Early B stars in NGC~6618 (M~17 cluster)
\label{tab:m17_OB}}

\tablehead{
\multicolumn{3}{c}{Name\tablenotemark{a}} &
\colhead{R.A.\ (J2000)\tablenotemark{b}} &
\colhead{Dec (J2000)\tablenotemark{b}} &
\colhead{Spectral} &
\colhead{ACIS} \\ \cline{1-3}
\colhead{CEN} &
\colhead{B} &
\colhead{OI} &
\colhead{} &
\colhead{} &
\colhead{Type\tablenotemark{c}} &
\colhead{detected?}
}

\startdata
37  & 174   &\nodata&  18 20 30.54 &   -16 10 53.3&   O3-O6   & yes\\
43  & 137   &\nodata&  18 20 33.14 &   -16 11 21.6&   O3-O4   & yes\\
 1  & 189   & 341   &  18 20 29.92 &   -16 10 45.5&   O4+O4\tablenotemark{d} & yes\\
102 & 305   & 780   &  18 20 23.05 &   -16 07 58.9&   O5      & no \\
 2  & 111   & 337   &  18 20 34.55 &   -16 10 12.1&   O5      & yes\\
\nodata& \nodata& 345& 18 20 27.52 &   -16 13 31.8&   O6      & yes\\
18  & 260   & 326   &  18 20 25.94 &   -16 08 32.3&   O7-O8   & yes\\
25  & 164   &\nodata&  18 20 30.92 &   -16 10 08.0&   O7-O8   & yes\\
34  & 358   & 715   &  18 20 21.48 &   -16 10 00.0&   O8      & no \\
16  & 311   & 258   &  18 20 22.76 &   -16 08 34.3&   O9-B2   & yes\\
61  & 181   &\nodata&  18 20 30.30 &   -16 10 35.2&   O9-B2   & yes\\
3\tablenotemark{e}& 98& 342& 18 20 35.47& -16 10 48.9&O9      & yes\\
31  & 289   &\nodata&  18 20 24.45 &   -16 08 43.3&   O9.5    & yes\\
51  & 243   &\nodata&  18 20 26.64 &   -16 10 03.7&   early B & no\tablenotemark{f}\\
92  & 331   & 770   &  18 20 21.77 &   -16 11 18.3&   B0      & no \\
28  & 150   &\nodata&  18 20 31.92 &   -16 11 38.6&   B0      & yes\\
57  & 269   & 736   &  18 20 25.59 &   -16 11 16.3&   B1      & yes\\
95  & 266   & 773   &  18 20 25.72 &   -16 08 59.1&   B1      & no \\
101 & 206   & 779   &  18 20 29.04 &   -16 11 10.5&   B1      & yes\\
97  & 204   & 775   &  18 20 29.31 &   -16 09 43.3&   B1      & yes\\
100 & 159   & 778   &  18 20 31.25 &   -16 09 29.9&   B1      & yes\\
45  &\nodata&\nodata&  18 20 35.6  &   -16 10 53  &   B1      & no \\
33  & 324   & 714   &  18 20 22.19 &   -16 09 19.5&   B2      & no \\
24  & 275   &\nodata&  18 20 25.11 &   -16 10 26.7&   B2      & no \\
49  & 268   &\nodata&  18 20 25.35 &   -16 10 19.2&   B2      & no \\
48  & 248   &\nodata&  18 20 26.31 &   -16 10 16.4&   B2      & no \\
90  & 241   & 768   &  18 20 26.91 &   -16 10 58.8&   B2      & yes\\
85  & 223   & 763   &  18 20 28.29 &   -16 10 49.8&   B2      & yes\\
84  & 207   & 762   &  18 20 29.01 &   -16 10 42.5&   B2      & no \\
96  &\nodata& 774   &  18 20 29.2  &   -16 09 39  &   B2      & no \\
89  & 190   & 767   &  18 20 29.88 &   -16 10 11.5&   B2      & no \\
99  & 152   & 777   &  18 20 31.83 &   -16 09 46.2&   B2      & yes\\
17  & 336   & 269   &  18 20 21.40 &   -16 11 40.7&   B3      & yes\\
93  & 337   & 771   &  18 20 21.43 &   -16 10 41.2&   B3      & yes\\
35  & 333   & 716   &  18 20 21.54 &   -16 09 39.4&   B3      & yes\\
75  & 308   & 754   &  18 20 22.98 &   -16 08 14.8&   B3      & no \\
94  & 299   & 772   &  18 20 23.76 &   -16 10 35.8&   B3      & no \\
91  & 282   & 769   &  18 20 24.76 &   -16 11 09.0&   B3      & yes\\
83  & 267   & 761   &  18 20 25.63 &   -16 10 54.1&   B3      & yes\\
14  & 254   & 703   &  18 20 26.09 &   -16 10 51.5&   B3      & no \\
26  & 253   &\nodata&  18 20 26.13 &   -16 11 05.1&   B3      & yes\\
74  & 245   & 753   &  18 20 26.77 &   -16 08 22.9&   B3      & no \\
46  & 230   & 726   &  18 20 27.89 &   -16 11 02.5&   B3      & yes\\
47  & 227   & 727   &  18 20 28.07 &   -16 11 09.3&   B3      & no \\
65  & 220   & 744   &  18 20 28.32 &   -16 09 58.5&   B3      & no \\
27  & 197   & 708   &  18 20 29.93 &   -16 11 37.6&   B3      & yes\\
44  &\nodata&\nodata&  18 20 30.4  &   -16 10 05  &   B3      & no \\
50\tablenotemark{g}&\nodata& 730   &  18 20 26.35 & -16 10 19.0 & B3  & no \\
58\tablenotemark{g}&\nodata&\nodata&  18 20 22.45 & -16 10 17.3 & B3  & no \\
81\tablenotemark{h}&\nodata&\nodata&  \nodata & \nodata & B3  & \nodata\\
82\tablenotemark{g}&\nodata& 760   &  18 20 25.56 & -16 10 55.1 & B3  & yes\\
\enddata

\tablenotetext{a}{CEN = \citet{Chini80}; B = \citet{Hanson97}
derived from \citet{Bumgardner92}; OI =  \citet{Ogura76}.}

\tablenotetext{b}{Positions are from \citet{Hanson97, Chini98}, or SIMBAD.}

\tablenotetext{c}{Spectral types from \citet{Hanson97} or
\citet{Chini80}; assumed to be Luminosity Class V.}

\tablenotetext{d}{The O4--O4 binary Kleinmann's Anonymous Star; see \S\ref{review_M17.sec}.}

\tablenotetext{e}{BD -16$^\circ$4818}

\tablenotetext{f}{Source lies in gap between ACIS CCD chips.}

\tablenotetext{g}{Positions provided by R.\ Chini, private comm.}

\tablenotetext{h}{Position unavailable, so no X-ray match can be made.}

\end{deluxetable}

\newpage

\begin{deluxetable}{cccllcc}
\centering
\tabletypesize{\scriptsize}
\tablewidth{0pt}
\tablecolumns{7}
\tablecaption{O and Early B stars in NGC~2244 (Rosette cluster)
\label{tab:rosette_OB}}

\tablehead{
\multicolumn{3}{c}{Name\tablenotemark{a}} &
\colhead{R.A.\ (J2000)\tablenotemark{b}} &
\colhead{Dec (J2000)\tablenotemark{b}} &
\colhead{Spectral Type\tablenotemark{c}} &
\colhead{ACIS}  \\ \cline{1-3}
\colhead{HD/BD} &
\colhead{J} &
\colhead{OI} &
\colhead{} &
\colhead{} &
\colhead{} &
\colhead{detected?} 
}

\startdata

46223           & 3     & 203     & 06 32 09.32          & +04 49 24.9        &   O4 V((f))$\heartsuit$   & yes         \\
46150           & 2     & 122     & 06 31 55.53          & +04 56 34.5        &   O5 V((f))$\heartsuit$   & yes         \\
46485		&\nodata& 387	  & 06 33 50.96$\bowtie$ & +04 31 31.6$\bowtie$&  O7 V $\diamond$	  & yes\tablenotemark{e}\\
46056           & 7     &  84     & 06 31 20.88          & +04 50 04.0        &   O8 V((f))               & yes         \\
46149           & 4     & 114     & 06 31 52.54          & +05 01 59.4        &   O8.5 V((f)              & yes         \\
258691          &\nodata& 376     & 06 30 33.33          & +04 41 27.9        &   O9 V((f))               & not in FoV  \\
46202           & 6     & 180     & 06 32 10.48          & +04 58 00.1        &   O9 V((f))               & yes         \\ 
259238		&19	& 239	  & 06 32 18.21$\diamond$& +05 03 21.7$\diamond$& B0 V $\diamond$	  & not in FoV	\\
46106           & 5     & 115     & 06 31 38.40          & +05 01 36.6        &   B0.2 V                  & yes         \\
$\star$         &\nodata& \nodata & 06 31 37.10          & +04 45 53.7        &   B0.5 V                  & not in FoV  \\
259135$\dagger$ & 8     & 200     & 06 32 00.62          & +04 52 41.1        &   B0.5 V\tablenotemark{d} & yes	        \\
$\star \spadesuit$&\nodata&\nodata& 06 33 37.51          & +04 48 47.0        &   B0.5 V                  & not in FoV  \\
259012          & 9A    &  80     & 06 31 33.48          & +04 50 40.0        &   B1 V                    & yes         \\
259105          & 10    & 128     & 06 31 52.02          & +04 55 57.5        &   B1 V                    & no          \\
+04$^\circ$1299s& 11    & 201     & 06 32 06.15          & +04 52 15.6        &   B1 III                  & yes         \\
46484		&\nodata& 389	  & 06 33 54.41$\bowtie$ & +04 39 44.6$\bowtie$&  B1 V $\diamond$	  & not in FoV	\\
+05$^\circ$1281B& 14	& 193	  & 06 31 58.95          & +04 55 40.1        &   B1.5 V                  & yes	        \\
259172          & 16    & 167     & 06 32 02.59          & +05 05 08.9        &   B2 V                    & not in FoV  \\
\nodata		&\nodata& 345	  & 06 33 06.58$\diamond$& +05 06 03.4$\diamond$& B2 $\Box$		  & not in FoV	\\
+04$^\circ$1295p&15     &  79     & 06 31 31.48          & +04 51 00.0        &   B2.5 V                  & no          \\
\nodata         & 27    & 130     & 06 31 47.90          & +04 54 18.3        &   B2.5 V$\diamond$        & no          \\ 
\nodata         & 23    & 190     & 06 31 58.92$\diamond$& +04 56 16.1$\diamond$& B2.5 Vn$\diamond$       & no          \\
\nodata         & 22    & 172     & 06 32 09.9$\triangle$& +05 02 14$\triangle$&   B2.5 V$\diamond$        & no          \\
$\star$         &\nodata& 274	  & 06 32 24.24          & +04 47 04.0        &   B2.5 V                  & no\tablenotemark{f}	\\
$\star$         &\nodata& 392     & 06 33 50.56          & +05 01 37.8        &   B2.5 V                  & not in FoV  \\
\nodata		&33	& 194	  & 06 32 15.48 	 & +04 55 20.2	      &   B3 $\Box$		  & yes		\\
$\star$         &\nodata&\nodata  & 06 32 22.49          & +04 55 34.4        &   B3 V                    & yes         \\
259268		&20	& 241	  & 06 32 23.05$\diamond$& +05 02 45.7$\diamond$& B3 $\Box$		  & not in FoV	\\
259300          & 17    & 253     & 06 32 29.40          & +04 56 56.3        &   B3 Vp$\diamond$         & yes         \\
\nodata		&\nodata& 334	  & 06 32 51.79$\diamond$& +04 47 16.1$\diamond$& B3 $\Box$		  & yes\tablenotemark{f}\\
$\star$         &\nodata&\nodata  & 06 33 10.16          & +04 59 50.2        &   B3 V                    & not in FoV  \\
\enddata

\tablenotetext{a}{J = \citet{Johnson62}; OI = \citet{Ogura81};
$\star$ = \citet{Massey95};  $\dagger$ = V578 Mon; $\spadesuit$ =
IRAS 06309+0450.}

\tablenotetext{b}{Positions from \citet{Massey95} except those marked
with $\bowtie$ from {\it Hipparcos}, $\diamond$ from \citet{Ogura81}, and 
$\triangle$ from \citet{Park02}.}

\tablenotetext{c}{Spectral types from \citet{Massey95} except those
marked with $\heartsuit$ from \citet{Walborn02}, $\diamond$ from \citet{Ogura81}, and $\Box$ from \citet{Kuznetsov00}.}

\tablenotetext{d}{Double star}

\tablenotetext{e}{Observed in Rosette Field 4.}

\tablenotetext{f}{Observed in Rosette Field 2.}

\end{deluxetable}

\end{document}